\documentclass[aps, pre,10pt,singlecolumn,superscriptaddress,amsmath,amssymb,nofootinbib]{revtex4-2} 
\pdfoutput=1
\usepackage{silence}
\usepackage{amsmath,natbib, amsfonts}
\usepackage{amssymb,color, float}
\usepackage{graphicx}
\usepackage[centerlast]{caption}
\usepackage{subfig}
\usepackage[normalem]{ulem}
\usepackage{tikz}
\usepackage{listings}
\usepackage[colorlinks=true, citecolor=blue, urlcolor = blue, linkcolor= red,bookmarks=true]{hyperref}
\usepackage{multirow}
\usepackage{diagbox}
\usepackage[normalem]{ulem}
\usepackage{multirow}
\newcommand{\hh}{\mathcal{H}}
\newcommand{\bs}[1]{\boldsymbol{#1}}
\usepackage[toc,page]{appendix}

\begin{document}

\title{Optimal constrained control for generally damped Brownian heat engines}
\author{Monojit Chatterjee}
\affiliation{Department of Physics, Indian Institute of Technology Delhi, Hauz Khas 110 016, New Delhi, India}
\author{Viktor Holubec}
\email{viktor.holubec@mff.cuni.cz} 
\affiliation{Department of Macromolecular Physics, Faculty of Mathematics and Physics, Charles University, 18000 Prague, Czech Republic}
\author{Rahul Marathe}
\email{maratherahul@physics.iitd.ac.in} \affiliation{Department of Physics, Indian Institute of Technology Delhi, Hauz Khas 110 016, New Delhi, India}

\date{\today}
\begin{abstract}
Optimization of cyclic stochastic heat engines, a topic spanning decades of research, commonly assumes fixed control or response parameters at discrete points in the cycle—a limitation that often leads to experimentally impractical protocols. We overcome this with a general algorithm, adapted from optimal control theory, that optimizes full-cycle dynamics under realistic constraints, such as stiffness and temperature bounds, across diverse systems. Unlike geometric or mass transport methods, which rely on fixed endpoints and are unsuitable for unconstrained cycles, our approach simultaneously tunes both cycle time and control variations. Applied to a generally damped Brownian particle in a harmonic potential—an experimentally relevant case—our method is validated in the overdamped regime and extended to arbitrary damping rates. As damping decreases, maximum power vanishes and cycle time diverges; at fixed cycle times, efficiency follows a similar trend, with optimal protocols exhibiting non-monotonic complexity. Notably, optimizing temperature profiles—often overlooked—significantly enhances efficiency in intermediate damping regimes. Our work establishes the first systematic framework for optimizing cyclic stochastic processes under experimental constraints, broadening the scope of power and efficiency optimization in nonequilibrium thermodynamics.
\end{abstract}

\maketitle

\section{Introduction}

The conversion of abundant thermal energy into useful work by heat engines has driven technological and societal advancements for over two centuries. While the macroscopic principles of heat engines are well established, recent experimental advancements have facilitated the development of classical~\cite{ColloidalReview} and quantum~\cite{Quantum_engines} microscopic heat engines, providing new opportunities to explore nonequilibrium energy conversion at the microscale. Best-known examples~\cite{micro_engine,Edgar2016,Sood2016} are cyclic colloidal engines based on 
single Brownian particles confined by harmonic potentials created by optical tweezers~\cite{optical,Deng_2007,optical_power_spectrum} and linear Paul traps ~\cite{SingelAtomHE2016, underdamped_realization_2025}.

In addition to enabling the study of fluctuations in thermodynamic quantities~\cite{Holubec_2022}, realizations of microscopic heat engines offer a remarkable degree of control, allowing for the precise optimization of their performance. This optimization has become a vibrant area of research, integrating well-established techniques in finite-time thermodynamics~\cite{berry2000thermodynamic, Bellman}, leveraging optimal transport theory~\cite{optimal_transport,opt_proto_transport}, and exploring innovative approaches such as thermodynamic geometric optimization~\cite{Crooks2012, Geometric_Optimization}, shortcuts to adiabaticity~\cite{Shortcuts_to_adiabaticity}, and machine-learning-based methods~\cite{control_active}. In simple scenarios, optimal protocols can also be deduced through intuitive geometric reasoning~\cite{Viktor2022}. Despite their microscopic scale, the average thermodynamic behavior of these engines adheres to the constraints of classical thermodynamics. Consequently, the thermodynamic performance of the resulting optimal cycles is likely close to the currently insufficiently explored theoretical bounds for general finite-time heat engines.

Without imposing constraints on control parameters, microscopic heat engines described by idealized models, such as overdamped Langevin dynamics~\cite{HolubecMaxPower2018} or quantum dots~\cite{Polettini_2017}, can theoretically operate near Carnot efficiency while achieving diverging output power in the limit of infinitely fast relaxation. Even though current experimental setups approach these idealized conditions~\cite{HolubecMaxPower2018}, achieving them exactly is impossible in practice. Furthermore, the idealized models often break down in the relevant parameter regimes. It is thus crucial to account for experimental limitations when optimizing the performance of a given engine.

Most of the available studies on the optimal control of cyclic microscopic heat engines assume a Carnot-type protocol for temperature, consisting of two isotherms and two adiabats, and further constrain the state of the working medium at the initial and final points of the isotherms~\cite{Schmiedl-Seifert2007}. This approach simplifies the optimization process by reducing it to two independent problems for the individual isotherms, which can be solved using optimal transport theory~\cite{optimal_transport, viscoelastic_transport_2024} or thermodynamic geometry~\cite{Crooks2012, Geometric_Optimization,Blaber_2023}. Analytical solutions for engines based on harmonically confined overdamped Brownian particles reveal that maximum power protocols feature infinitely fast variations in the stiffness of the confining potential~\cite{Schmiedl-Seifert2007, Then2008}. These jumps appear to be a universal characteristic of optimal control protocols in microscopic engines. 
\begin{figure}[t]
    \centering
    \includegraphics[width=0.7\linewidth]{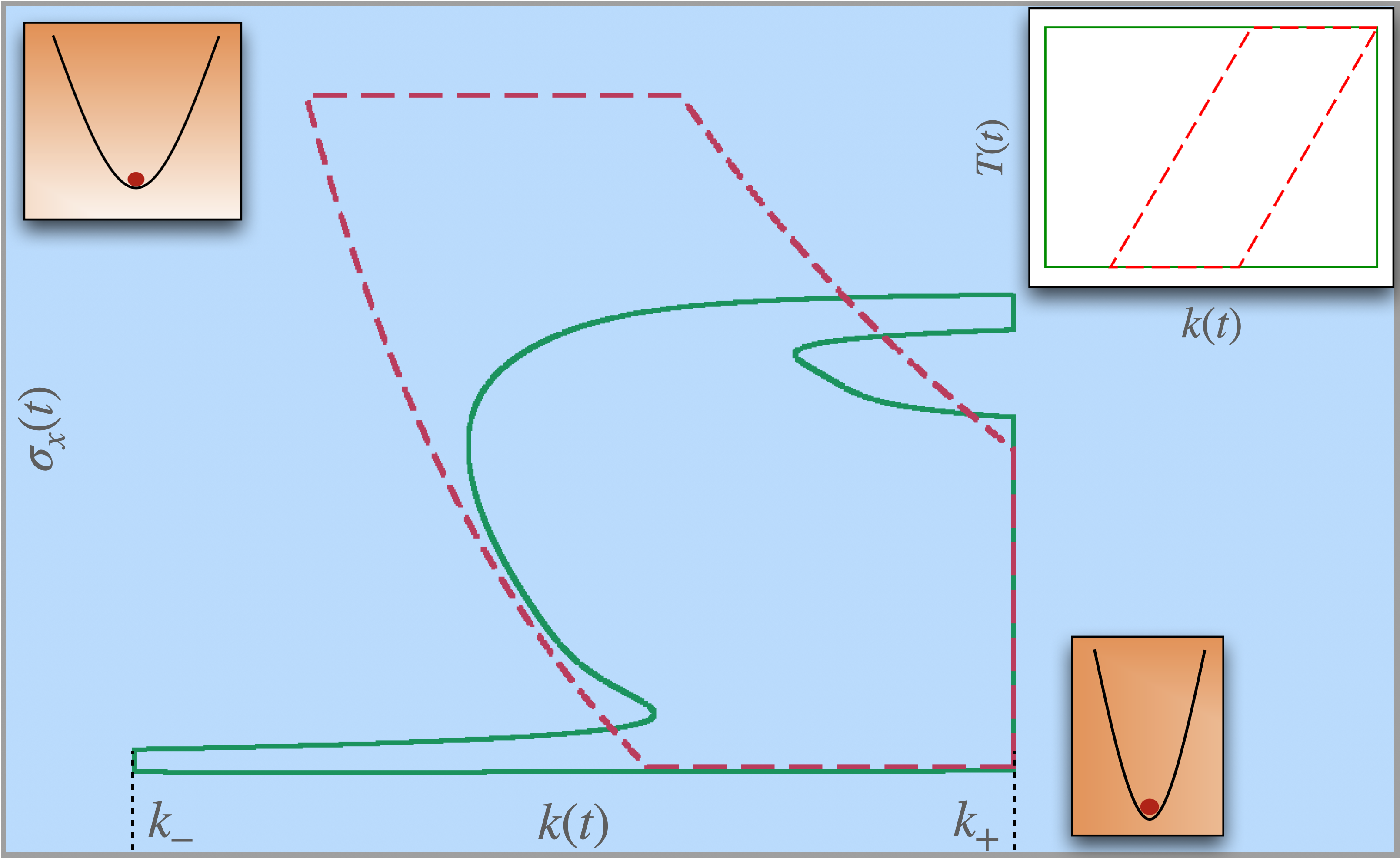}
    \caption{\textbf{Brownian heat engine.} The working medium of the engine is a generally damped Brownian particle with mass $m$, subjected to a harmonic potential with time-periodic stiffness, $k(t)$, and immersed in an equilibrium bath at a time-periodic temperature, $T(t)$. We calculate the optimal cycle time $\tau$ and protocols for $k(t)$ and $T(t)$ that maximize the output power or efficiency of the engine under the constraints $k(t) \in [k_-, k_+]$ and $T(t) \in [T_-, T_+]$. The figure shows the stiffness-variance and stiffness-temperature (inset) diagrams corresponding to the maximum power cycles with $\tau=4$ in the generally damped regime ($\gamma=2$, green solid line) and in the strongly overdamped regime (red dashed line). The work done by the engine is represented by the area enclosed by the stiffness-variance diagram. The strongly non-convex shape of the diagram in the general damping regime is a universal feature that enables optimal work extraction in this regime. For more details regarding the two cycles, see Figs.~\ref{fig:over_pow} and~\ref{full_pow}.
}
\label{fig:model} 
\end{figure}

However, while both involve jumps, optimal protocols derived under experimentally relevant constraints—such as limits on temperature differences or stiffness strengths~\cite{Plata2019, Viktor2022, MAJUMDAR2025130278}—often differ significantly from those based on fixed states~\cite{Schmiedl-Seifert2007, Then2008}. For overdamped Brownian and single-spin heat engines, both the maximum efficiency protocol and, for small allowed variations in the control parameter $k_+ - k_-$, the maximum power protocol are piecewise constant~\cite{Viktor2022, MAJUMDAR2025130278}, with the corresponding efficiency given by $1 - k_-/k_+$. An exception is the maximum power protocol for potential stiffness in extremely underdamped Brownian heat engines, which is exponential and achieves the Curzon-Ahlborn efficiency, irrespective of the constraints~\cite{Dechant_2017}.

A notable exception of a work where the reservoir temperature protocol and system state have not been fixed is Ref.~\cite{optimal_geodesic_brownian}, where the authors optimized efficiency of finite-time generally damped Brownian heat engines under the constraint that temperature and stiffness are specified at four distinct points during the cycle. These constraints still allow for the application of geometric methods. While the authors employed a non-standard definition of efficiency, which measures the departure of the working medium from equilibrium rather than the efficiency of heat-to-work conversion~\cite{Brandner2020}, it is notable that the resulting optimized temperature profiles were not Carnot-type.

In the current manuscript, we employ standard functional optimization techniques~\cite{berry2000thermodynamic, Bellman} to numerically find the functional forms and periods of cyclic maximum power and maximum efficiency protocols for trap stiffness and temperature in Brownian heat engines based on a harmonically confined, generally damped Brownian particle. The optimization is performed under the constraints of minimum and maximum values for trap stiffness and bath temperature, as shown in Fig.~\ref{fig:model}. We thus do not fix the state of the system or control parameters at any given time instant during the cycle. This means the optimization problem solved here does not lend itself to geometric or optimal transport methods, which are typically designed to optimize protocols that map one distribution to another.

Our work conveys four main messages. First, we demonstrate that optimization techniques routinely applied to macroscopic heat engines~\cite{Craun2015, symmetry2021} can also solve practical optimization problems for microscopic engines, even under experimentally relevant constraints that render widely used methods, such as optimal mass transport or thermodynamic geometry, unfeasible. For the first, albeit limited, such application of the method to optimization of a model of a microscopic spin engine, see Ref.~\cite{MAJUMDAR2025130278}. Second, we use these powerful techniques to explore, for the first time without relying on approximations, how the optimal protocols and the corresponding performance of the engine change as the damping of the Brownian particle varies between the deeply overdamped and underdamped regimes~\cite{Dechant_2017, Viktor2022}. In this respect, we find that both maximum power and efficiency decrease as the system enters the underdamped regime by reducing the damping rate. This shows that the recently described enhancement in the performance of Brownian collision heat engines in the underdamped regime, due to the employment of resonances~\cite{collisional_engines}, is not possible in cyclically driven heat engines, where both overdamped~\cite{micro_engine, Edgar2016, Sood2016} and underdamped~\cite{cantilever} regimes are experimentally realizable.
Third, our findings emphasize the importance of optimizing the temperature protocol to maximize power and, in particular, efficiency in the regime of general damping.
Finally, we emphasize the influence of the employed constraints on the resulting optimal protocol. This is particularly important because related studies often do not explicitly specify the conditions under which their optimizations are performed, making it difficult to assess to which experimental situations their results apply.

The rest of the manuscript is structured as follows: In Sec.~\ref{sec:model}, we describe the model and its corresponding dynamical equations, along with the numerical values of the model parameters used throughout the paper. Section~\ref{sec:thermo} presents the definitions of output power and efficiency of the engine in the fully overdamped and generally damped regimes. In Sec.~\ref{sec:opt_procedure}, we review the employed optimization procedure and the corresponding numerical method, as described in Refs.~\cite{Craun2015, symmetry2021}. As a proof of concept, in Sec.~\ref{sec:verification}, we demonstrate that the maximum power and maximum efficiency protocols obtained from our numerical method in the fully overdamped regime agree with the corresponding analytical results \cite{Viktor2022}. In Sec.~\ref{sec:underdamped}, we present our main results, i.e., maximum power and maximum efficiency protocols for the generally damped regime, along with dependencies of maximum power and efficiency on the friction coefficient. We conclude in Sec.~\ref{sec:conclusion}.

\section{Model Dynamics and Investigated Parameter Regime}
\label{sec:model}
We consider a heat engine based on a generally damped Brownian particle with mass $m$, confined by a harmonic trap with time-periodic stiffness and immersed in a heat bath with time-periodic temperature. The cycle time is $\tau$. Our goal is to determine the optimal time variation of stiffness and temperature, including the cycle time, to maximize the output power or efficiency of the engine. To achieve this technically, it is convenient to measure time in units of $\tau$. In dimensionless time $t$, the trap stiffness $k(t)$ and temperature $T(t)$ are periodic functions with period one. The value of a variable $A(t\tau)$ at the standard time $t\tau$ (measured in seconds) is expressed in terms of its representation in the dimensionless time as $\tilde{A}(t)$.

Due to the symmetry of the problem, it suffices to model the particle in one dimension, considering its position, $x(t)$, and velocity, $v(t)$. In the dimensionless time, these coordinates obey the system of Langevin equations 
\begin{eqnarray}
\tau^{-1} \dot{x}(t) &=& v(t),\label{langevinx}\\
m\tau^{-1} \dot v(t) &=& -m\gamma v(t) - k(t) x(t) + \sqrt{2m\gamma \tau^{-1} k_B T(t)} ~\xi(t),\label{langevinv}
\end{eqnarray}
where $\gamma$ is the damping rate, $k_B$ Boltzmann constant, and $\xi(t)$ is a unit variance, unbiased, Gaussian white noise. 

As we will show in the next section, the average thermodynamic performance of the engine depends solely on the moments 
$\sigma_x(t) = \langle x^2(t) \rangle$, $\sigma_v(t) = \langle v^2(t) \rangle$, and the cross-correlation $\sigma_{xv}(t) = \langle x(t)v(t) \rangle$ of position and velocity. The corresponding dynamical equations can be derived from \eqref{langevinx} and \eqref{langevinv} by multiplying the equations by $x(t)$ or $v(t)$ and using the results $\langle x(t) \xi(t) \rangle = 0$ and $\langle v(t) \xi(t) \rangle = 2 \gamma k_B T(t)/m$~\cite{Dechant_2017}. Introducing a column vector $\boldsymbol{\sigma}(t) = \left[\sigma_x(t), \sigma_{xv}(t), \sigma_v(t) \right]^\intercal$, the equations read
\begin{equation}
   \dot{ \bs{ \sigma}}(t) =\tau \begin{bmatrix}2 \sigma_{xv}(t) \\\sigma_v(t)-k(t)\sigma_x(t)/m-\gamma\sigma_{xv}(t)\\
    -2\gamma\sigma_v(t)-2k(t)\sigma_{xv}(t)/m + 2\gamma k_B T(t)/m\end{bmatrix}.
\label{eq:Sigma}
\end{equation}

We assume that the heat engine operates in a time-periodic steady state, attained upon the periodic driving at long times, starting from arbitrary initial conditions. In this state, the moments in \( \bs{\sigma}(t) \) are also time-periodic functions with period one.

Neglecting the stochastic noise term, Eqs.~\eqref{langevinx} and \eqref{langevinv} describe a damped harmonic oscillator. The solutions are linear combinations of $\exp(- \lambda_\pm t)$, with the eigenvalues 
\begin{equation}
\lambda_\pm = \frac{\tau}{\tau_v} \left(1 \pm \sqrt{1 - \omega^2 \tau_v^2} \right).
\label{eq:eigenvalues}
\end{equation}
Above, $\tau_v = 2 / \gamma$ is the velocity relaxation time in the limit of shallow potentials ($k \to 0$), and $\omega = \sqrt{k/m}$ is the natural angular frequency of oscillations in the absence of damping ($\gamma \to 0$). The working medium's dimensionless relaxation times are given by the real parts of $\lambda_\pm$, i.e., by $\tau_v/\tau$ for $\omega^2 \tau_v^2 > 1$ and by $\lambda_\pm^{-1}$ otherwise.

Most experimental realizations of Brownian heat engines have been achieved in the colloidal domain~\cite{ColloidalReview}, corresponding to the regime $\omega \tau_v \ll 1$ of strongly overdamped oscillations. In this limit, one can assume that velocity degrees of freedom relax so fast that they are in thermodynamic equilibrium corresponding to the instantaneous bath temperature at all times, i.e., $\sigma_v(t) = k_B T(t)/m$. Imposing these conditions in the system~\eqref{eq:Sigma}, we find that the cross moment $\sigma_{xv}(t)$ relaxes on time scale $2 \omega \tau_v \ll 1$ to $[k_B T(t) - k(t) \sigma_x(t)]/(\gamma m)$ and thus
\begin{equation}
    m \gamma \tau^{-1} \dot \sigma_x(t)=-2 k(t) \sigma_x(t) + 2 k_B T(t).
    \label{eq:over0}
\end{equation}
We will use these equations to validate our optimization algorithm against the known optimal protocols in this limit~\cite{Viktor2022}. Optimal protocols for the opposite limit $\omega \tau_v \gg 1$, where the oscillations are strongly underdamped, are analytically derived in Ref.~\cite{Dechant_2017}. 

While we do not treat the strongly underdamped limit explicitly in this work, we study the behavior of maximum power and maximum efficiency obtained by our optimization procedure as the complete model~\eqref{eq:Sigma} transition from the overdamped to the underdamped regime. Since this transition is controlled by the parameter $\omega \tau_v \propto \sqrt{k/m \gamma^2}$, there are three physically distinct ways to achieve it. The first involves varying the drag coefficient. According to Stokes' law, $m \gamma = 6 \pi \mu R$, this can be achieved by altering the particle radius $R$ or the dynamic viscosity $\mu$ of the surrounding fluid. For example, the deeply overdamped and underdamped regimes correspond to the viscosities of water and a rarefied gas, respectively. 
The second method is to modify the trap stiffness, $k$. However, this approach has limitations, which actually motivated the constraint $k \in [k_-,k_+]$ on stiffness we impose in this paper. When the potential is realized using optical tweezers: high trap stiffness requires intense laser power, which may lead to evaporation of the trapped particles. Conversely, low stiffness values can result in particles escaping the trap due to thermal motion, as the infinite harmonic well only approximates real, finite-size traps near their minima. The third method involves changing the particle's mass, $m$, while keeping the drag coefficient constant. This is the least versatile option, as the range of accessible material densities is limited, and changing the particle's volume would simultaneously alter $m \gamma$. In the present paper, we use this last option as it ensures a well-defined reference point in the overdamped dynamics. 

Specifically, we set \(m\gamma\) and \(k_B / m\gamma\) numerically equal to one. Dividing Eq.~\eqref{langevinv} by $m\gamma R$ to obtain,  
\begin{equation}
\frac{\tau^{-1}}{\gamma} \dot{v}(t) = - v(t) - \frac{k(t)}{m\gamma} x(t) + \sqrt{2 \tau^{-1} \frac{k_B T(t)}{m\gamma R^2}} ~\xi(t),
\end{equation}
where the chosen length unit is the particle radius \(R\), this amounts to reducing the stiffness by \(m\gamma\) and the temperature by \(m\gamma R^2 / k_B\). The reduced temperature \(\frac{k_B T(t)}{m\gamma R^2} = \frac{k_B T(t)}{6\pi \mu R^3}\)\ and stiffness $\frac{k}{m\gamma}$ have units of inverse time (since we fixed $k_B$ to one, and we vary cycle time $\tau$, stiffness $k$, and damping rate $\gamma$ throughout the study, there is no natural free parameter that would allow us to make the control parameters dimensionless).

From now on, we will denote the reduced control parameters simply as \(T\) and \(k\). 
For a Brownian particle with a radius of \(0.6 \, \mu \text{m}\) in water with a dynamic viscosity of \(\mu \approx 10^{-3} \, \text{Pa\,s}\), the value of the reduced temperature \(T = 1\) corresponds to \(T \approx 290 \, \text{K}\), which is approximately the room temperature. For the same parameters, the reduced stiffness \(k = 1\) corresponds to \(k \approx 0.02 \, \text{pN}/\mu\text{m}\), which is within the range achievable by optical tweezers.  
Since we set the Boltzmann constant to unity, we express energy in units of temperature throughout the paper. In all subsequent figures, we show the reduced temperatures and stiffness. Length and time are measured in units of the particle radius, $R$, and the cycle time, $\tau$, respectively. The cycle time is measured in seconds. In the numerical examples below, without loss of generality, we always set the range of allowed temperatures to $[T_-, T_+] = [1,4]$. For stiffness, we either use the tight range $[k_-, k_+] = [0.4,0.45]$ or the range $[k_-, k_+] = [0.2,0.8]$.  

~\\
\section{Thermodynamic definitions}\label{sec:thermo}
Let us now define the output power and efficiency of the engine. We use the standard stochastic thermodynamic definitions of heat and work, based on the decomposition of the change in the internal energy of the particle into two components~\cite{Jarzynski1997}. The component proportional to the change in the externally controlled parameter (stiffness) is identified as the work flux into the system, while the remainder is identified as the heat flux into the system. 

The internal energy of a particle in a harmonic trap is given by the sum of its (average) potential and kinetic energy: $U = \frac{1}{2}k(t) \sigma_x(t) + \frac{1}{2} m \sigma_v(t)$. Its time derivative is given by $\dot{U} = \frac{1}{2} \dot{k}(t) \sigma_x(t) + \frac{1}{2}k(t) \dot{\sigma}_x(t) + \frac{1}{2} m \dot{\sigma}_v(t)$. Consequently, the average work done by the engine per unit time is identified as
\begin{equation}
    \dot W =-\frac{1}{2} \dot k(t) \sigma_x (t) .\label{work_defn}
\end{equation}
The average heat flux into the engine then reads
\begin{align}
    \dot Q = \frac{1}{2}  k(t) \dot{\sigma}_x (t) + \frac{1}{2}  m \dot{\sigma}_v (t) = \tau(\gamma T(t) - m \gamma \sigma_v(t)), \label{eq:heat_defn_under}
\end{align}
where we used Eq.~\eqref{eq:Sigma} to obtain the last equality. \footnote{From Eqs.~\eqref{langevinx} and \eqref{langevinv}, it follows that the total force exerted by the bath on the Brownian particle is given by \( F_{\rm B} = -m\gamma v(t) + \sqrt{2m \gamma k_B T}~\xi(t) = m\dot{v}(t) + k(t) x(t) \). The stochastic power delivered to the Brownian particle by the heat bath---commonly referred to as the heat flux---is thus \( F_{\rm B} v(t) = \left[ m\dot{v}(t) + k(t) x(t) \right] v(t) = \frac{d}{dt} \left[ \frac{1}{2} m v(t)^2 + \frac{1}{2} k(t) x(t)^2 \right] - \frac{1}{2} \dot{k}(t) x(t)^2 \). Here, the quantity in parentheses represents the total energy of the Brownian particle, while the last term denotes the energy flux extracted from the system by the time-dependent external driving, commonly referred to as the output work flux. Taking ensemble averages of these expressions yields our definitions of work flux and heat flux in Eqs.~\eqref{work_defn} and \eqref{eq:heat_defn_under}.}

The average work done by the engine per cycle is given by
\begin{equation}
W = -\frac{1}{2} \int_0^1 dt\, \dot{k}(t) \sigma_x(t).
\label{eq:Wout}
\end{equation}
and equals the output power of the engine measured in dimensionless time. The output work does not depend on the chosen unit of time. We will always show the output power in physical units of energy per second. It reads 
\begin{equation}
P = \frac{W}{\tau}.
\label{eq:power}
\end{equation}

We define the efficiency $\eta$ of the engine 
as the ratio of work output, $W$, to the heat absorbed from the bath, $Q_+$,
\begin{equation}
\eta = \frac{W}{Q_+}.
\label{eq:efficiency}
\end{equation}
The efficiency $\eta$ thus measures how much of the disordered energy taken from the hot bath is transformed into work. In our definition, we exclude possible recuperation of energy released to the bath during some parts of the cycle~\cite{wiese2024,Ginot2024}. Without energy recuperation, the heat absorbed by the working medium from the bath is given by  
\begin{equation}  
    Q_+ = \int_0^1 dt \, \theta(\dot{Q}) ~\dot{Q},
    \label{eq:heat_it}
\end{equation}  
where the Heaviside function \( \theta(x) \) ensures integration over the portions of the cycle when the heat flux \( \dot{Q} \) into the system is positive.

When studying overdamped Brownian heat engines~\cite{Schmiedl-Seifert2007, micro_engine, wiese2024}, the kinetic contribution $m \dot{\sigma}_v(t)/2$ to the heat flux~\eqref{eq:heat_defn_under} is often neglected. In such cases, the input heat flux is approximated as  
\begin{align}
    \dot{Q}_o = \frac{1}{2} k(t) \dot{\sigma}_x(t). \label{eq:heat_defn_over}
\end{align}  
The corresponding efficiency, $\eta_{OD}$, is then calculated by substituting \( \dot{Q}_o \) for \( \dot{Q} \) in Eqs.~\eqref{eq:efficiency}, and \eqref{eq:heat_it}.

Assuming that heat flows into the system during the part of the cycle when the temperature increases from \( T_- \) to \( T_+ \), as in Carnot-type cycles, the heat absorbed from the bath calculated from the ``overdamped'' definition~\eqref{eq:heat_defn_over} differs in the overdamped regime~\eqref{eq:over0} from the correct heat~\eqref{eq:heat_it} by the change in the particle's average kinetic energy between the two temperatures~\cite{Schmiedl-Seifert2007}:  
\begin{equation}  
    Q_{\mathrm{leak}} = \frac{T_+ - T_-}{2}.  
\end{equation}  
This amount of heat is effectively wasted or ``leaked" in the overdamped limit~\cite{heat_leakage}, as the velocity does not influence the particle's position.

In Sec.~\ref{sec:verification}, we use the overdamped approximation of the dynamics along with the corresponding overdamped definition of efficiency to validate the optimization procedure introduced in the next section. Outside this section, we employ the correct definition of efficiency~\eqref{eq:efficiency}, which includes the kinetic contribution.

\section{Optimization Procedure}\label{sec:opt_procedure}

Let us now present the optimization procedure we employ. Consider a dynamical system described by the equation \( \dot{\bs{\sigma}}(t) = \bs{f}\left[\bs{\sigma}(t), \bs{u}(t)\right] \). The parameters \( \bs{\sigma}(t) \) represent the state of the system, while the parameters \( \bs{u}(t) \) are externally controlled. As in our case, we assume that the control parameters are cyclically modulated with period one and that the system operates in a time-periodic steady state, i.e., \( \bs{\sigma}(t) = \bs{\sigma}(t+1) \). Our aim is to maximize the functional 
$J = \int_0^1 \xi[\bs{\sigma}(t), \bs{u}(t)] \, dt$.
Applied to our heat engine model, \( \bs{\sigma} \) will be the vector of coordinate moments, \( \bs{u}(t) = [k(t), T(t)]^\intercal \), and \( J \) will represent the output power. For efficiency, the algorithm needs to be generalized for functionals involving a fraction of two integrals~\cite{symmetry2021}. We present this generalization in Sec.~\ref{overdamped_efficiency}.

According to the calculus of variations~\cite{Brunt_COV,Horn1967,Craun2015}, maximizing a functional \( J \) under the constraints $\dot{\bs{\sigma}} = \bs{f}$ and \( \bs{\sigma}(t) = \bs{\sigma}(t+1) \) can be achieved by maximizing the functional 
$\mathcal{J} = \int_0^1 \left( \xi - \bs{\lambda} \cdot (\dot{\bs{\sigma}} - \bs{f}) \right) dt$,
where \( \bs{\lambda} = \bs{\lambda}(t) \) is a vector of Lagrange multipliers. Variation of the functional \( \mathcal{J} \) with respect to \( \bs{\sigma} \) and \( \bs{u} \) reads $\delta \mathcal{J} = \int_0^1 dt\, \frac{\partial \hh}{\partial \boldsymbol{u}} \cdot \bs{\delta u} + \int_0^1 dt\, \left(\boldsymbol{\dot{\lambda}} + \frac{\partial \mathcal{H}}{\partial \boldsymbol{\sigma}} \right) \cdot \bs{\delta \sigma} - \left(\bs{\lambda}(1) - \bs{\lambda}(0)\right) \cdot \bs{\sigma}(0)$ with the ``Hamiltonian'' $\hh = \xi + \boldsymbol{\lambda} \cdot \boldsymbol{f}(t)$. 
Under the given conditions, $\delta \mathcal{J} = 0$ when
\begin{align}
    \boldsymbol{\dot{\sigma}}&=\frac{\partial \mathcal{H}}{\partial \boldsymbol{\lambda}}, \label{eq:opt0}\\
    \boldsymbol{\dot{\lambda}}&=-\frac{\partial \mathcal{H}}{\partial \boldsymbol{\sigma}}, \label{eq:opt1}\\
    0 &= \frac{\partial \mathcal{H}}{\partial \boldsymbol{u}}. \label{eq:optx}
\end{align}
The first equation follows from the constraint that $\bs{\sigma}(t)$ has to obey the dynamical equation $\dot{\bs{\sigma}} = \bs{f}$ and the remaining equations follow from the condition on the vanishing variation of $\mathcal{J}$. The first two equations have to be solved under the conditions $\bs{\sigma}(0) = \bs{\sigma}(1)$ and $\bs{\lambda}(0) = \bs{\lambda}(1)$. Interestingly, they describe a Hamiltonian flow induced by the Hamiltonian $\hh$ in the space of generalized position $\bs{\sigma}$ and momentum $\bs{\lambda}$.

Assuming that $\xi$ or $\bs{f}$ depend on additional parameters, such as the cycle time $\tau$ in our case, and we want to optimize $J$ also with respect to those, we get additional equations of the form
\begin{equation}
    \frac{\partial{\mathcal{J}}}{\partial \tau} = \int_0^1 \frac{\partial \hh}{\partial \tau} dt = 0.  
    \label{eq:Jtau}
\end{equation}

Solving Eqs.~\eqref{eq:opt0}--\eqref{eq:Jtau} would yield the unconstrained optimal time variation of control parameters $\bs{u}$ that maximize $J$. To find the optimal protocol under constraints imposed on $\bs{u}$, one can adjust the functional $J$ so that it becomes small whenever the control parameters violate the desired constraints~\cite{symmetry2021}. Instead, we solve the problem using a variant of the gradient descent algorithm~\cite{Craun2015,symmetry2021}, imposing the constraints on $\bs{u}$ by modifying components of the gradient that would guide the control parameters outside the allowed region. Specifically, we employ the following iterative algorithm:

\begin{enumerate}
   \item Initialize $\tau$, and then $\bs{u}(t)$, and $\bs{\lambda}(t)$ as arbitrary periodic functions obeying the prescribed constraints.
    \item Solve Eq.~\eqref{eq:opt0} for a time-periodic solution $\bs{\sigma}(t)$ by forward integration from an initial condition, given the current $\bs{u}(t)$ and $\bs{\lambda}(t)$.
    \item Solve Eq.~\eqref{eq:opt1} for a time-periodic solution $\bs{\lambda}(t)$ by backward integration~\cite{symmetry2021}, given the current $\bs{u}(t)$ and $\bs{\sigma}(t)$.
    \item Update the control parameters $\bs{u}(t)$ using the relation $\bs{u}(t) \to \bs{u}(t) + \varepsilon \delta\bs{ u}(t)$, where $\delta\bs{ u}(t) = \frac{\partial \hh}{\partial \boldsymbol{u}}$. This increases both $\mathcal{J}$ and $J$. In practice, it is useful to apply different learning rates $\varepsilon>0$ to different components of $\bs{u}$ to improve the convergence speed. If any component of $\bs{u}(t)$ exceeds its imposed bounds at certain times $t$, it is manually corrected to the nearest boundary value. For instance, if $u_1(t)$ must remain below $u_+$, we set $u_1(t) \to \min[u_1(t), u_+]$. Similar corrections are applied to other components of $\bs{u}$ using their constraints.
    \item Update the cycle duration $\tau$ using $\tau \to \tau + \varepsilon_2 \frac{\partial \hh}{\partial \tau}$, $\varepsilon_2>0$, ensuring a positive change in $J$.
    \item Repeat steps $2$--$5$ until the change in $J$ falls below a prescribed threshold. In our implementation, the algorithm is considered to have converged when the change in \( J \) is smaller than \( 10^{-8} \).
\end{enumerate}

In principle, the above algorithm might yield a local maximum of $J$. To mitigate this risk, we verify that the algorithm reaches the same value of $J$ for different initial conditions in step $1$.

In the following sections, we apply this general algorithm to specific problems. First, in the next section, we verify that it yields known optimal protocols for power and efficiency in the overdamped regime~\cite{Viktor2022}. In Sec.~\ref{sec:underdamped}, we apply the algorithm to obtain our main results: maximum power and efficiency protocols for temperature and stiffness in the generally damped situation.

\section{Verification of the Technique: Overdamped regime}
\label{sec:verification}

In the deeply overdamped regime, where the dynamics of the working medium of the engine can be approximated by Eq.~\eqref{eq:over0}, analytical predictions are available for optimal protocols for power and efficiency under constraints on the allowable variations of stiffness and temperature~\cite{Viktor2022}. Specifically, it has been found that, when the allowed variation of stiffness is small, the maximum power protocols for stiffness and temperature are piecewise constant, switching abruptly between the allowed maximum and minimum values at the same time instant. This same protocol also yields the maximum `overdamped' efficiency (the kinetic term in Eq.~\eqref{eq:heat_it} is neglected), regardless of the allowed ranges of the control parameters. In this section, we use these analytical results to verify the algorithm introduced above. We begin by optimizing the power output of the engine.

\subsection{Power Optimization}
\label{subsec:W_and_P}

When the functional $J$ in the algorithm of Sec.~\ref{sec:opt_procedure} is output power $P$ in Eq.~\eqref{eq:power}, we have $J = P=\int_0^1 \xi_{P} \, dt$, with 
\begin{equation}
\xi_{P} \equiv \xi_{P}[\sigma_x(t), k(t), T(t)] =\frac{\dot{W}}{\tau} = -\frac{1}{2\tau}\dot{k}(t)\sigma_x(t) =k(t)\left[T(t)-k(t)\sigma_x(t)\right],
\end{equation}
where we have used the definition of output work~\eqref{eq:Wout} and the overdamped dynamical equation~\eqref{eq:over0} for the position variance $\sigma_x(t)$. 

Our aim is now to find optimal time variation of temperature $T(t) \in [T-,T+]$ and stiffness $k(t) \in [k_-, k_+]$ in time interval $[0,\tau]$ as well as the optimal cycle duration $\tau$ that will maximize $P$ under the overdamped dynamics. We will achieve this by following the algorithm in the previous section. 

The Hamiltonian $\hh$ now reads
\begin{align}
    \hh = \xi_{P} + 2\tau \lambda(t) [T(t)-k(t)\sigma_x(t)] = [k(t)+2\tau\lambda(t)]~[T(t)-k(t) \sigma_x(t)]
\end{align}
and yields the following set of equations which needs to be solved iteratively according to the algorithm above to find the maximum power protocol:
\begin{align}
\dot \sigma_x(t) &= 2\tau[T(t)-k(t)\sigma_x(t)], \label{eq:over_opt_pow_sig}\\
\dot \lambda(t) &= k^2(t) + 2\tau \lambda(t)k(t), \\
\delta k(t) &= \varepsilon~[T(t) -2k(t)\sigma_x(t) - 2 \tau \lambda(t)\sigma_x(t)], \\
\delta T(t) &= \varepsilon'~[k(t) + 2\tau \lambda(t)], \label{eq:over_opt_pow_T}\\
d\tau &= \varepsilon'' \int_0^1 \lambda(t)[T(t)-k(t)\sigma_x(t)]~ dt. \label{eq:over_opt_pow_tau}
\end{align}
Here, $\varepsilon, \varepsilon'$, and $\varepsilon''$ act as learning rate parameters, which should be small enough for a proper convergence of the algorithm. We have used the values $\varepsilon=0.01,\varepsilon'=0.01, \varepsilon''=0.001$.

\begin{figure}[!tbp]
\centering
\begin{tabular}{ccc}
\hspace{-1cm}
\parbox[c][0.3\textwidth][t]{0.35\textwidth}{
    \raggedright \textbf{(a)}\\
    \includegraphics[width=\linewidth]{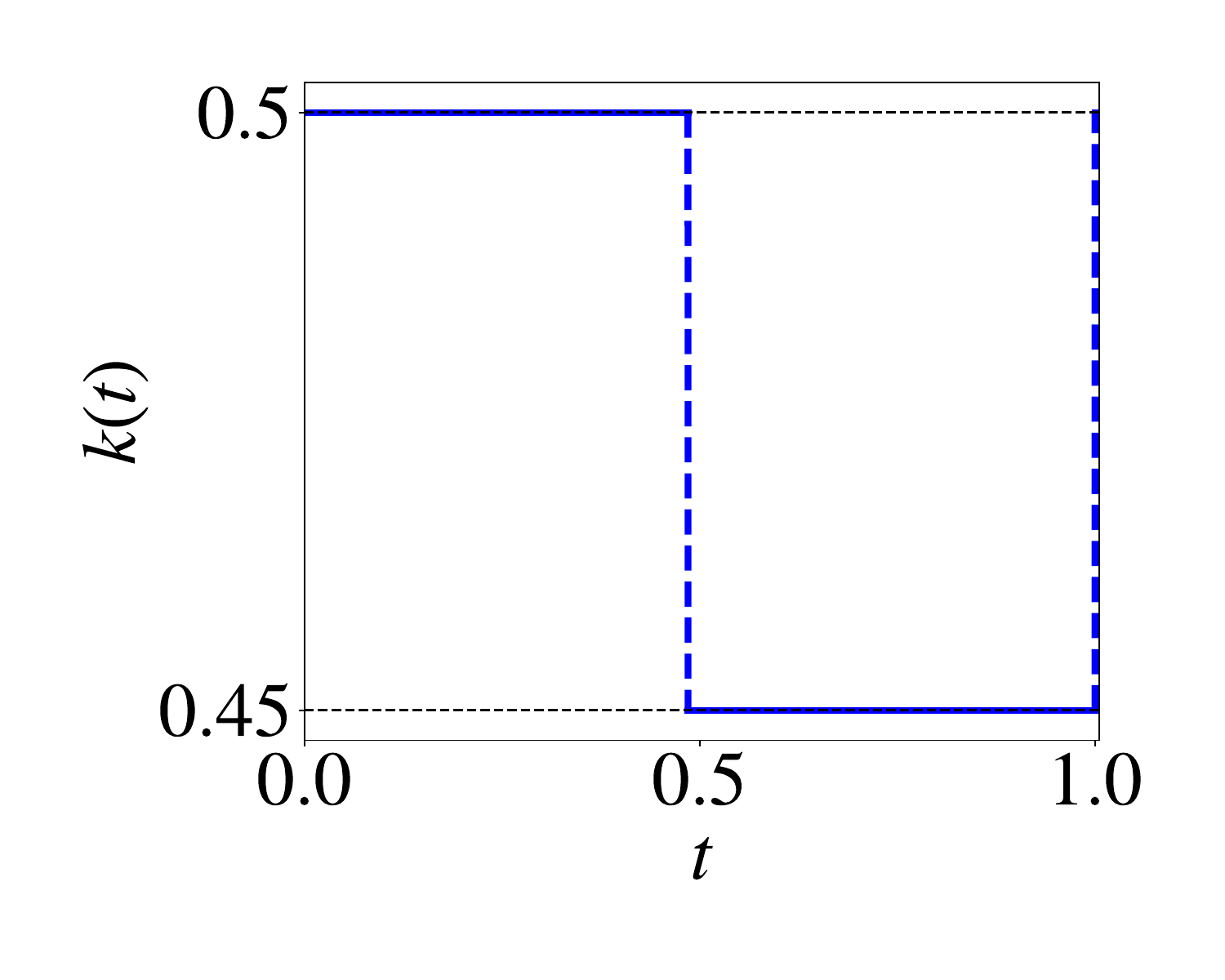}} &
\parbox[c][0.3\textwidth][t]{0.35\textwidth}{
    \raggedright \textbf{(b)}\\
    \includegraphics[width=\linewidth]{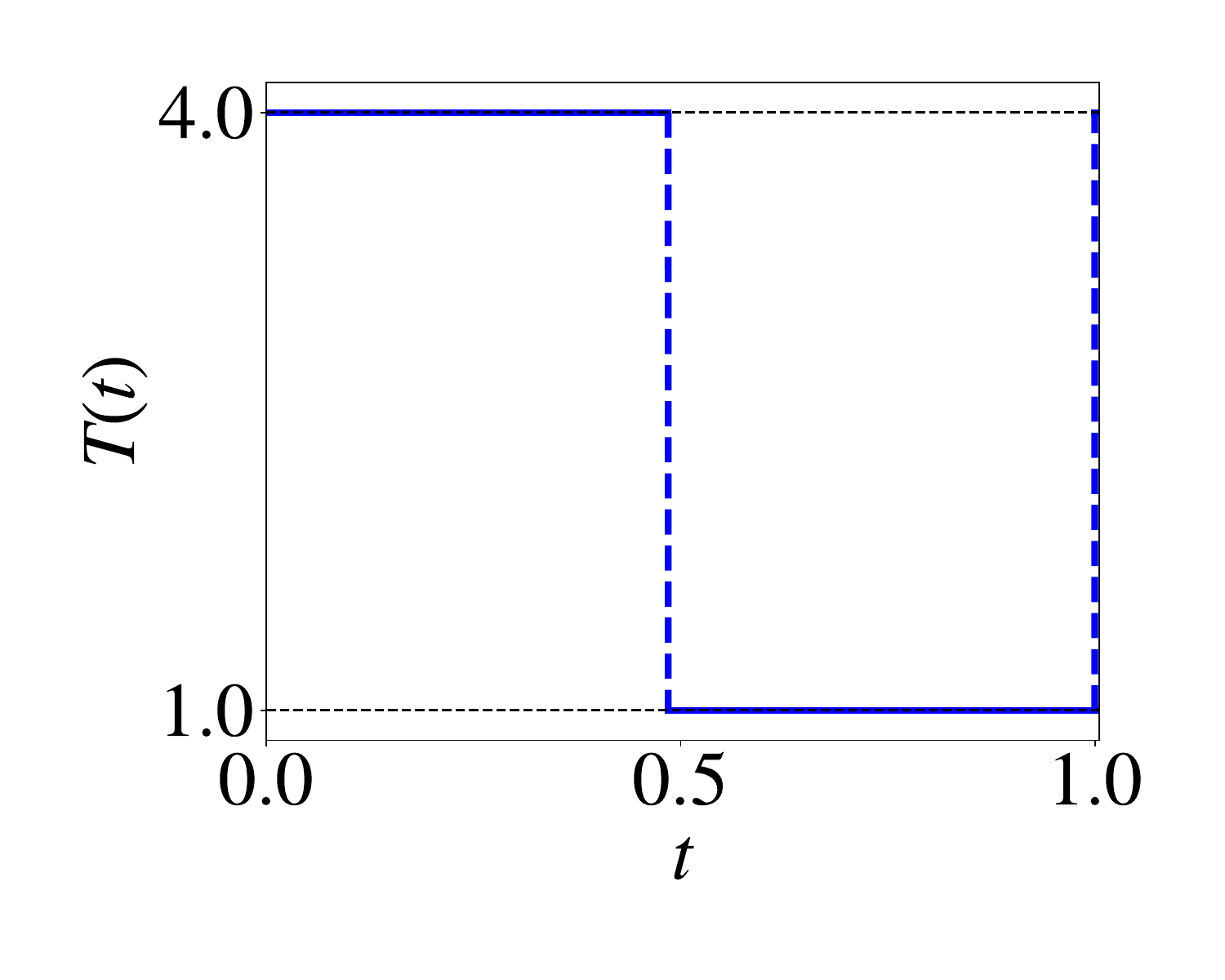}} &
\parbox[c][0.3\textwidth][t]{0.35\textwidth}{
    \raggedright \textbf{(c)}\\
    \includegraphics[width=\linewidth]{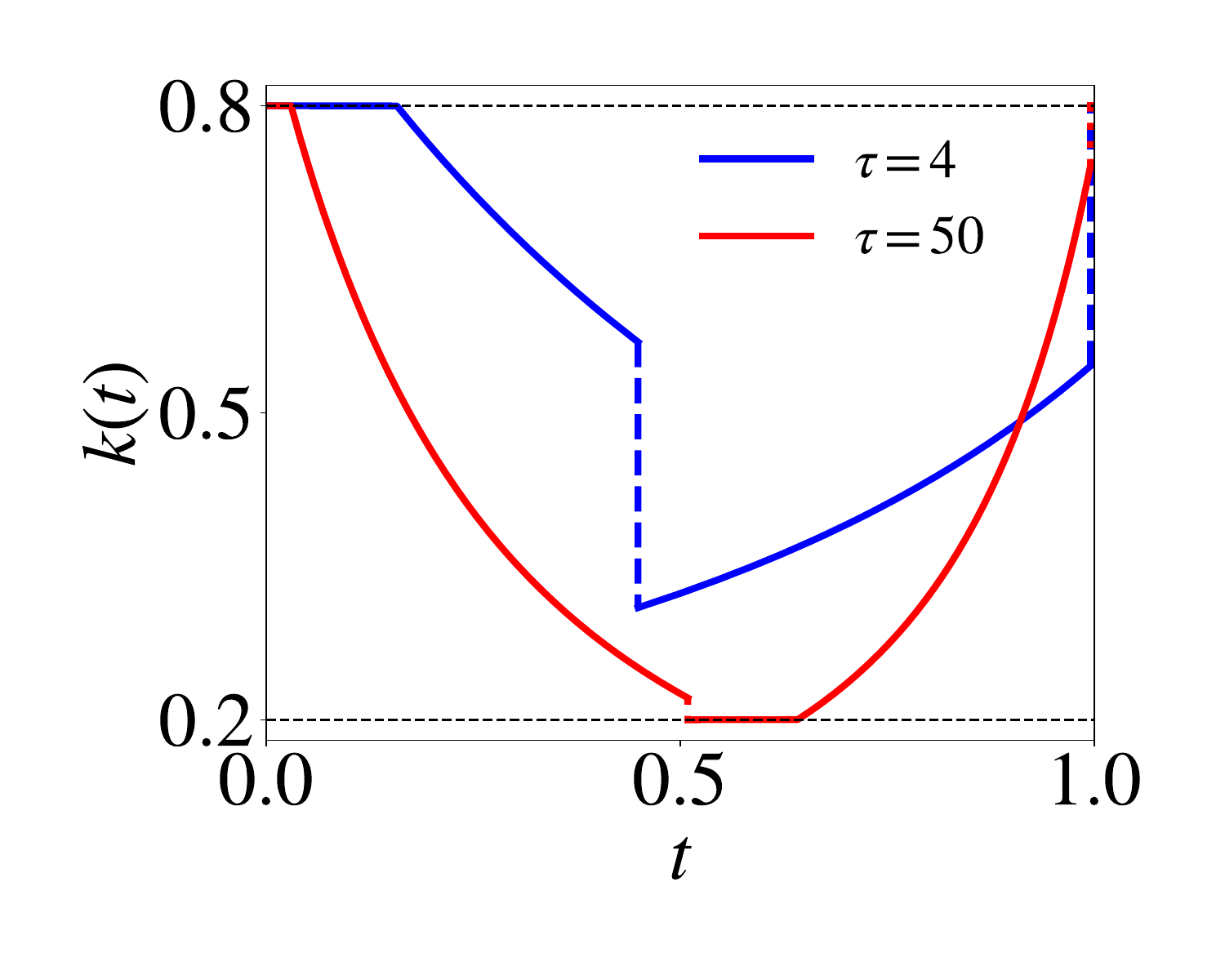}} \\
\end{tabular}
\vspace{6pt}
\begin{tabular}{ccc}
\hspace{-1cm}
\parbox[c][0.3\textwidth][t]{0.35\textwidth}{
    \raggedright \textbf{(d)}\\
    \includegraphics[width=\linewidth]{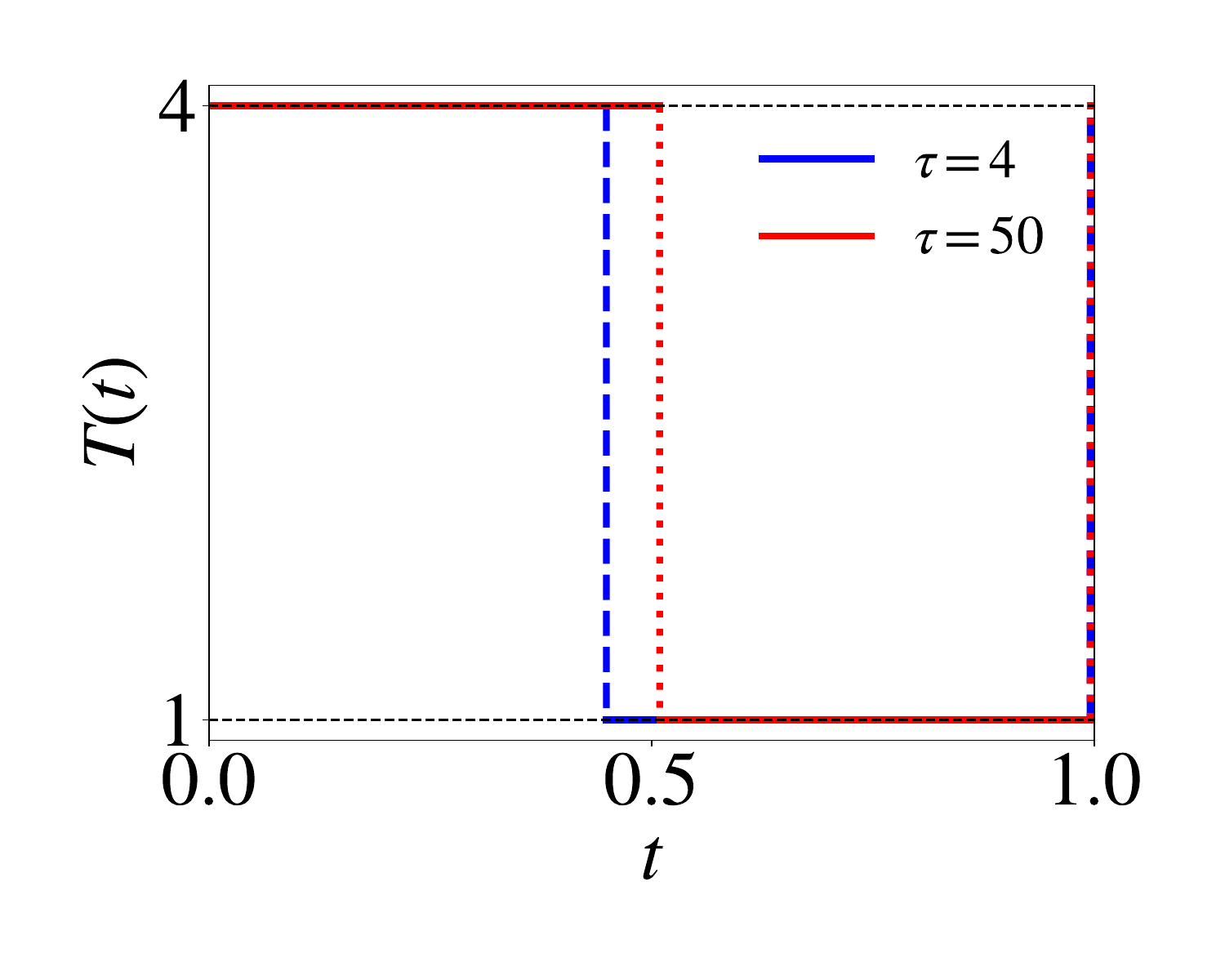}} &
\parbox[c][0.3\textwidth][t]{0.35\textwidth}{
    \raggedright \textbf{(e)}\\
    \includegraphics[width=\linewidth]{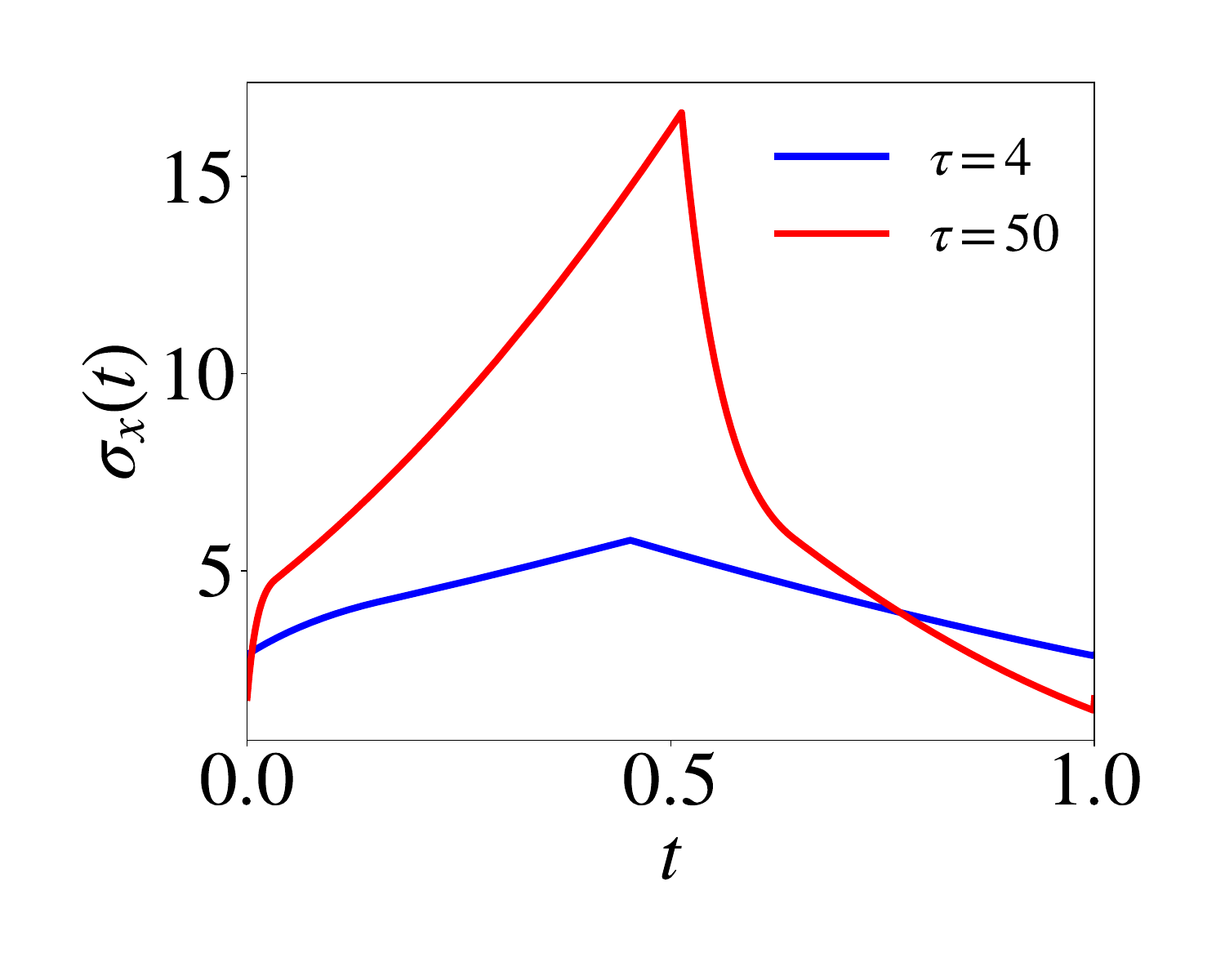}} &
\parbox[c][0.3\textwidth][t]{0.35\textwidth}{
    \raggedright \textbf{(f)}\\
    \includegraphics[width=0.35\textwidth]{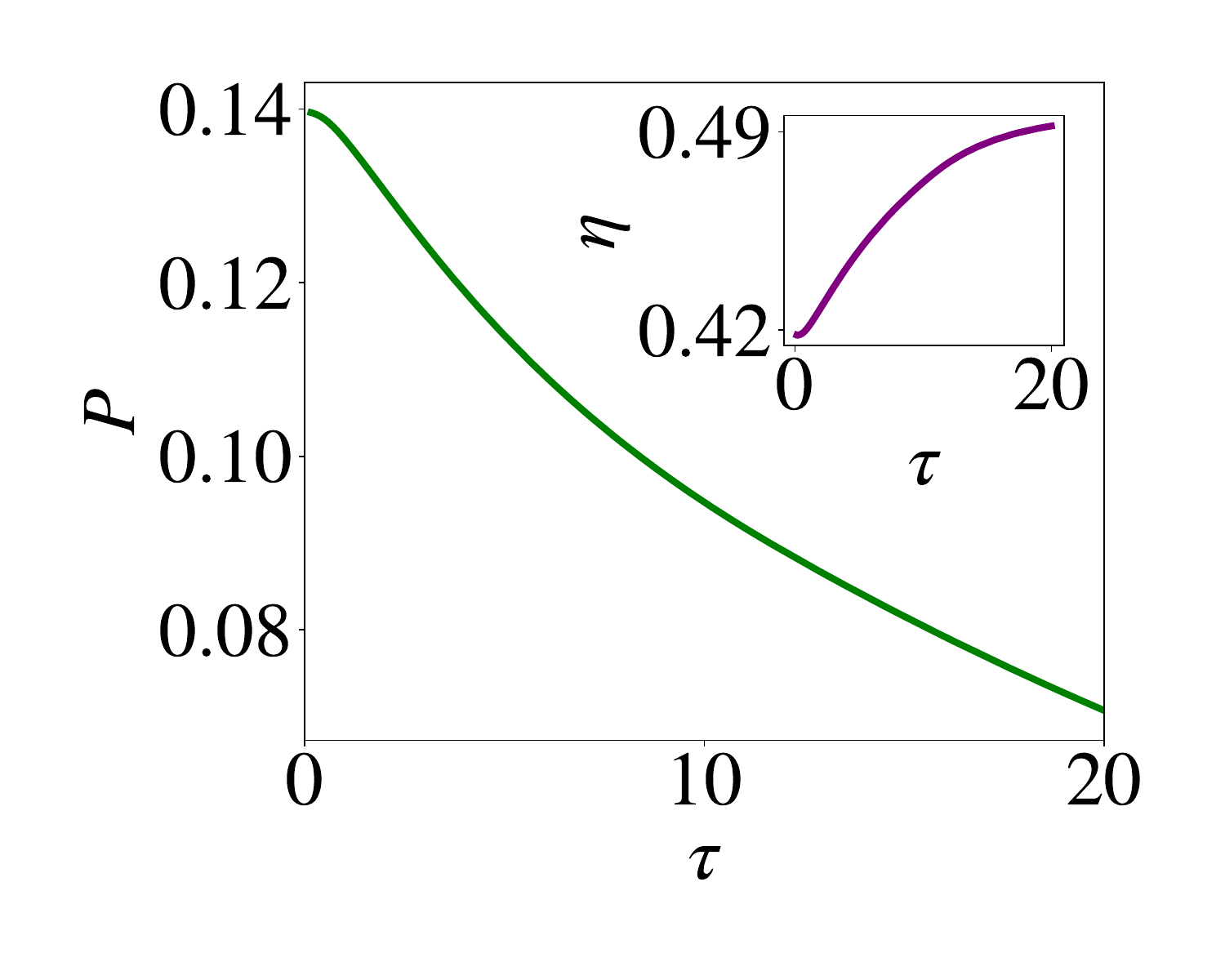}} \\
\end{tabular}
\vspace{-0.5cm}
\caption{\textbf{Maximum power protocols in the overdamped regime.} 
(a) Optimal time variation of stiffness \( k(t) \) for cycle time \( \tau = 4.0 \) and \( k(t) \in [0.45, 0.5] \). The dashed lines indicate instantaneous jumps. 
(b) The corresponding \( T(t) \) protocol. 
(c) Optimal \( k(t) \) protocols for \( k(t) \in [0.2, 0.8] \) and \( \tau = 4 \) (blue) as well as \( \tau = 50 \) (red). 
(d) The corresponding optimal temperature protocol is asymmetric for \( \tau = 4 \) and symmetric for \( \tau = 50 \). 
(e) The corresponding position variances. 
(f) Maximum power as a function of cycle time saturates as \( \tau \) approaches zero. The corresponding efficiency in the inset approaches the Curzon-Ahlborn efficiency of \( \eta_{\rm CA} = 1 - \sqrt{T_- / T_+} = 0.5 \) as the cycle time increases.}
\label{fig:over_pow}
\end{figure}

To verify the validity of the algorithm, we first consider the narrow range of allowed stiffness values $[k_-, k_+]= [0.45, 0.5]$ and fix the cycle time $\tau=4.0$. Vaguely speaking, the maximum output work is achieved by maximizing the product of variations $\Delta \sigma_x$ and $\Delta k$ of the variance $\sigma_x(t)$ and the stiffness $k(t)$ during the cycle~\cite{Viktor2022}. 
For a small allowed variation of stiffness,
one can prove analytically~\cite{Viktor2022} that this is achieved by piecewise constant protocols 
for $\{k(t),T(t)\}$, jumping between the maximum allowed values. In the present case, the optimal cycle thus should consist of two finite-time isotherms-isochores connected by two instantaneous adiabats. In Fig.~\ref{fig:over_pow}(a) and \ref{fig:over_pow}(b), we show the optimal protocols for stiffness and temperature obtained using Eqs.~\eqref{eq:over_opt_pow_sig}-\eqref{eq:over_opt_pow_T} of our algorithm. These results are in accord with the theory~\cite{Viktor2022}. In addition, the numerical calculation shows that the hot isotherm-isochor should be slightly shorter than the cold one, to achieve the optimal relaxation of $\sigma_x(t)$. This universal feature of maximum power and maximum efficiency protocols, observable in all numerical examples presented in this work, was not predicted in~\cite{Viktor2022} and is in accord with the finding that \( \sigma_x(t) \) relaxes faster at hot temperatures than at cold temperatures~\cite{heating_cooling_asymmetry}. We obtained $W=0.107$, $P=0.027$, and $\eta= 1 - k_-/k_+=0.100$ in this case.

After verifying the algorithm, let us now test it for fixed cycle times $\tau = 4$ and $\tau = 50$ and a larger window of allowed stiffnesses, $[k_-, k_+]= [0.2, 0.8]$, where no analytical results are known. The maximum power protocols for stiffness $k(t)$ and temperature $T(t)$, obtained from our algorithm, are shown in Figs.~\ref{fig:over_pow}(c) and ~\ref{fig:over_pow}(d), respectively, and the corresponding position variance is shown in \ref{fig:over_pow}(e). For $\tau = 4$, the temperature protocol quenches between a shorter hot isotherm at $T_+$, and a longer cold isotherm at $T_-$. For $\tau = 50$, the quench occurs at $~\tau/2$. The stiffness protocols are strongly asymmetric, with quenches at the same time-instants as quenches in $T(t)$. Interestingly, $k(t)$ exhibits a flat portion at $k_+$ at the beginning of the cycle, and stays well above the minimal allowed value at the end of the cycle. This finding is universal in our numerical experiments and reflects the fact that $\sigma_x$ dimensionless relaxation time $1/2 k \tau$ increases with decreasing stiffness. This disqualifies small stiffness values, as the system cannot relax to the desired small value of \( \sigma_x \) fast enough unless \( \tau \) is very large. A similar protocol has been reported for optimal variation of a spectral gap in a microscopic spin engine~\cite{MAJUMDAR2025130278}, where the temperature protocol has been fixed, and thus the quenches occurred at times $\tau/2$ and $\tau$. The output work, efficiency, and power obtained for $\tau=4$ were about $W= 0.485$, $\eta=0.448$, and $P= 0.121$. For $\tau=50$, we got $W=1.817$, $P=0.036$, and $\eta=0.495$, which is close to the Curzon-Ahlborn efficiency of $\eta_{\rm CA} = 1 - \sqrt{T_-/T_+} =0.5$~\cite{Seifert_CA_eff}.

Finally, we used the same broad stiffness window $[k_-, k_+]= [0.2, 0.8]$ as above and optimized output power also with respect to the cycle time $\tau$. In Fig.~\ref{fig:over_pow}(f), we plot the maximum power, obtained using our algorithm with Eqs.~\eqref{eq:over_opt_pow_sig}--\eqref{eq:over_opt_pow_T}, along with the corresponding efficiency as functions of the cycle time. The maximum power of $0.14$ is reached for a vanishing cycle time and corresponds to the minimum efficiency of $0.423$. This result is qualitatively in accord with the findings of Ref.~\cite{low_dissipation_2020} for general low-dissipation heat engines.
As the cycle time decreases, the power increases monotonically, while the corresponding efficiency decreases from a value close to the Curzon-Ahlborn efficiency of $0.5$.

\subsection{Efficiency Optimization.}\label{overdamped_efficiency}
Considering the overdamped definition of efficiency, $\eta_{OD}$, the optimal protocols for $k(t)$ and $T(t)$ bounded by $k_-,~T_-$ and $k_+, T_+$ are piecewise constant functions jumping between the allowed boundary values~\cite{Holubec_2022}. 
Let us now use this result to verify our algorithm for deriving maximum efficiency protocols. In this case, the form of the functional $J$ in the algorithm of Sec.~\ref{sec:opt_procedure} is more complicated than considered therein because it is given by a ratio of two integrals. The efficiency variation is given by
\begin{equation}
    \delta \eta = \frac{\delta W}{Q_+} - \frac{W}{Q_+^2}\delta Q_+,
\end{equation}
where, for the sake of this section, $Q_+$ is the input heat in Eq.~\eqref{eq:heat_it} but without taking into account the kinetic contribution (i.e., when setting $m=0$ in Eq.~\eqref{eq:heat_defn_under}). Rewriting the above equation in terms of the rates, we find
\begin{equation}
    \delta \eta = \frac{1}{Q_+} \delta \int_0^1 \dot W dt - \frac{W}{Q_+^2} \delta \int_0^1 \dot Q_+ dt = \delta \left [ \int_0^1 \left (\frac{\dot W}{Q_+} - \frac{W}{Q_+^2}\dot Q_+ \right )dt\right ]. \label{del_eta}
\end{equation}
From now on, it is possible to proceed according to Sec.~\ref{sec:opt_procedure} with $\xi = \frac{\dot W}{Q_+} - \frac{W}{Q_+^2}\dot Q_+$. Since $\int_0^1 \left[\frac{\dot W(t)}{Q_+} - \frac{W}{Q_+^2}\dot Q_+(t)\right]\,dt = 0$, the rate $\xi$ needs to be corrected to $\xi_{\eta}  = \xi + W/Q_+$ to yield the efficiency through the formula $\eta=\int_0^1 \xi_{\eta} dt$. The maximum efficiency protocol can be calculated by applying our algorithm either to $\xi$ or to
\begin{figure}[!t]
    \centering \includegraphics[width=0.4\linewidth]{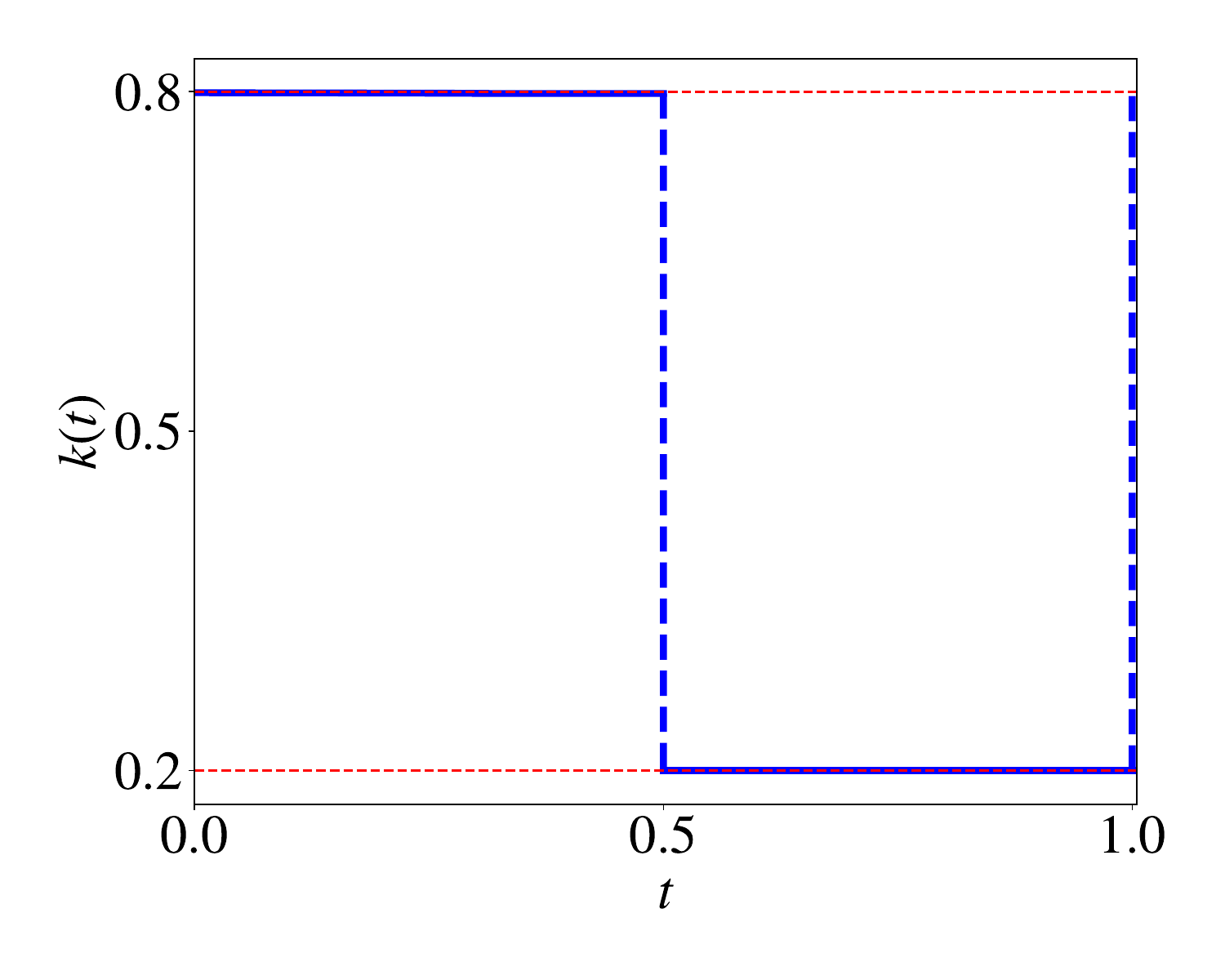}
    \caption{Protocol for stiffness that yields maximum overdamped efficiency.}
    \label{fig:over_eff}
\end{figure} 

 \begin{equation}
      \xi_{\eta} = \frac{W}{Q_+} + \frac{\dot W}{Q_+} - \frac{W}{Q_+^2}\dot Q_+.\label{eq:OD_eff_target}
 \end{equation}
Substituting the work flux~\eqref{work_defn} and overdamped heat flux in Eq.~\eqref{eq:heat_defn_over} for $\dot{W}$ and $\dot{Q}_+$ in this last equation, we get 
\begin{align}
    \xi_{\eta,OD} = \frac{W}{Q_+} + \frac{\tau k(T-k\sigma_x)}{Q_+} - \frac{\tau W}{Q_+^2} k(T-k\sigma_x) \theta(T-k\sigma_x) .
\end{align}
The Hamiltonian in Eqs.~\eqref{eq:opt0}--\eqref{eq:optx} thus reads
\begin{align}
    \hh =\xi_{\eta,OD} + 2\tau\lambda(T-k\sigma_x).
\end{align}

While applying our numerical algorithm in Sec.~\ref{sec:opt_procedure} using this Hamiltonian, we encountered convergence issues when we allowed the algorithm to optimize also the temperature protocol. We were not able to decipher the source of these numerical problems. Nevertheless, we found that, using the temperature protocol
\begin{equation}T(t)=\begin{cases} T_+ & 0\leq t<0.5 \\ T_- & 0.5\leq t<1\end{cases},\label{temp_protocol} 
\end{equation}
the algorithm converges to analytically predicted~\cite{Viktor2022} maximum overdamped efficiency of $1-k_-/k_+$. 
The maximum efficiency is under the given constraints independent of the cycle time and, without adjustments, the cycle time obtained from our algorithm diverges. We have thus additionally fixed the cycle duration in our optimization. Specifically, we have iteratively solved the equations 
\begin{align}
\dot \sigma_x(t) &= 2\tau[T(t)-k(t)\sigma_x(t)], \\
\dot \lambda(t) &= \left [ \frac{\tau k^2}{Q_+} - \frac{\tau W k^2}{Q_+^2} \theta(T-k\sigma_x) + 2\tau\lambda k\right], \\
\delta k(t) &= \epsilon \left [ \frac{\tau(T-2k\sigma_x)}{Q_+} - \frac{\tau W (T-2 k \sigma_x)}{Q_+^2} \theta(T-k\sigma_x) - 2\tau\lambda\sigma_x \right],
\end{align}
following from our algorithm for fixed temperature profile and cycle time. In our implementation of the algorithm, the efficiency exhibited oscillatory behavior, requiring very small learning rates to achieve convergence, particularly when the maximal efficiency was close to Carnot's efficiency. To speed up convergence, we therefore used adaptive learning rates, starting at \( \varepsilon_i = 0.0002\), which were progressively decreased by 10\% after every 5000 iterations if oscillatory behavior was detected.

In Fig.~\ref{fig:over_eff}, we show the resulting maximum efficiency protocol for stiffness for $[k_-, k_+] = [0.2, 0.799]$ and $\tau = 4$. Using $k_+ = 0.8$, when $k_-/k_+ = T_-/T_+$ and thus the theoretically predicted maximum efficiency reaches Carnot's efficiency, leads to numerical instabilities.
In agreement with theoretical prediction, the protocol in the figure is piecewise constant, and the corresponding efficiency is $0.749 = 1 - k_-/k_+ < 0.75 = 1 - T_-/T_+$.

 \section{Main Results: Generally Damped Regime}
 \label{sec:underdamped}

Now that we have validated the algorithm of Sec.~\ref{sec:opt_procedure} in the overdamped regime, we move on to the regime of arbitrary damping. There, we validated the algorithm by checking that maximum power and maximum overdamped efficiency protocols obtained from the full model converge with increasing damping rate $\gamma$ to the results of the previous section. In all numerical examples in this section, we set $[k_-,k_+] = [0.2, 0.8]$. We start by presenting our results for maximum power protocols.

\subsection{Power optimization}
In the generally damped regime, the output work $W$ and power are still given by Eqs.~\eqref{work_defn} and \eqref{eq:power}. What changes is the dynamical equation for the position variance $\sigma_x(t)$.  The functional $J$ from Sec.~\ref{sec:opt_procedure} to be optimized is thus the same as in Sec.~\ref{subsec:W_and_P}: $J=P=\int_0^1 \xi_P~dt$ with
\begin{align}
    \xi_P=\frac{\dot W}{\tau}=-\frac{1}{2\tau}\dot k\sigma_x=k\sigma_{xv}, \label{eq: full_work}
\end{align}
where we have used the first of the dynamical equations in Eq.~\eqref{eq:Sigma}.

Following the procedure in Sec.~\ref{sec:opt_procedure}, we find that the equations that yield upon iterative solutions
stiffness and temperature protocols and cycle time that maximize the output power are
 \begin{align}
    \boldsymbol{\dot \sigma}(t) &= \boldsymbol{f}(t), \label{eq:pu1}\\
    \boldsymbol{\dot \lambda}(t) &= \begin{bmatrix}
        k\tau\lambda_1/m \\
        -k -2\tau\lambda_0 + \tau\gamma\lambda_1+2\tau k \lambda_2/m \\
        -\tau\lambda_1+2\gamma\tau \lambda_2
    \end{bmatrix} ,\label{eq:pu2}\\
    \delta k(t) &= \epsilon_1(\sigma_{xv} -\tau\lambda_1\sigma_x/m - 2\tau\lambda_2\sigma_{xv}/m), \label{eq:pu3} \\
    d\tau &= \epsilon_2 \int_0^1\frac{\boldsymbol{\lambda(t)} \cdot \boldsymbol{f(t)}}{\tau} dt, \label{eq:pu4}\\
    \delta T(t) &= \epsilon_3 (2\gamma \tau \lambda_2(t)/m).
		\label{eq:pu5}
\end{align}
Above, $\bs{f}(t)$ represents the right-hand side of Eq.~\eqref{eq:Sigma}.
In our numerical studies, we used the learning rates $\epsilon_1=\epsilon_2=0.01$ and $\epsilon_3=0.001$. 

\begin{figure}[!t]
  \centering
\begin{tabular}{cc}
\parbox[c][0.33\textwidth][t]{0.35\textwidth}{
    \raggedright \textbf{(a)}\\
    \includegraphics[width=\linewidth]{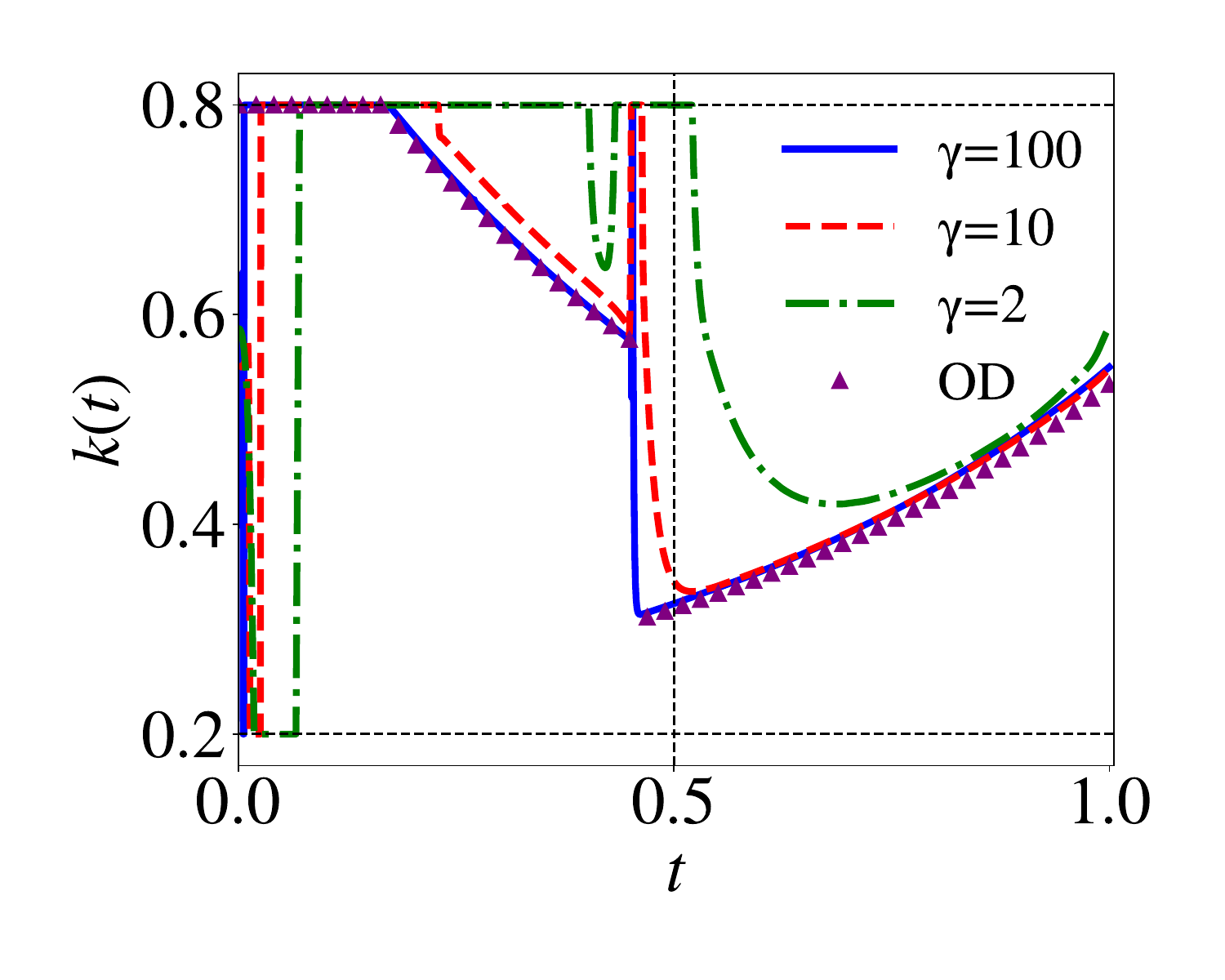}} &
\parbox[c][0.33\textwidth][t]{0.35\textwidth}{
    \raggedright \textbf{(b)}\\
    \includegraphics[width=\linewidth]{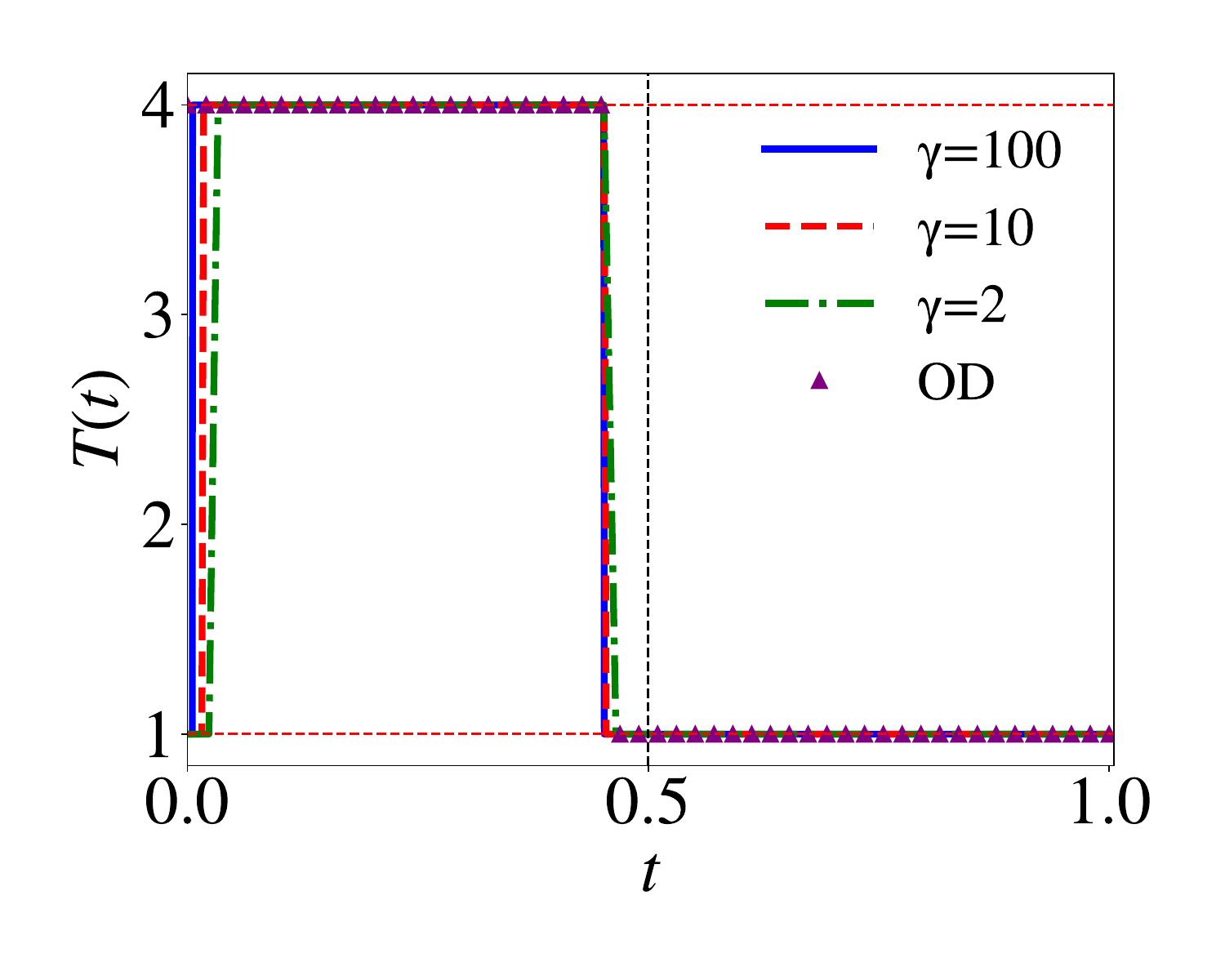}}
\end{tabular}
\vspace{6pt}
\begin{tabular}{ccc}
\hspace{-1cm}
\parbox[c][0.33\textwidth][t]{0.35\textwidth}{
    \raggedright \textbf{(c)}\\
    \includegraphics[width=\linewidth]{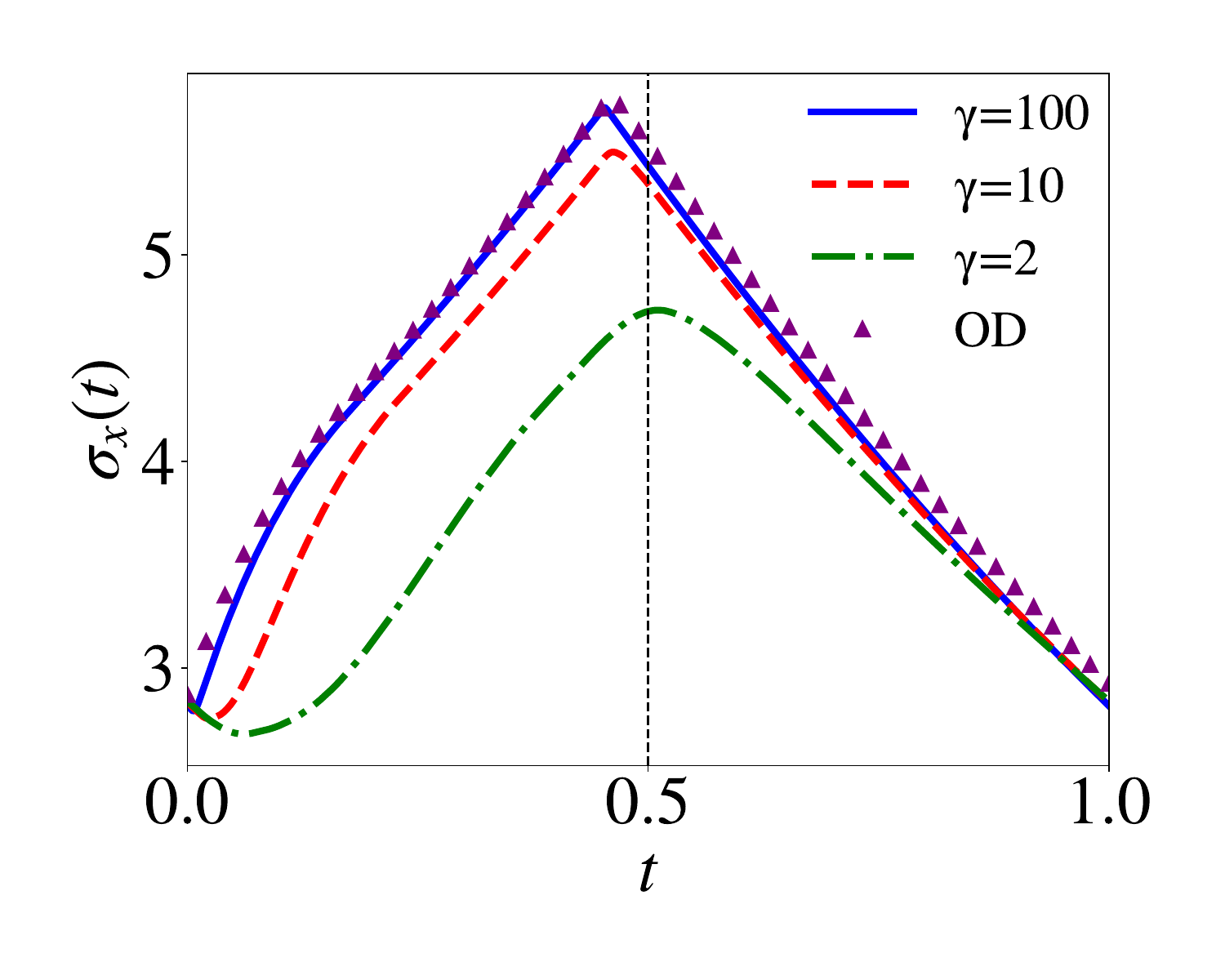}} &
\parbox[c][0.33\textwidth][t]{0.35\textwidth}{
    \raggedright \textbf{(d)}\\\includegraphics[width=\linewidth]{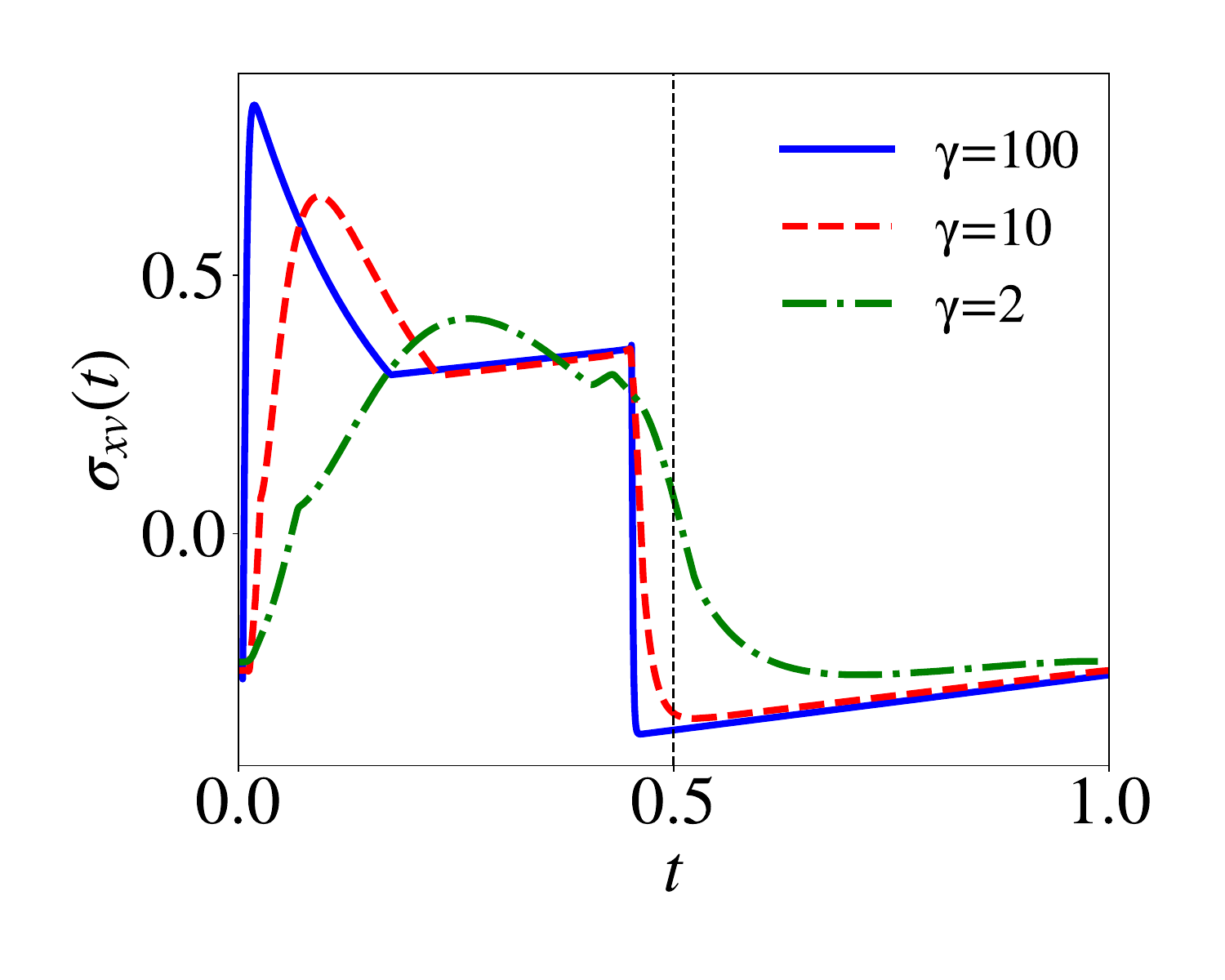}} &
\parbox[c][0.33\textwidth][t]{0.35\textwidth}{
    \raggedright \textbf{(e)}\\
    \includegraphics[width=\linewidth]{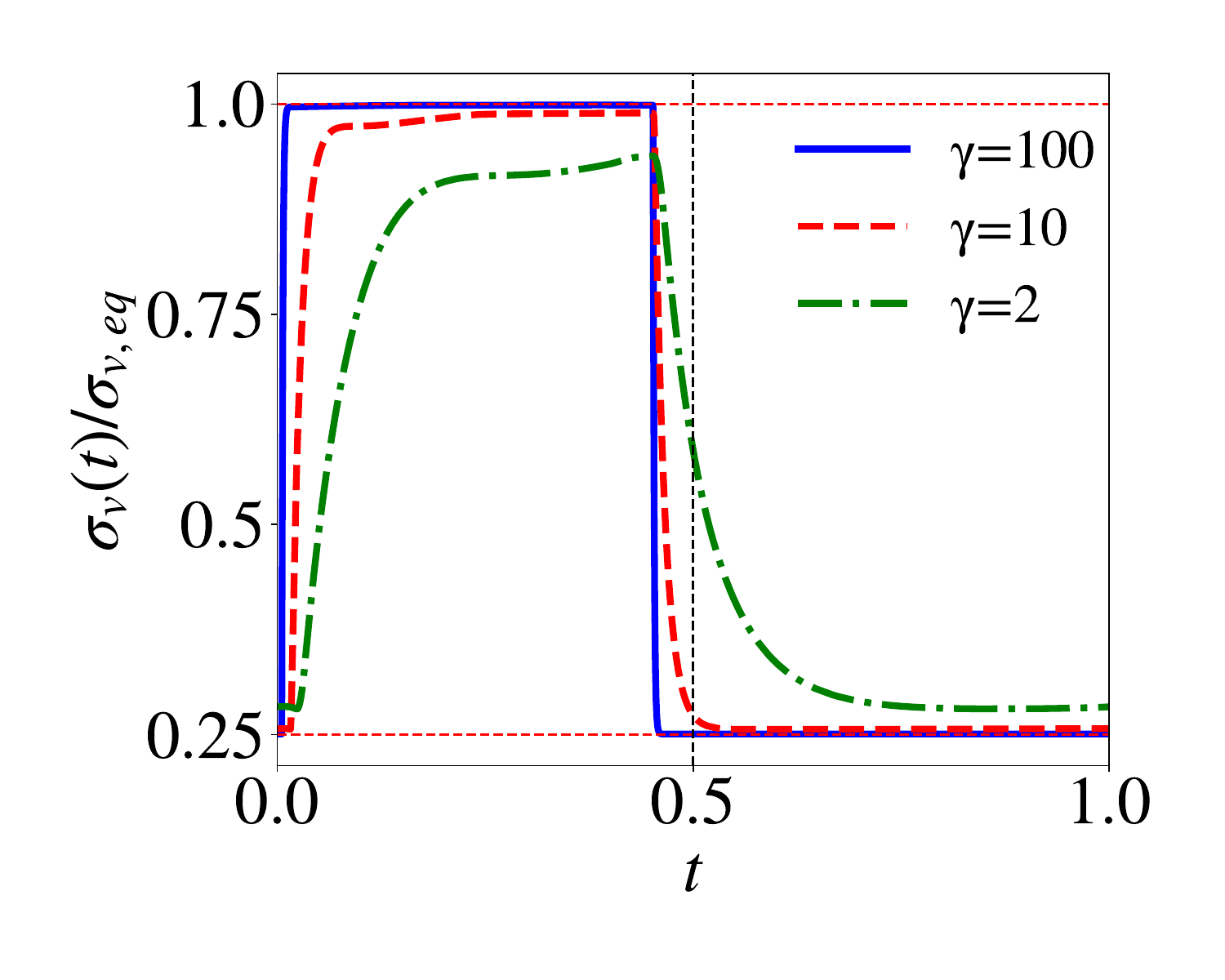}} \\
\end{tabular}
\vspace{-1cm}
\caption{\textbf{Maximum power protocols in the generally damped regime.} (a) and (b) The optimal protocols for stiffness, $k(t)$, and temperature, $T(t)$ for $\tau = 4$ and $\gamma = 2, 10, 100$. For comparison, we also replot the overdamped optimal protocols (OD) from Fig.~\ref{fig:over_pow} for $\tau = 4$. (c), (d), and (e) The corresponding response functions $\sigma_x(t)$, $\sigma_{xv}(t)$, and $\sigma_v(t)/\sigma_{v,eq}$, with $\sigma_{v,eq} = T_+/m = T_+ \gamma$. The corresponding heat absorbed and work done until time $t$ during the cycle are plotted in Figs.~\ref{fig:power_vs_gamma}(a) and (b).
}
  \label{full_pow}
\end{figure}

For a comparison with the overdamped case in Sec.~\ref{subsec:W_and_P}, we first present in Fig.~\ref{full_pow}(a-b) maximum power protocols for stiffness and temperature for three values of the damping rate $\gamma=2, 10, 100$, and fixed cycle time $\tau = 4$. To obtain these results, we have used Eqs.~\eqref{eq:pu1}--\eqref{eq:pu3} and Eq.~\eqref{eq:pu5}. The remaining panels (c-e) in Fig.~\ref{full_pow} depict the corresponding response functions $\boldsymbol{\sigma}(t)$. To show the convergence of the protocols to those valid in a fully overdamped regime with increasing $\gamma$, we give in the figures also the overdamped optimal control and response functions found in Sec.~\ref{subsec:W_and_P}. 
For all the damping rates used, the cycles are far from quasistatic execution, as can be seen by comparing \( \sigma_x(t) \) in Fig.~\ref{full_pow}(c) with the quasistatic variance \( T(t)/k(t) \). The cross-correlation shown in Fig.~\ref{full_pow}(d) is proportional to the slope of \( \sigma_x(t) \), as follows from Eq.~\eqref{eq:Sigma}.
When the variance increases, the Brownian particle's average velocity tends to point outward from the trap center, and vice versa. Figure~\ref{full_pow}(e) demonstrates that the velocity variance reaches its quasistatic value \( T(t)/m \) as the damping rate \( \gamma \) increases. 

Interestingly, for all values of \( \gamma \), the maximum power protocol for stiffness exhibits fast up-down transitions around the time when the temperature drops from \( T_+ \) to \( T_- \), during which \( k(t) \) increases to the upper bound, \( k_+ \), and then returns to its original value. Additionally, similar portions, where the stiffness swiftly drops to \( k_- \) and then increases again, are observed at the time when the temperature increases from \( T_- \) to \( T_+ \). The widths of these portions decrease as the damping rate \( \gamma \) increases.  They are completely missing in the overdamped approximation, where the isothermal variations of stiffness are monotonous, probably due to the first-order nature of the corresponding differential equation. The temperature protocols are qualitatively the same as in the overdamped case, almost instantaneously transitioning between the boundary values $T_\pm$, with the speed of the transitions decreasing with decreasing damping $\gamma$. The main benefit of optimizing power with respect to the temperature protocol is thus the optimal allocation of cycle time between the hot and cold branches.

\begin{figure}[!t]
  \centering
\begin{tabular}{cc}
\parbox[c][0.35\textwidth][t]{0.4\textwidth}{
    \raggedright \textbf{(a)}\\\includegraphics[width=\linewidth]{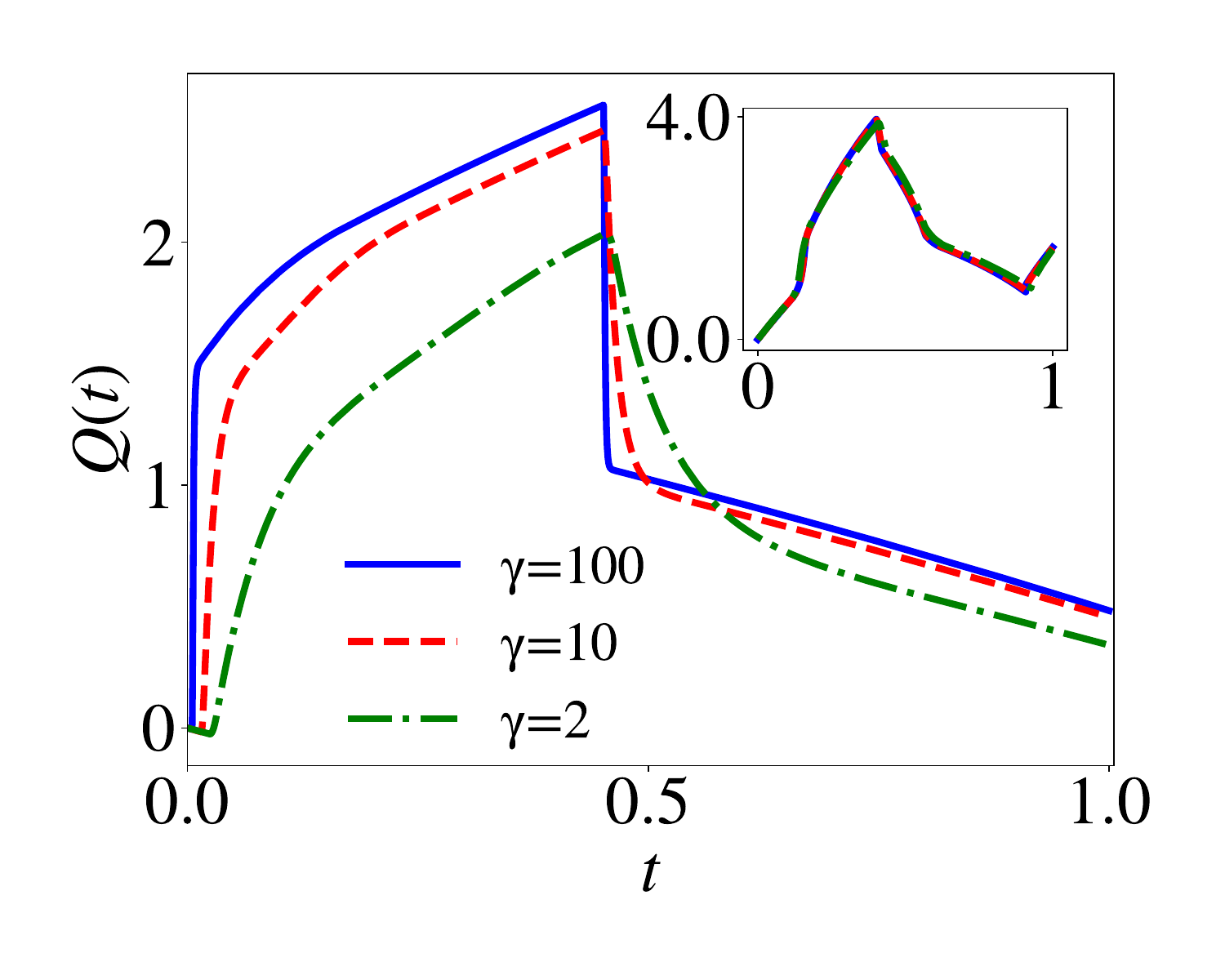}} &
\parbox[c][0.35\textwidth][t]{0.4\textwidth}{
    \raggedright \textbf{(b)}\\\includegraphics[width=\linewidth]{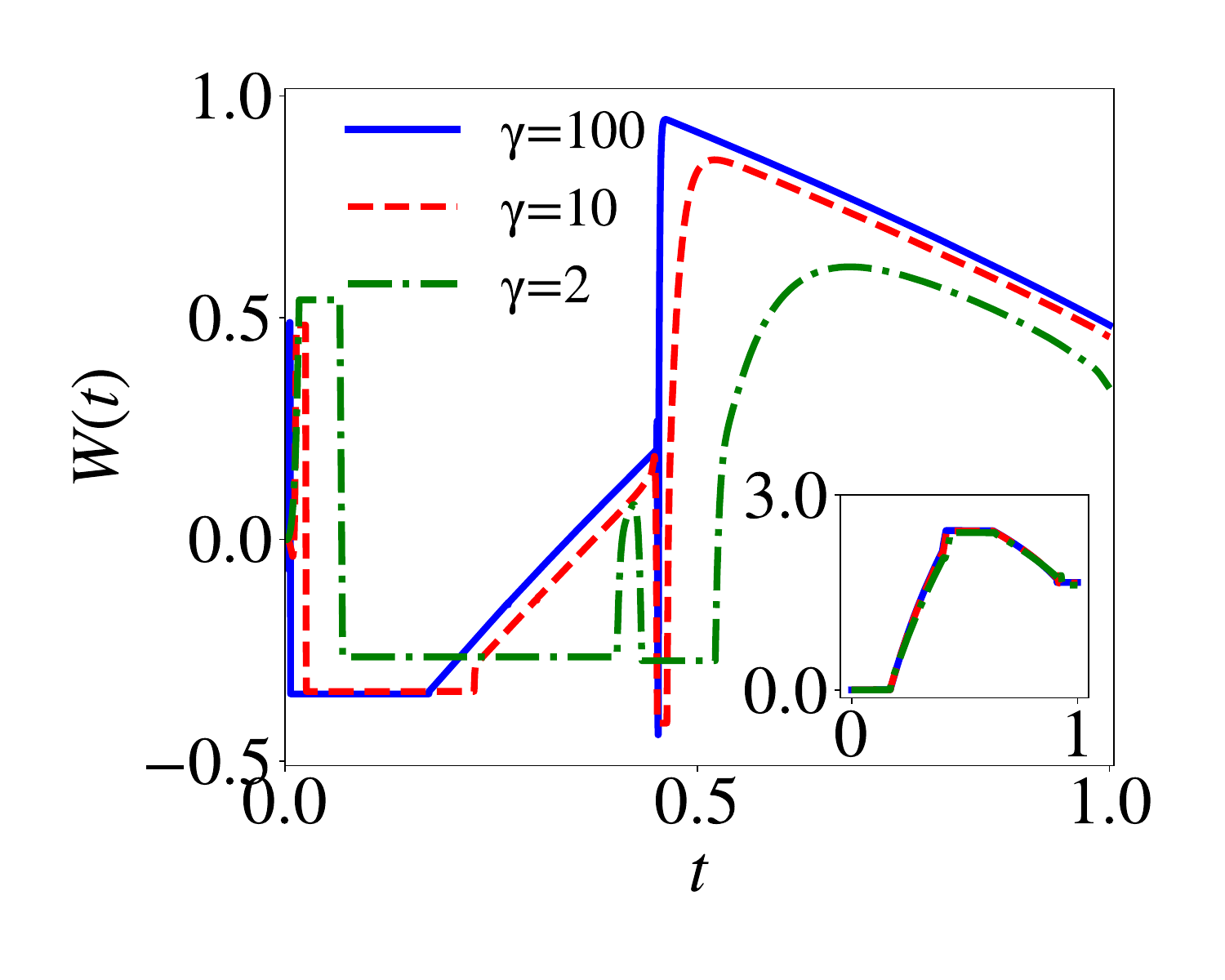}}
\end{tabular}
\vspace{-\baselineskip}
\begin{tabular}{ccc}
\hspace{-1cm}
\parbox[c][0.5\textwidth][t]{0.35\textwidth}{
    \raggedright \textbf{(c)}\\
\includegraphics[width=\linewidth]{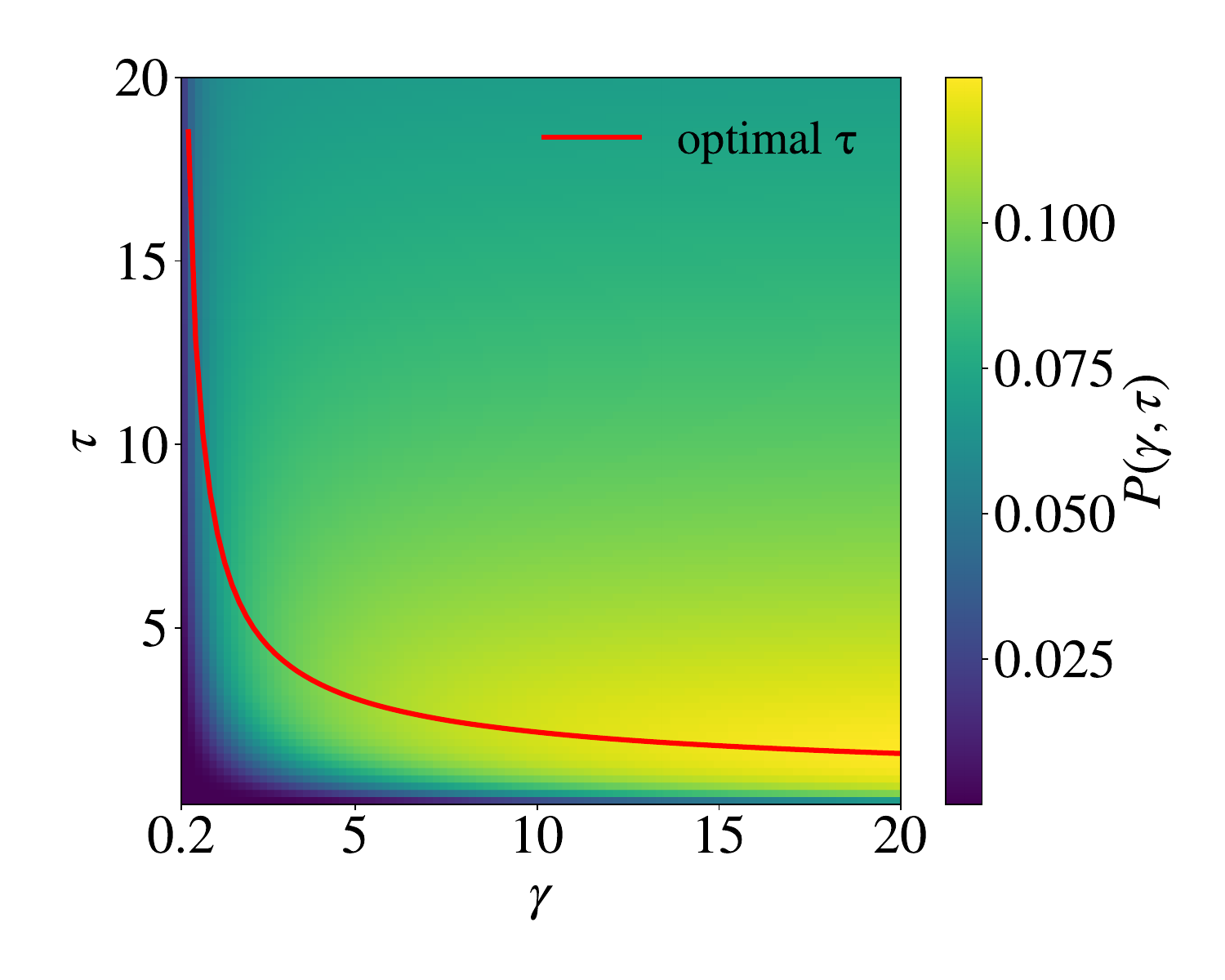}}
\parbox[c][0.5\textwidth][t]{0.35\textwidth}{
    \raggedright \textbf{(d)}\\
\includegraphics[width=\linewidth]{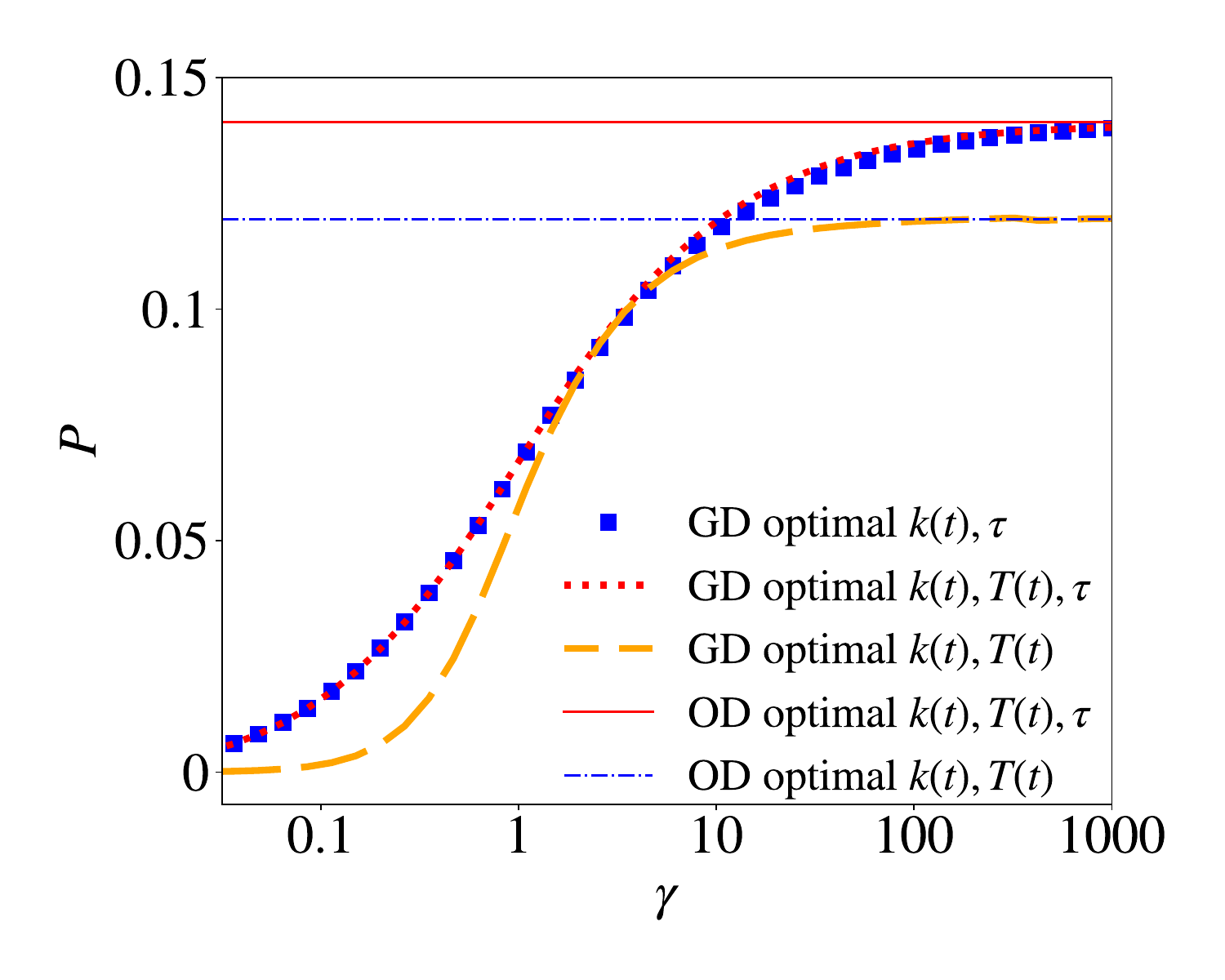}}
    \parbox[c][0.5\textwidth][t]{0.35\textwidth}{
    \raggedright \textbf{(e)}\\
\includegraphics[width=\linewidth]{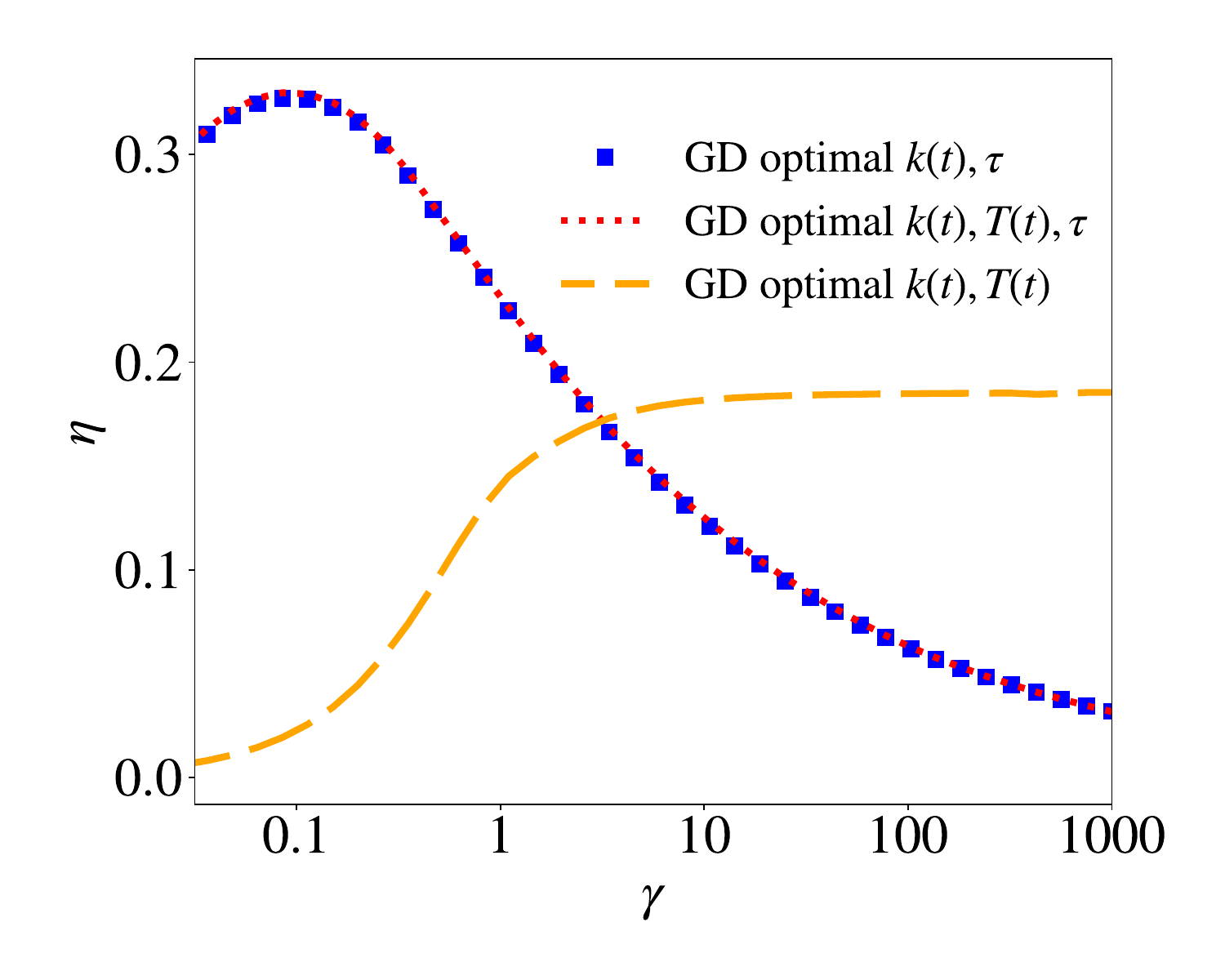}}
\vspace{-3cm}
\end{tabular}
\caption{(a) and (b) Total heat absorbed and work done until time $t$, $Q(t)$ and $W(t)$, using the optimal protocols shown in Fig.~\ref{full_pow}. The insets additionally show these variables for a long cycle time of $\tau=50$ with the same type of optimization.
(c)--(e) \textbf{Maximum power and the corresponding efficiency as functions of the damping rate.}
(c) The variation of maximum power as a function of damping rate $\gamma$ and cycle time $\tau$. The cycle time corresponding to maximum power for a given damping rate can be approximately described by $1.078 + 7.754\gamma^{-0.865}$. 
(d) Maximum power $P$ in the generally damped regime (GD) as a function of the damping rate $\gamma$, obtained upon optimization with respect to: (i) temperature $T(t)$ and stiffness $k(t)$, keeping $\tau = 4$; (ii) cycle time and stiffness with $T(t)$ in \eqref{temp_protocol}; and (iii) all three variables. The horizontal lines show the values $0.119$ and $0.140$ of maximum power obtained from overdamped optimization with respect to: (i) $T(t)$ and $k(t)$, keeping $\tau = 4$; and (ii) $T(t)$, $k(t)$, and $\tau$.
(e) The variation of the efficiencies corresponding to the maximum power values shown in (d).
}
  \label{fig:power_vs_gamma}
\end{figure}

Let us now attempt to intuitively understand the shapes of the optimal protocols. When the stiffness, \( k(t) \), decreases, the working medium performs work proportional to both the magnitude of the decrease and the current position variance, \( \sigma_x(t) \). Conversely, increasing the stiffness requires input work proportional to the magnitude of the increase and the current position variance.
To provide better intuition on the connection between control, response, and work generation, we present in Figs.~\ref{fig:power_vs_gamma}(a) and (b) the total heat absorbed and work done by the working medium up to time \( t \) during the cycle, for the generally damped optimal protocols shown in Fig.~\ref{fig:over_pow}. The insets additionally show the behavior of these quantities for a long cycle time of \( \tau=50 \), where the curves for different damping coefficients collapse, as the system response becomes quasistatic and thus independent of kinematic quantities.

To summarize, maximizing the work output requires achieving the largest possible \( \sigma_x(t) \) during phases of the cycle when the stiffness decreases, and the smallest possible \( \sigma_x(t) \) during phases when the stiffness increases.
For this reason, the strongest transitions from low to high stiffness values occur just after connecting the working medium to the hottest bath, when the position variance, \( \sigma_x(t) \), is at its smallest during the cycle due to prior contact with the coldest bath. For \( \gamma = 100 \), where inertia is negligible, this transition happens immediately after the bath is heated. As the role of inertia increases, the change in the trend of \( \sigma_x(t) \) in Fig.~\ref{full_pow}(c) begins to lag behind the change in temperature. Consequently, the rapid increase in \( k(t) \) is delayed after the temperature rise, with the delay being greater for lower \( \gamma \). A similar situation occurs when the temperature and stiffness are decreased near the time when \( \sigma_x(t) \) reaches its maximum.

Other variations in \( k(t) \) are fine-tuned to balance changes in position variance and work extraction or injection prior to these rapid stiffness transitions. For instance, decreasing stiffness induces work extraction and increases the quasistatic position variance, \( T(t)/k(t) \), toward which the system relaxes. To achieve the desired large \( \sigma_x(t) \) before decreasing stiffness when the reservoir is cooled, it seems optimal to start  reducing \( k(t) \) while the system is still in contact with the hot bath. However, in the overdamped regime, decreasing stiffness also increases the relaxation time required to reach the quasistatic variance. Additionally, decreasing \( k(t) \) before the position variance has sufficiently increased due to heating by the hot bath results in less work being performed than is otherwise possible.

For \( \gamma = 2 \) and \( k > 0.5 \), the system's relaxation rates in Eq.~\eqref{eq:eigenvalues} become complex, and the system's relaxation time is independent of stiffness. Hence the stiffness stays at its largest value for most of the hot part of the cycle in Fig.~\ref{full_pow}(a) due to the last point above.  Because of the slow relaxation for \( \gamma = 2 \), \( \sigma_x(t) \) is not able to reach even the smallest quasistatic variance $T_+/k_+ = 1/2$ attainable at the hot temperature. Therefore, systematically reducing stiffness before the position variance peaks would be wasteful. For \( \gamma > 3.2 \), the dynamics is overdamped, and the position variance relaxes faster with increased stiffness. For \( \gamma = 10 \) and \( 100 \), it is thus advantageous to decrease stiffness slightly during the hot part of the cycle to extract work as the system relaxes towards variances higher than \( 1/2 \) before the temperature drop.

During the cold part of the cycle, the main difference between the optimal stiffness protocols arises from the shift in the stiffness decrease, caused by the delayed reaction of the variance to temperature change, as described above. When the position variance loses memory of these different initial conditions at \( t \approx 0.8 \), the three optimal stiffness protocols converge. The protocol for \( \gamma = 2 \) departs from the other two at the instant when the stiffness \( k(t) \) exceeds 0.5, and the working medium dynamics become underdamped.

The portions of the optimal protocols that are hardest to explain intuitively are the brief up-down and down-up changes in stiffness near the time instants of abrupt changes in bath temperature, as described above. We conjecture that these portions accelerate the change in the average particle velocity, thereby facilitating the switch between the heat-accepting phase (\( 2\dot{Q} = k(t){\dot \sigma_x}(t) + m{\dot \sigma_v}(t) > 0 \)) and the heat-releasing phase (\( \dot{Q} < 0 \)) during changes in bath temperature, as shown in Fig.~\ref{fig:power_vs_gamma}(a).
Specifically, increasing the stiffness during the up-down transition before decreasing the temperature induces a reorientation of the Brownian particle's velocity toward the trap center, whereas the down-up transition before raising the bath temperature has the opposite effect, as shown in Fig.~\ref{full_pow}(d), where the correlation changes sign. For \( \gamma = 10 \) and \( 100 \), the stiffness changes occur so rapidly that \( \sigma_x(t) \) remains approximately constant during these transitions, making the corresponding work done in Fig.~\ref{fig:power_vs_gamma}(b) negligible. In contrast, for \( \gamma = 2 \), the duration of the down-up variation is non-negligible, leading to a negative work increment in Fig.~\ref{fig:power_vs_gamma}(b), as the down transition occurs at a lower position variance than the subsequent up transition.
To demonstrate that these non-intuitive features indeed induce additional work extraction and also improve the engine's energy efficiency, we have, for comparison, calculated the output work and efficiency for the same protocols but with artificially smoothed up-down and down-up portions. The ``smoothed" protocols thus take the form of the maximum overdamped power protocol in Fig.~\ref{fig:over_pow}(c). For example, for \( \gamma = 100 \), the optimal protocol yields \( W = 0.475 \), \( P = 0.119 \), \( \eta = 0.185 \), while the smoothed protocol gives \( W = 0.460 \), \( P = 0.115 \), and \( \eta = 0.178 \). That is, the work (and hence power) and efficiency are increased by about \( 3\% \) and \( 4\% \), respectively.

After analyzing how the shapes of the optimal protocols change with the damping rate, we now focus on the \(\gamma\)-dependence of maximum power for \( \gamma \in (1/100, 1000) \), ranging from the deeply underdamped regime to the deeply overdamped regime.

In Fig.~\ref{fig:power_vs_gamma}(c), we show the resulting dependence of the optimized output power on \( \gamma \) and \( \tau \). As the working medium reaches the deeply underdamped regime, the optimal cycle time diverges, and the corresponding maximum power decreases to zero.
The decrease in power is caused by the decrease of the working medium relaxation time $\propto \gamma^{-1}$, which hinders heat exchange between the working medium and the bath. Using the positivity of the velocity variance $\sigma_v(t)$, the heat absorbed by the working medium per cycle can be estimated as $Q_+=\tau [\gamma T_+/2 - \int_0^{0.5}\sigma_v(t)dt] < \tau \gamma T_+/2$. The first law in the form $\eta = W/Q_+ \le 1$ then implies that $P = W/\tau < Q_+/\tau \le \gamma T_+/2$ vanishes as $\gamma \to 0$. In the opposite deeply overdamped limit, both optimal cycle time and corresponding power saturate at finite values, calculable using the overdamped dynamics. 

These observations are confirmed by Fig.~\ref{fig:power_vs_gamma}(d), where we present results for maximum power obtained from our algorithm under the following conditions: (i) using the temperature protocol in Eq.~\eqref{temp_protocol} and optimizing with respect to stiffness and cycle time, (ii) fixing the cycle time \( \tau = 4 \) and optimizing with respect to stiffness and temperature, and (iii) optimizing with respect to all control parameters. The corresponding efficiencies are shown in Fig.~\ref{fig:power_vs_gamma}(e). At low values of \( \gamma \), the power obtained using all of these techniques converges to zero, as discussed above. With increasing damping, the individual curves for maximum power monotonically converge to the corresponding values obtained from the overdamped approximation, as calculated using the algorithm in Sec.~\ref{sec:verification}.
 As expected, the largest power is achieved by optimizing over all control parameters. However, using the fixed temperature protocol in Eq.~\eqref{temp_protocol} yields only slightly smaller results, particularly for weaker damping. The worst performance is observed when the cycle time is fixed. 

The efficiency corresponding to power optimization with a fixed cycle time monotonically decreases with the damping coefficient, from its $\gamma$-independent overdamped value to zero in the underdamped regime. The former can be understood by noting that the overdamped working medium relaxation time $\propto \gamma m/k$ is, for fixed $\gamma m$, independent of $\gamma$. The latter follows from the fact that the underdamped relaxation time $\propto 1/\gamma$ diverges as $\gamma \to 0$, leading to vanishing energy exchange within a finite cycle time. Efficiencies corresponding to power optimizations involving cycle time exhibit a single maximum in the underdamped regime for low $\gamma \approx 0.1$. This results from a tradeoff between decreasing output work (Fig.~\ref{fig:power_vs_gamma}(b)) and increasing optimal cycle time (Fig.~\ref{fig:power_vs_gamma}(c)) as $\gamma$ decreases.

It is appropriate to compare Fig.~\ref{fig:power_vs_gamma}(d) with a similar figure presented in Ref.~\cite{Dechant_2017}, where the depicted power was calculated using optimal protocols for stiffness analytically derived in the deeply underdamped~\cite{Dechant_2017} and overdamped~\cite{Schmiedl-Seifert2007} regimes under different constraints. In Ref.~\cite{Dechant_2017}, the power remains finite in both the limits of low and high damping, which seemingly contradicts our results. However, in Ref.~\cite{Dechant_2017}, the underdamped regime is achieved by increasing the stiffness \( k \) of the confining potential at a fixed damping \( \gamma \). This approach to reaching the underdamped regime is not applicable in our present setting, where the boundary values of stiffness are fixed. 

\begin{figure}[!t]
  \centering
\begin{tabular}{cc}
\parbox[c][0.33\textwidth][t]{0.35\textwidth}{
    \raggedright \textbf{(a)}\\
    \includegraphics[width=\linewidth]{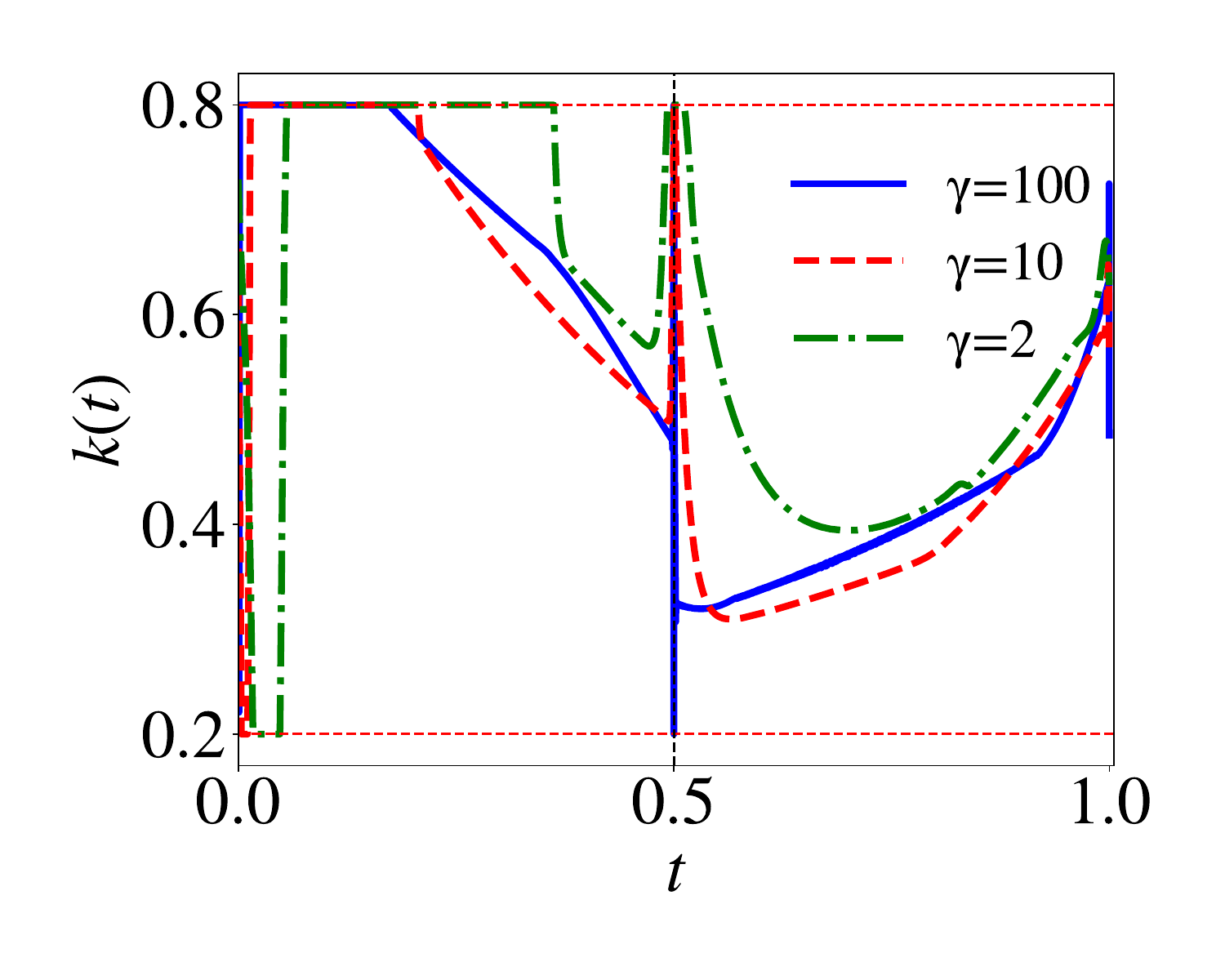}} &
\parbox[c][0.33\textwidth][t]{0.35\textwidth}{
    \raggedright \textbf{(b)}\\
    \includegraphics[width=\linewidth]{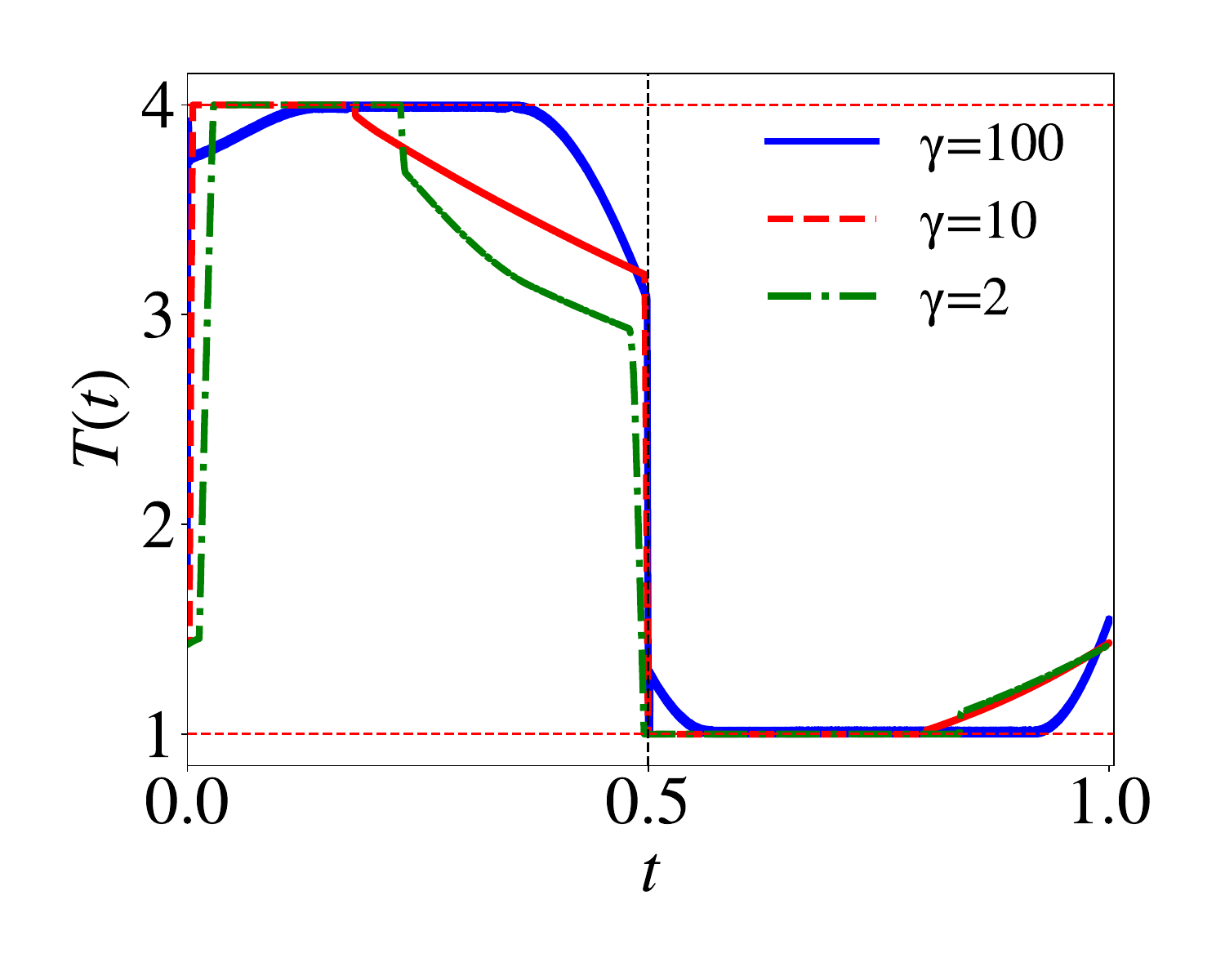}}
\end{tabular}
\vspace{6pt}
\begin{tabular}{ccc}
\hspace{-1cm}
\parbox[c][0.33\textwidth][t]{0.35\textwidth}{
    \raggedright \textbf{(c)}\\
    \includegraphics[width=\linewidth]{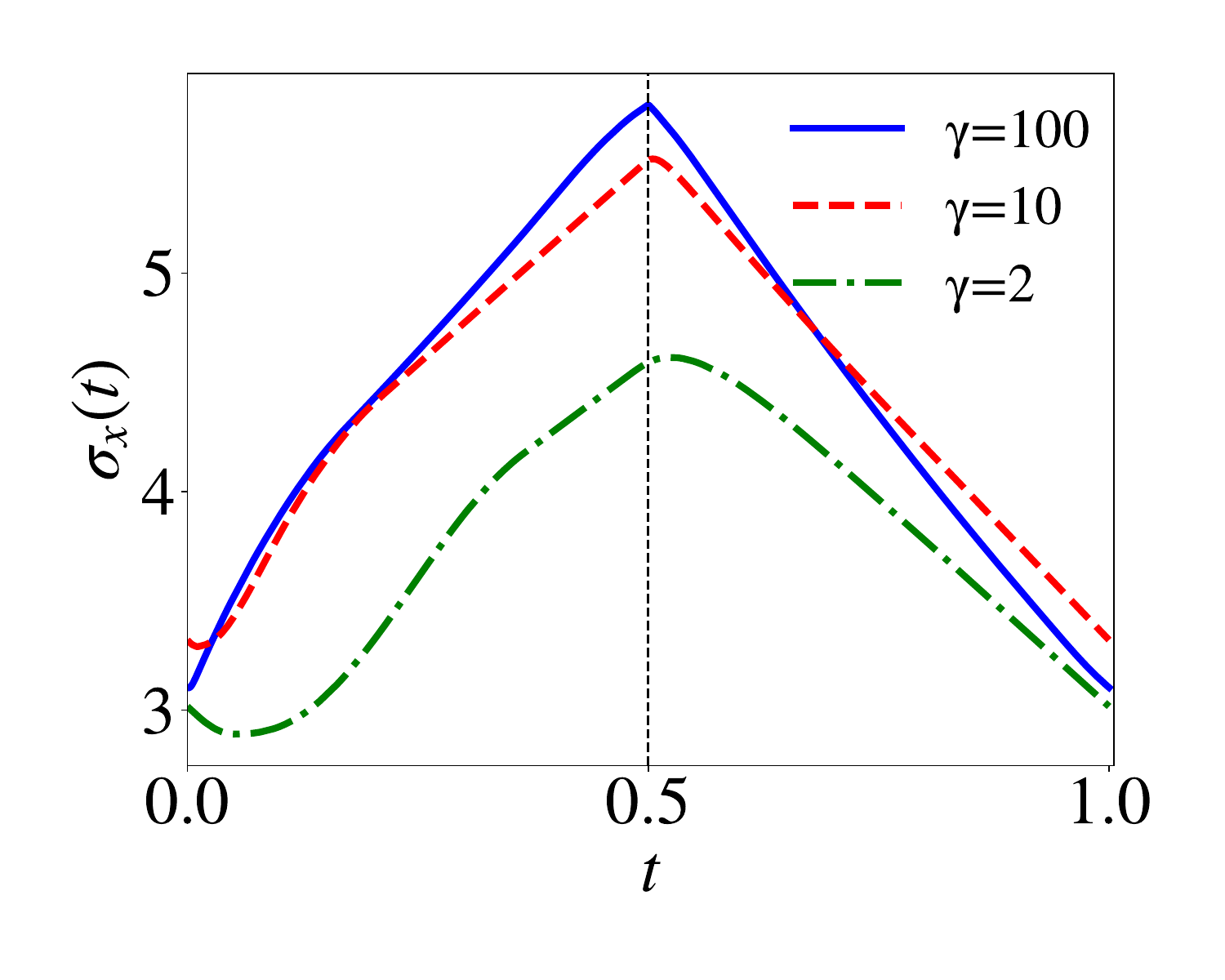}} &
\parbox[c][0.33\textwidth][t]{0.35\textwidth}{
    \raggedright \textbf{(d)}\\\includegraphics[width=\linewidth]{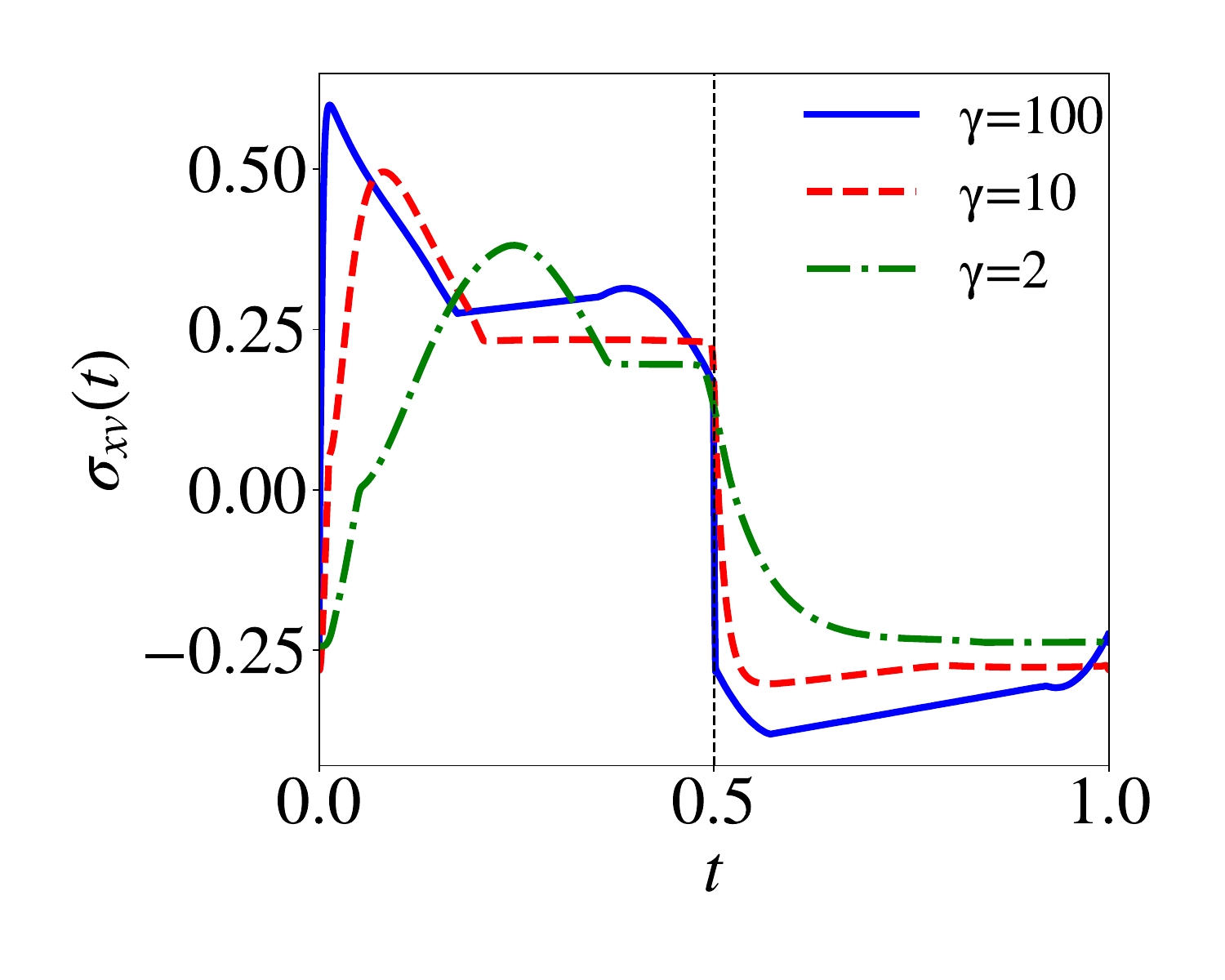}} &
\parbox[c][0.33\textwidth][t]{0.35\textwidth}{
    \raggedright \textbf{(e)}\\
    \includegraphics[width=\linewidth]{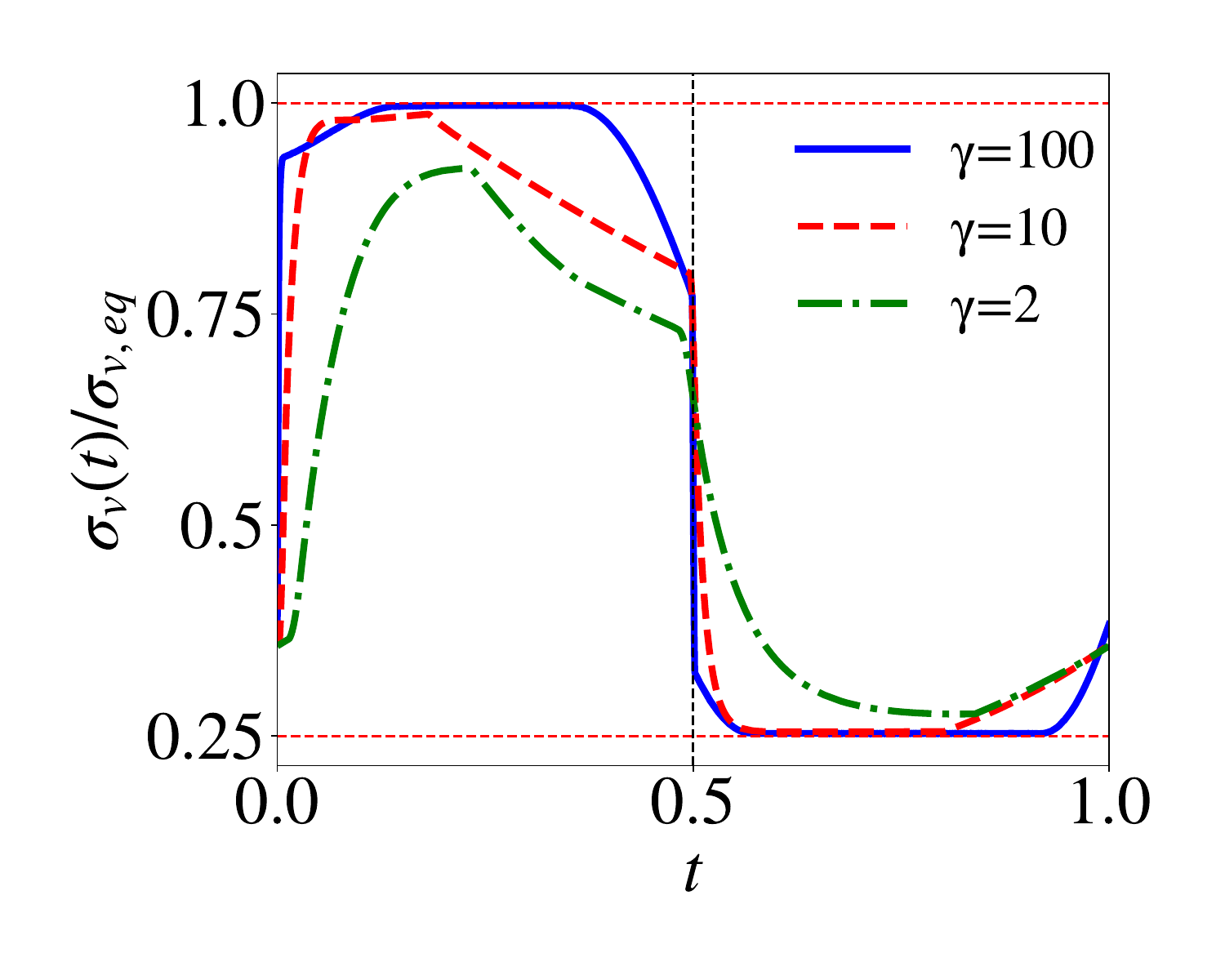}} \\
\end{tabular}
\vspace{-1.0cm}
\caption{\textbf{Maximum efficiency protocols in the generally damped regime.} 
(a) and (b) Optimal stiffness and temperature protocols, $k(t)$ and $T(t)$, for $\tau = 4$ and $\gamma=2$, $10$, and $100$. (c)--(e) The corresponding response functions $\sigma_x(t)$, $\sigma_{xv}(t)$, and $\sigma_v(t)/\sigma_{v,eq}$, with $\sigma_{v,eq} = T_+/m = T_+ \gamma$.}
  \label{fig:full_eff}
\end{figure}

\subsection{Efficiency optimization.}

As the final application of our algorithm, we aim to determine the maximum efficiency protocols for temperature and stiffness. To achieve this, we fix the cycle time, as maximum efficiency is naturally attained in the reversible limit where the cycle time diverges. In this section, we use the correct definition of efficiency~\eqref{eq:efficiency}, which takes into account the heat leakage through momentum degrees of freedom, rather than the overdamped approximation of efficiency.

The current target functional $\eta=\int_0^1 \xi_\eta dt$ can be written using $\xi_\eta$ in Eq.~\eqref{eq:OD_eff_target} with $\dot{W}$ and $\dot{Q}_+$ given by Eqs.~\eqref{work_defn} and \eqref{eq:heat_defn_under}. The integrand thus reads
\begin{align}
    \xi_\eta= \frac{W}{Q_+}+\frac{\tau k \sigma_{xv}}{Q_+}-\frac{\tau W}{Q_+^2}(\gamma T-\sigma_v) \theta(\gamma T-\sigma_v).
\end{align}

Using the results of Sec.~\ref{sec:opt_procedure}, the expression for \( \xi_\eta \) leads to the following equations, which provide maximum efficiency protocols for stiffness and temperature upon iterative solution:
\begin{align}
\boldsymbol{\dot X}(t) &= \boldsymbol{f}(t), \\
    \boldsymbol{\dot \lambda} &= \begin{bmatrix}
        \tau\lambda_1 k/m \\
        -\tau k/Q_+ -2\tau\lambda_0 + \tau \gamma \lambda_1 + 2\tau \lambda_2 k/m \\
        -\frac{W\tau}{Q_+^2}\theta(\gamma T-\sigma_v) - \tau \lambda_1 + 2\tau \gamma \lambda_2
    \end{bmatrix}, \\    
    \delta k &= \epsilon_3\left( \frac{\tau \sigma_{xv}}{ Q_+} - \frac{\tau \lambda_1 \sigma_x}{m} - \frac{2\tau \lambda_2 \sigma_{xv}}{m}\right), \\
    \delta T(t) &= \epsilon_4 \left(-\frac{W\tau \gamma}{Q_+^2}\theta(\gamma T-\sigma_v) + 2\tau\gamma \lambda_2/m \right ).
\end{align}

\begin{figure}[!t]
  \centering
\begin{tabular}{cc}
\parbox[c][0.5\textwidth][t]{0.5\textwidth}{
    \raggedright \textbf{(a)}\\
    \includegraphics[width=\linewidth]{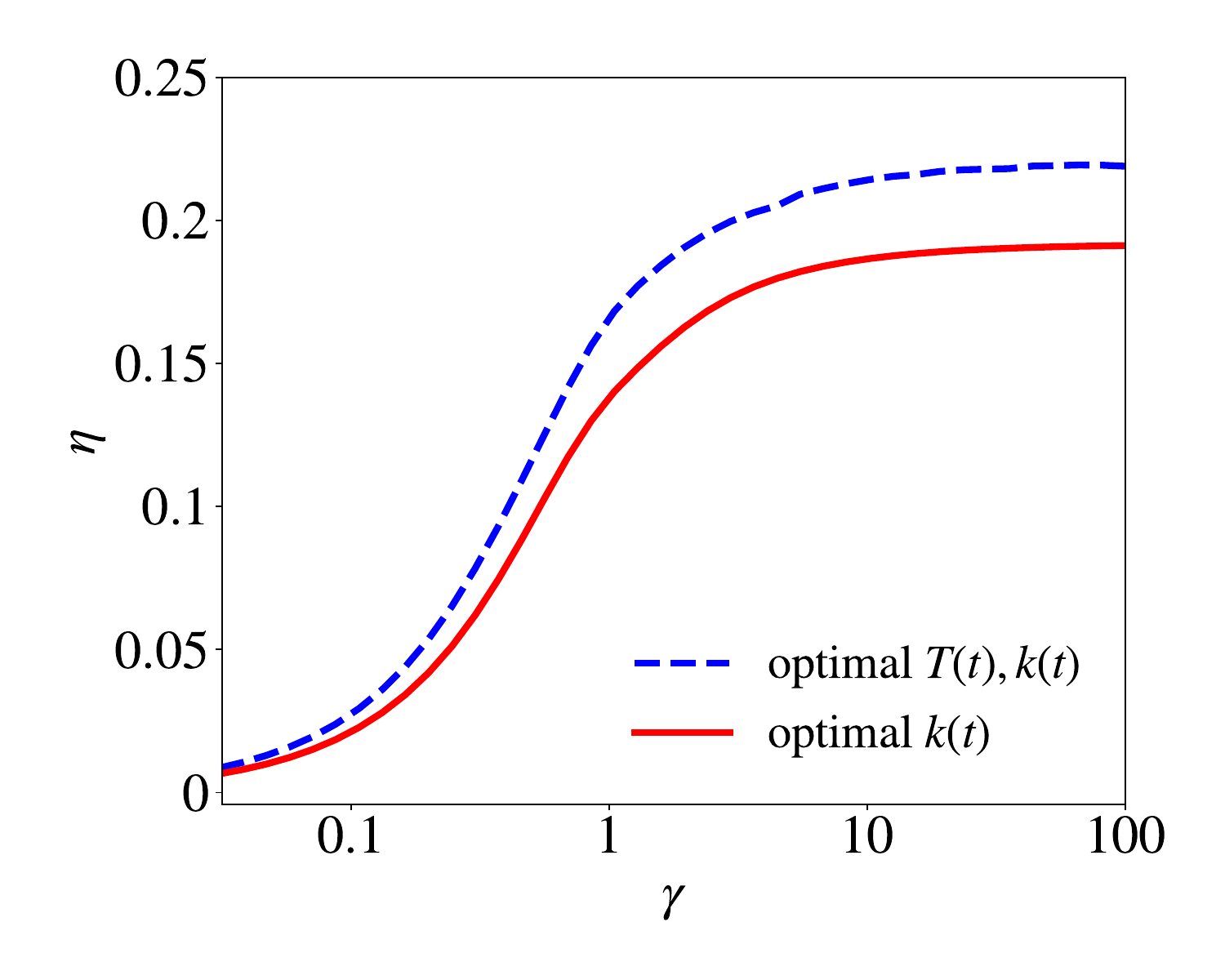}}&
\parbox[c][0.5\textwidth][t]{0.5\textwidth}{
    \raggedright \textbf{(b)}\\
    \includegraphics[width=\linewidth]{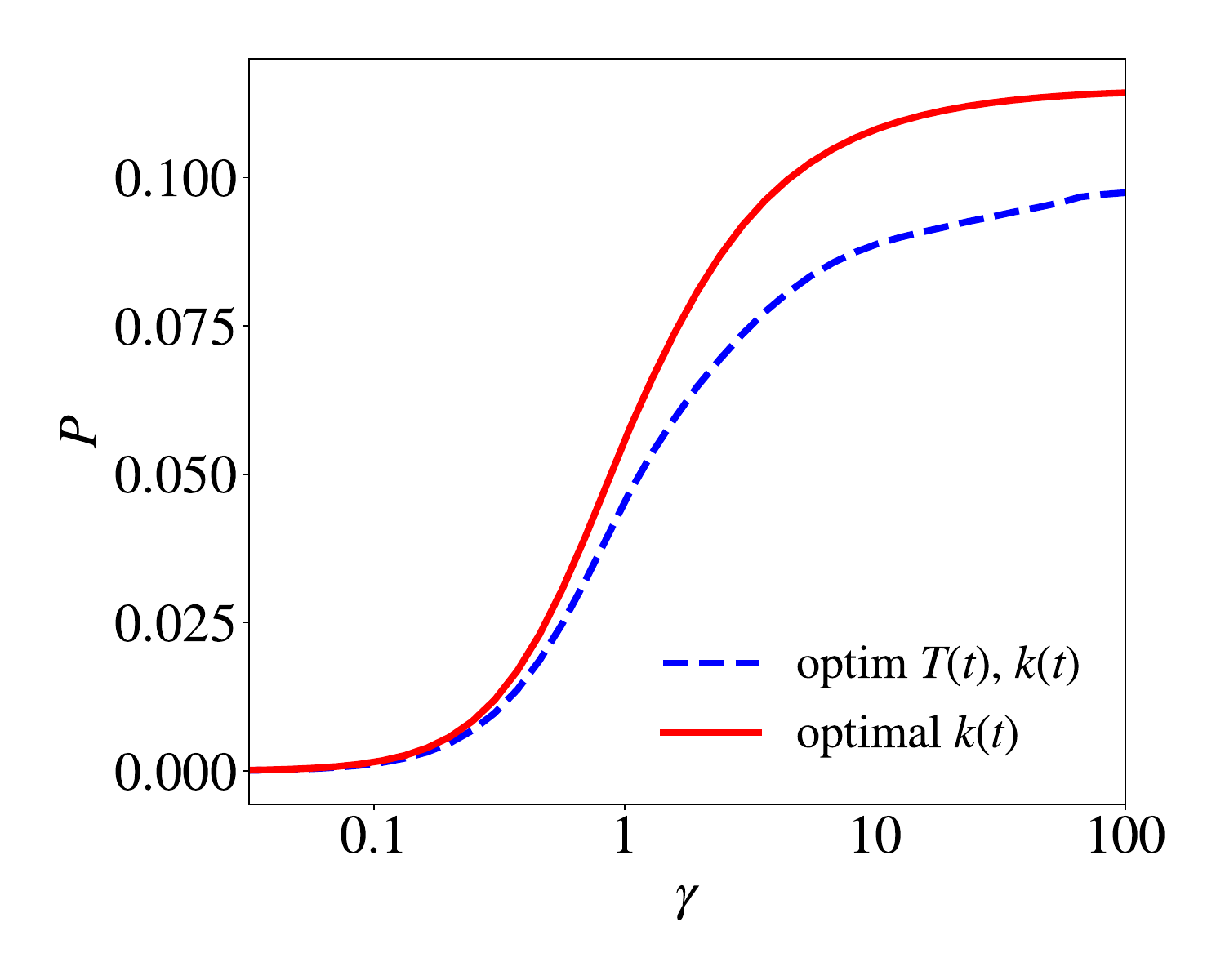}}
\end{tabular}
\vspace{-1cm}
\caption{\textbf{Maximum efficiency and the corresponding power as functions of the damping rate.} (a) The efficiency optimized with respect to (i) stiffness $k(t)$ with temperature $T(t)$ in \eqref{temp_protocol}; and (ii) both stiffness and temperature, as functions of $\gamma$ for $\tau = 4.0$. (b) The corresponding powers.}
\label{eta_vs_gamma}
\end{figure}

In Fig.~\ref{fig:full_eff}, we present the resulting maximum efficiency protocols for \( k(t) \) and \( T(t) \) with a cycle time of \( \tau = 4.0 \). The optimal stiffness protocol in Fig.~\ref{fig:full_eff}(a) exhibits the same qualitative features as the corresponding maximum power protocol in Fig.~\ref{full_pow}(a). However, due to the significant role of heat leakage through momentum degrees of freedom in efficiency, the temperature protocols are very different. Specifically, the temperatures in Fig.~\ref{fig:full_eff}(b) vary between the allowed boundary values much more smoothly than in the case of maximum power or maximum overdamped efficiency. The temperature protocols still include relatively fast transitions; however, due to the otherwise smooth variation, these correspond to smaller changes in temperature than in the previous sections. This is important because the amount of heat leaked through momentum degrees of freedom during rapid heating of the working medium in the time interval \( [t_i, t_f] \), can be roughly estimated using the equipartition theorem as \(\left[ \sigma_v(t_f) - \sigma_v(t_i) \right] / 2m \approx \left[ T(t_f) - T(t_i) \right] / 2 \). As a result, this heat leakage increases with the magnitude of fast temperature changes, where the momentum degrees of freedom cannot keep up with the rapid temperature variation, leading to significant entropy production. The fact that the temperature protocols dwell longer at \( T_- \) than at \( T_+ \) is again a manifestation of the heating and cooling asymmetry~\cite{heating_cooling_asymmetry} discussed in Sec.~\ref{subsec:W_and_P}.

To demonstrate the importance of temperature optimization, we show in  Fig.~\ref{eta_vs_gamma}(a) the maximum efficiency obtained by optimizing stiffness and temperature and partially optimal efficiency obtained just by stiffness optimization and using the fixed temperature protocol in Eq.~\eqref{temp_protocol} as functions of the damping rate $\gamma$ for $\tau = 4$. The corresponding power is plotted in \ref{eta_vs_gamma}(b). The temperature optimization significantly improves the efficiency, particularly for larger values of $\gamma$. The efficiency monotonically decreases with decreasing \( \gamma \) due to the increasing relaxation time, \( \propto 1/\gamma \), of the working medium, which limits the heat exchanged within the fixed cycle time in our calculations. More precisely, the low-damping limit for a fixed \( \tau \) corresponds to a fast-driving regime, where the system is unable to react to the details of the stiffness and temperature protocols but instead relaxes to an equilibrium state corresponding to the average values of \( k(t) \) and \( T(t) \) over the cycle. Thermodynamically, the system effectively interacts with a single bath and, therefore, cannot produce output work, as required by the second law of thermodynamics.
 This also provides an alternative explanation for why power \( P \) vanishes as \( \gamma \to 0 \) in Fig.~\ref{fig:power_vs_gamma}.

For large $\gamma$ the efficiencies in Fig.~\ref{eta_vs_gamma}(a) converge to finite maximum values. We have checked that they can be predicted using the overdamped dynamics in Eq.~\eqref{eq:over0} together with the definition of efficiency
\begin{align}
    \eta =\frac{\int_0^1 -\frac{1}{2} \dot k \sigma_x dt}{\int_0^1 [ \frac{1}{2} m \dot \sigma_v + \frac{1}{2}k\dot \sigma_x ] \theta(m \dot \sigma_v + k\dot \sigma_x)dt},
    \label{eq:eff_over_ok}
\end{align}
with $m \sigma_v(t)=T(t)$.

\section{Discussion and outlook}
\label{sec:conclusion}
In this work, we applied an established algorithm for optimal periodic control~\cite{Craun2015, symmetry2021} to determine the maximum power and efficiency protocols for cyclic Brownian heat engines operating under constraints on potential stiffness and bath temperature. These engines, based on a generally damped Brownian particle confined in a harmonic potential, were subject to experimentally relevant bounds on stiffness and temperature, reflecting practical realizations. Notably, this optimization problem lies beyond the scope of traditional geometric and mass transport methods, which require fixed control or response parameters at specific points in the cycle.

We validated the algorithm against known results for the deeply overdamped regime and subsequently applied it to study optimal protocols for general damping. The resulting maximum power protocols for temperature were approximately piecewise constant, as is often assumed in studies of maximum power. However, upon closer inspection, the apparent jumps in these protocols were linear, with slopes that decreased as the damping rate increased. In contrast to the maximum power stiffness protocols derived for the deeply overdamped~\cite{Schmiedl-Seifert2007, Viktor2022} and underdamped~\cite{Dechant_2017} regimes, the optimal stiffness protocols in the generally damped regime were strongly non-monotonic, even during the isothermal parts of the cycle. These protocols exhibited rapid up-down and down-up variations, with durations that increased as the damping rate decreased. We conjecture that these non-monotonicities aid in the transition between the heat-accepting and heat-releasing phases during rapid bath temperature changes by facilitating the reorientation of the Brownian particle's velocity.

\begin{table}[!t]
\centering
\renewcommand{\arraystretch}{1.3} 
\setlength{\tabcolsep}{8pt} 
\begin{tabular}{|c|||c|c||c|c|||c|c||c|c|}
\hline
\multirow{2}{*}{\diagbox{Model}{Target Function}}
& \multicolumn{2}{|c||}{$\max_{\{k(t),T(t)\}} P$} 
& \multicolumn{2}{|c|||}{$\max_{k(t)} P$} 
& \multicolumn{2}{|c||}{$\max_{\{k(t),T(t)\}} \eta$} 
& \multicolumn{2}{|c|}{$\max_{k(t)} \eta$} \\
\cline{2-9}
& $P$ & $\eta$ & $P$ & $\eta$ & $P$ & $\eta$ & $P$ & $\eta$ \\
\hline \hline
Overdamped Eq.~\eqref{eq:over0}, $\eta_{\rm OD}$ 
& 0.121  & 0.448  & 0.119 & 0.443 & ---  & --- & 0.0005 & 0.749 \\
\hline
Overdamped Eq.~\eqref{eq:over0}, $\eta$ 
& 0.121  & 0.189 & 0.119 & 0.186 & ---  & --- & 0.0005 & 0.005 \\
\hline
General Eq.~\eqref{eq:Sigma}, $\gamma=100$ 
& 0.119  & 0.184 & 0.118 & 0.185 & 0.095 & 0.219 & 0.114 & 0.191  \\
\hline
General Eq.~\eqref{eq:Sigma}, $\gamma=0.5$ 
& 0.030  & 0.102 & 0.026 & 0.094 & 0.021  & 0.117 & 0.026 & 0.099 \\
\hline
\end{tabular}
\caption{Maximum powers and corresponding efficiencies (first four columns) and maximum efficiencies and corresponding powers (last four columns) obtained from our algorithm under various levels of optimization with cycle time $\tau = 4$. For example, $\max_{\{k(t), T(t)\}} P$ means that power has been optimized with respect to both temperature $T(t) \in [1, 4]$ and stiffness $k(t) \in [0.2, 0.8]$. The second row corresponds to the overdamped working medium, but uses the definition of efficiency~\eqref{eq:eff_over_ok}, which takes into account heat leakage through the momentum degrees of freedom. For overdamped dynamics, our algorithm did not converge when optimizing efficiency with respect to temperature, which is why the corresponding entries in the table are marked with dashes.
}
\label{The_second_table}
\end{table}

While the maximum efficiency stiffness protocols were similar to those for maximum power, the maximum efficiency temperature protocols were strongly non-monotonic. This behavior mitigated the detrimental effects of heat leakage through momentum degrees of freedom on efficiency. Notably, temperature optimization—often overlooked when deriving optimal protocols for microscopic heat engines—significantly enhanced the resulting efficiency. To illustrate these improvements, we have summarized exemplary results for optimized power and efficiency in Table \ref{The_second_table}. The significant increase in power due to temperature optimization observed for \( \gamma = 0.5 \) in the table can be attributed to the adjustment of the relative durations of the hot and cold isotherms, as required by the heating-cooling time asymmetry~\cite{heating_cooling_asymmetry}. However, the substantial increase in efficiency can only be ascribed to the non-monotonicity of the temperature protocols.

Besides studying the features of the optimal protocols, we also investigated the variation of maximum power and efficiency as functions of the damping rate. Both quantities monotonically increase with the damping rate. In the strongly underdamped regime, both power and efficiency vanish due to slow system dynamics. In the strongly overdamped regime, these quantities saturate at values determined by the overdamped dynamics. 

These predictions regarding the general damping regime can be experimentally tested in platforms such as linear Paul traps~\cite{SingelAtomHE2016, underdamped_realization_2025}, underdamped mechanical cantilevers~\cite{CilibertoPRE2025}, and noisy electric circuits~\cite{Luchinsky1999,CircuitsEsposito2020}.

The main strength of the presented optimization algorithm lies in its universality with respect to the dynamical equations of the working medium and the constraints on the control parameters. Our study can therefore serve as a foundation for optimizing engines with nonlinear dynamics arising from more complex potentials than the harmonic case, or from the inclusion of feedback~\cite{Quantum_engines}. The method can also be readily extended to target performance measures beyond power and efficiency. Notable examples include power fluctuations~\cite{min_fluctuations2018, HolubecMaxPower2018} and trade-offs between power output and its fluctuations~\cite{HolubecMaxPower2018,Holubec_2022}, both of which are crucial figures of merit in the operation of microscopic heat engines. Furthermore, the algorithm is applicable to the optimization of heat engines coupled to nonequilibrium (or active) baths~\cite{Sood2016, Saha_2019, ViktorPRR2020, Rahul2020, Rahul2024}, a domain currently under intense investigation~\cite{control_active, wiese2024}. 

The approach can also be extended to systems simultaneously in contact with multiple heat baths, a scenario of practical relevance since different parts of a nanomachine may experience distinct thermal environments. Additionally, optimizing information erasure represents another compelling application, and our method can be adapted to such contexts as well~\cite{HYParkPRL22, CilibertoPRE2025}. Finally, the algorithm can support optimal control of other cyclic processes, such as those operating in refrigerators or heat pumps.


\section*{Acknowledgements}
M.C. gratefully acknowledges the University Grants Commission (UGC), India for financial support through the Junior Research Fellowship number $231610187141$.  V.H. acknowledges the support of Charles University through project PRIMUS/22/SCI/009. 
\bibliography{references}

\begin{thebibliography}{53}%
\makeatletter
\providecommand \@ifxundefined [1]{%
 \@ifx{#1\undefined}
}%
\providecommand \@ifnum [1]{%
 \ifnum #1\expandafter \@firstoftwo
 \else \expandafter \@secondoftwo
 \fi
}%
\providecommand \@ifx [1]{%
 \ifx #1\expandafter \@firstoftwo
 \else \expandafter \@secondoftwo
 \fi
}%
\providecommand \natexlab [1]{#1}%
\providecommand \enquote  [1]{``#1''}%
\providecommand \bibnamefont  [1]{#1}%
\providecommand \bibfnamefont [1]{#1}%
\providecommand \citenamefont [1]{#1}%
\providecommand \href@noop [0]{\@secondoftwo}%
\providecommand \href [0]{\begingroup \@sanitize@url \@href}%
\providecommand \@href[1]{\@@startlink{#1}\@@href}%
\providecommand \@@href[1]{\endgroup#1\@@endlink}%
\providecommand \@sanitize@url [0]{\catcode `\\12\catcode `\$12\catcode
  `\&12\catcode `\#12\catcode `\^12\catcode `\_12\catcode `\%12\relax}%
\providecommand \@@startlink[1]{}%
\providecommand \@@endlink[0]{}%
\providecommand \url  [0]{\begingroup\@sanitize@url \@url }%
\providecommand \@url [1]{\endgroup\@href {#1}{\urlprefix }}%
\providecommand \urlprefix  [0]{URL }%
\providecommand \Eprint [0]{\href }%
\providecommand \doibase [0]{https://doi.org/}%
\providecommand \selectlanguage [0]{\@gobble}%
\providecommand \bibinfo  [0]{\@secondoftwo}%
\providecommand \bibfield  [0]{\@secondoftwo}%
\providecommand \translation [1]{[#1]}%
\providecommand \BibitemOpen [0]{}%
\providecommand \bibitemStop [0]{}%
\providecommand \bibitemNoStop [0]{.\EOS\space}%
\providecommand \EOS [0]{\spacefactor3000\relax}%
\providecommand \BibitemShut  [1]{\csname bibitem#1\endcsname}%
\let\auto@bib@innerbib\@empty
\bibitem [{\citenamefont {Mart\'inez}\ \emph {et~al.}(2017)\citenamefont
  {Mart\'inez}, \citenamefont {Rold\'an}, \citenamefont {Dinis},\ and\
  \citenamefont {Rica}}]{ColloidalReview}%
  \BibitemOpen
  \bibfield  {author} {\bibinfo {author} {\bibfnamefont {I.~A.}\ \bibnamefont
  {Mart\'inez}}, \bibinfo {author} {\bibfnamefont {E.}~\bibnamefont
  {Rold\'an}}, \bibinfo {author} {\bibfnamefont {L.}~\bibnamefont {Dinis}},\
  and\ \bibinfo {author} {\bibfnamefont {R.~A.}\ \bibnamefont {Rica}},\
  }\bibfield  {title} {\bibinfo {title} {Colloidal heat engines: a review},\
  }\href {https://doi.org/10.1039/C6SM00923A} {\bibfield  {journal} {\bibinfo
  {journal} {Soft Matter}\ }\textbf {\bibinfo {volume} {13}},\ \bibinfo {pages}
  {22} (\bibinfo {year} {2017})}\BibitemShut {NoStop}%
\bibitem [{\citenamefont {Cangemi}\ \emph {et~al.}(2024)\citenamefont
  {Cangemi}, \citenamefont {Bhadra},\ and\ \citenamefont
  {Levy}}]{Quantum_engines}%
  \BibitemOpen
  \bibfield  {author} {\bibinfo {author} {\bibfnamefont {L.~M.}\ \bibnamefont
  {Cangemi}}, \bibinfo {author} {\bibfnamefont {C.}~\bibnamefont {Bhadra}},\
  and\ \bibinfo {author} {\bibfnamefont {A.}~\bibnamefont {Levy}},\ }\bibfield
  {title} {\bibinfo {title} {Quantum engines and refrigerators},\ }\href
  {https://doi.org/https://doi.org/10.1016/j.physrep.2024.07.001} {\bibfield
  {journal} {\bibinfo  {journal} {Physics Reports}\ }\textbf {\bibinfo {volume}
  {1087}},\ \bibinfo {pages} {1} (\bibinfo {year} {2024})}\BibitemShut
  {NoStop}%
\bibitem [{\citenamefont {Blickle}\ and\ \citenamefont
  {Bechinger}(2012)}]{micro_engine}%
  \BibitemOpen
  \bibfield  {author} {\bibinfo {author} {\bibfnamefont {V.}~\bibnamefont
  {Blickle}}\ and\ \bibinfo {author} {\bibfnamefont {C.}~\bibnamefont
  {Bechinger}},\ }\bibfield  {title} {\bibinfo {title} {Realization of a
  micrometre-sized stochastic heat engine},\ }\href
  {https://doi.org/10.1038/nphys2163} {\bibfield  {journal} {\bibinfo
  {journal} {Nature Physics}\ }\textbf {\bibinfo {volume} {8}},\ \bibinfo
  {pages} {143} (\bibinfo {year} {2012})}\BibitemShut {NoStop}%
\bibitem [{\citenamefont {Mart\'inez}\ \emph {et~al.}(2016)\citenamefont
  {Mart\'inez}, \citenamefont {Rold\'an}, \citenamefont {Dinis}, \citenamefont
  {Petrov}, \citenamefont {Parrondo},\ and\ \citenamefont {Rica}}]{Edgar2016}%
  \BibitemOpen
  \bibfield  {author} {\bibinfo {author} {\bibfnamefont {I.~A.}\ \bibnamefont
  {Mart\'inez}}, \bibinfo {author} {\bibfnamefont {E.}~\bibnamefont
  {Rold\'an}}, \bibinfo {author} {\bibfnamefont {L.}~\bibnamefont {Dinis}},
  \bibinfo {author} {\bibfnamefont {D.}~\bibnamefont {Petrov}}, \bibinfo
  {author} {\bibfnamefont {J.~M.~R.}\ \bibnamefont {Parrondo}},\ and\ \bibinfo
  {author} {\bibfnamefont {R.~A.}\ \bibnamefont {Rica}},\ }\bibfield  {title}
  {\bibinfo {title} {Brownian carnot engine},\ }\href
  {https://doi.org/10.1038/nphys3518} {\bibfield  {journal} {\bibinfo
  {journal} {Nat. Phys.}\ }\textbf {\bibinfo {volume} {12}},\ \bibinfo {pages}
  {67} (\bibinfo {year} {2016})}\BibitemShut {NoStop}%
\bibitem [{\citenamefont {Krishnamurthy}\ \emph {et~al.}(2016)\citenamefont
  {Krishnamurthy}, \citenamefont {Ghosh}, \citenamefont {Chatterji},
  \citenamefont {Ganapathy},\ and\ \citenamefont {Sood}}]{Sood2016}%
  \BibitemOpen
  \bibfield  {author} {\bibinfo {author} {\bibfnamefont {S.}~\bibnamefont
  {Krishnamurthy}}, \bibinfo {author} {\bibfnamefont {S.}~\bibnamefont
  {Ghosh}}, \bibinfo {author} {\bibfnamefont {D.}~\bibnamefont {Chatterji}},
  \bibinfo {author} {\bibfnamefont {R.}~\bibnamefont {Ganapathy}},\ and\
  \bibinfo {author} {\bibfnamefont {A.~K.}\ \bibnamefont {Sood}},\ }\bibfield
  {title} {\bibinfo {title} {A micrometre-sized heat engine operating between
  bacterial reservoirs},\ }\href {https://doi.org/10.1038/nphys3870} {\bibfield
   {journal} {\bibinfo  {journal} {Nature Physics}\ }\textbf {\bibinfo {volume}
  {12}},\ \bibinfo {pages} {1134} (\bibinfo {year} {2016})}\BibitemShut
  {NoStop}%
\bibitem [{\citenamefont {Wei}\ and\ \citenamefont {Chiou}(2005)}]{optical}%
  \BibitemOpen
  \bibfield  {author} {\bibinfo {author} {\bibfnamefont {M.-T.}\ \bibnamefont
  {Wei}}\ and\ \bibinfo {author} {\bibfnamefont {A.}~\bibnamefont {Chiou}},\
  }\bibfield  {title} {\bibinfo {title} {Three-dimensional tracking of brownian
  motion of a particle trapped in optical tweezers with a pair of orthogonal
  tracking beams and the determination of the associated optical force
  constants},\ }\href {https://doi.org/10.1364/OPEX.13.005798} {\bibfield
  {journal} {\bibinfo  {journal} {Opt. Express}\ }\textbf {\bibinfo {volume}
  {13}},\ \bibinfo {pages} {5798} (\bibinfo {year} {2005})}\BibitemShut
  {NoStop}%
\bibitem [{\citenamefont {Deng}\ \emph {et~al.}(2007)\citenamefont {Deng},
  \citenamefont {Bechhoefer},\ and\ \citenamefont {Forde}}]{Deng_2007}%
  \BibitemOpen
  \bibfield  {author} {\bibinfo {author} {\bibfnamefont {Y.}~\bibnamefont
  {Deng}}, \bibinfo {author} {\bibfnamefont {J.}~\bibnamefont {Bechhoefer}},\
  and\ \bibinfo {author} {\bibfnamefont {N.~R.}\ \bibnamefont {Forde}},\
  }\bibfield  {title} {\bibinfo {title} {Brownian motion in a modulated optical
  trap},\ }\href {https://doi.org/10.1088/1464-4258/9/8/S20} {\bibfield
  {journal} {\bibinfo  {journal} {Journal of Optics A: Pure and Applied
  Optics}\ }\textbf {\bibinfo {volume} {9}},\ \bibinfo {pages} {S256} (\bibinfo
  {year} {2007})}\BibitemShut {NoStop}%
\bibitem [{\citenamefont {Berg-Sørensen}\ and\ \citenamefont
  {Flyvbjerg}(2004)}]{optical_power_spectrum}%
  \BibitemOpen
  \bibfield  {author} {\bibinfo {author} {\bibfnamefont {K.}~\bibnamefont
  {Berg-Sørensen}}\ and\ \bibinfo {author} {\bibfnamefont {H.}~\bibnamefont
  {Flyvbjerg}},\ }\bibfield  {title} {\bibinfo {title} {Power spectrum analysis
  for optical tweezers},\ }\href {https://doi.org/10.1063/1.1645654} {\bibfield
   {journal} {\bibinfo  {journal} {Review of Scientific Instruments}\ }\textbf
  {\bibinfo {volume} {75}},\ \bibinfo {pages} {594} (\bibinfo {year}
  {2004})}\BibitemShut {NoStop}%
\bibitem [{\citenamefont {Roßnagel}\ \emph {et~al.}(2016)\citenamefont
  {Roßnagel}, \citenamefont {Dawkins}, \citenamefont {Tolazzi}, \citenamefont
  {Abah}, \citenamefont {Lutz}, \citenamefont {Schmidt-Kaler},\ and\
  \citenamefont {Singer}}]{SingelAtomHE2016}%
  \BibitemOpen
  \bibfield  {author} {\bibinfo {author} {\bibfnamefont {J.}~\bibnamefont
  {Roßnagel}}, \bibinfo {author} {\bibfnamefont {S.~T.}\ \bibnamefont
  {Dawkins}}, \bibinfo {author} {\bibfnamefont {K.~N.}\ \bibnamefont
  {Tolazzi}}, \bibinfo {author} {\bibfnamefont {O.}~\bibnamefont {Abah}},
  \bibinfo {author} {\bibfnamefont {E.}~\bibnamefont {Lutz}}, \bibinfo {author}
  {\bibfnamefont {F.}~\bibnamefont {Schmidt-Kaler}},\ and\ \bibinfo {author}
  {\bibfnamefont {K.}~\bibnamefont {Singer}},\ }\bibfield  {title} {\bibinfo
  {title} {A single-atom heat engine},\ }\href
  {https://doi.org/10.1126/science.aad6320} {\bibfield  {journal} {\bibinfo
  {journal} {Science}\ }\textbf {\bibinfo {volume} {352}},\ \bibinfo {pages}
  {325} (\bibinfo {year} {2016})},\ \Eprint
  {https://arxiv.org/abs/https://www.science.org/doi/pdf/10.1126/science.aad6320}
  {https://www.science.org/doi/pdf/10.1126/science.aad6320} \BibitemShut
  {NoStop}%
\bibitem [{\citenamefont {Message}\ \emph {et~al.}(2025)\citenamefont
  {Message}, \citenamefont {Cerisola}, \citenamefont {Pritchett}, \citenamefont
  {O'Flynn}, \citenamefont {Ren}, \citenamefont {Rashid}, \citenamefont
  {Anders},\ and\ \citenamefont {Millen}}]{underdamped_realization_2025}%
  \BibitemOpen
  \bibfield  {author} {\bibinfo {author} {\bibfnamefont {M.}~\bibnamefont
  {Message}}, \bibinfo {author} {\bibfnamefont {F.}~\bibnamefont {Cerisola}},
  \bibinfo {author} {\bibfnamefont {J.~D.}\ \bibnamefont {Pritchett}}, \bibinfo
  {author} {\bibfnamefont {K.}~\bibnamefont {O'Flynn}}, \bibinfo {author}
  {\bibfnamefont {Y.}~\bibnamefont {Ren}}, \bibinfo {author} {\bibfnamefont
  {M.}~\bibnamefont {Rashid}}, \bibinfo {author} {\bibfnamefont
  {J.}~\bibnamefont {Anders}},\ and\ \bibinfo {author} {\bibfnamefont
  {J.}~\bibnamefont {Millen}},\ }\href {https://arxiv.org/abs/2501.03677}
  {\bibinfo {title} {Extreme-temperature single-particle heat engine}},\
  \bibinfo {howpublished} {\url{https://arxiv.org/abs/2501.03677}} (\bibinfo
  {year} {2025})\BibitemShut {NoStop}%
\bibitem [{\citenamefont {Holubec}\ and\ \citenamefont
  {Ryabov}(2021)}]{Holubec_2022}%
  \BibitemOpen
  \bibfield  {author} {\bibinfo {author} {\bibfnamefont {V.}~\bibnamefont
  {Holubec}}\ and\ \bibinfo {author} {\bibfnamefont {A.}~\bibnamefont
  {Ryabov}},\ }\bibfield  {title} {\bibinfo {title} {Fluctuations in heat
  engines},\ }\href {https://doi.org/10.1088/1751-8121/ac3aac} {\bibfield
  {journal} {\bibinfo  {journal} {Journal of Physics A: Mathematical and
  Theoretical}\ }\textbf {\bibinfo {volume} {55}},\ \bibinfo {pages} {013001}
  (\bibinfo {year} {2021})}\BibitemShut {NoStop}%
\bibitem [{\citenamefont {Berry}\ \emph {et~al.}(2000)\citenamefont {Berry},
  \citenamefont {Kazakov}, \citenamefont {Sieniutycz}, \citenamefont {Szwast},\
  and\ \citenamefont {Tsirlin}}]{berry2000thermodynamic}%
  \BibitemOpen
  \bibfield  {author} {\bibinfo {author} {\bibfnamefont {R.}~\bibnamefont
  {Berry}}, \bibinfo {author} {\bibfnamefont {V.}~\bibnamefont {Kazakov}},
  \bibinfo {author} {\bibfnamefont {S.}~\bibnamefont {Sieniutycz}}, \bibinfo
  {author} {\bibfnamefont {Z.}~\bibnamefont {Szwast}},\ and\ \bibinfo {author}
  {\bibfnamefont {A.}~\bibnamefont {Tsirlin}},\ }\href
  {https://books.google.cz/books?id=ZPqOQgAACAAJ} {\emph {\bibinfo {title}
  {Thermodynamic Optimization of Finite-Time Processes}}}\ (\bibinfo
  {publisher} {Wiley},\ \bibinfo {year} {2000})\BibitemShut {NoStop}%
\bibitem [{\citenamefont {Martino~Bardi}(2009)}]{Bellman}%
  \BibitemOpen
  \bibfield  {author} {\bibinfo {author} {\bibfnamefont {I.~C.-D.}\
  \bibnamefont {Martino~Bardi}},\ }\href
  {https://doi.org/10.1007/978-0-8176-4755-1} {\emph {\bibinfo {title} {Optimal
  Control and Viscosity Solutions of Hamilton-Jacobi-Bellman Equations}}}\
  (\bibinfo  {publisher} {Birkhäuser Boston, MA},\ \bibinfo {year} {2009})\
  pp.\ \bibinfo {pages} {97--110}\BibitemShut {NoStop}%
\bibitem [{\citenamefont {Chen}\ \emph {et~al.}(2021)\citenamefont {Chen},
  \citenamefont {Georgiou},\ and\ \citenamefont {Pavon}}]{optimal_transport}%
  \BibitemOpen
  \bibfield  {author} {\bibinfo {author} {\bibfnamefont {Y.}~\bibnamefont
  {Chen}}, \bibinfo {author} {\bibfnamefont {T.~T.}\ \bibnamefont {Georgiou}},\
  and\ \bibinfo {author} {\bibfnamefont {M.}~\bibnamefont {Pavon}},\ }\bibfield
   {title} {\bibinfo {title} {Optimal transport in systems and control},\
  }\href
  {https://doi.org/https://doi.org/10.1146/annurev-control-070220-100858}
  {\bibfield  {journal} {\bibinfo  {journal} {Annual Review of Control,
  Robotics, and Autonomous Systems}\ }\textbf {\bibinfo {volume} {4}},\
  \bibinfo {pages} {89} (\bibinfo {year} {2021})}\BibitemShut {NoStop}%
\bibitem [{\citenamefont {Aurell}\ \emph {et~al.}(2011)\citenamefont {Aurell},
  \citenamefont {Mej\'{\i}a-Monasterio},\ and\ \citenamefont
  {Muratore-Ginanneschi}}]{opt_proto_transport}%
  \BibitemOpen
  \bibfield  {author} {\bibinfo {author} {\bibfnamefont {E.}~\bibnamefont
  {Aurell}}, \bibinfo {author} {\bibfnamefont {C.}~\bibnamefont
  {Mej\'{\i}a-Monasterio}},\ and\ \bibinfo {author} {\bibfnamefont
  {P.}~\bibnamefont {Muratore-Ginanneschi}},\ }\bibfield  {title} {\bibinfo
  {title} {Optimal protocols and optimal transport in stochastic
  thermodynamics},\ }\href {https://doi.org/10.1103/PhysRevLett.106.250601}
  {\bibfield  {journal} {\bibinfo  {journal} {Phys. Rev. Lett.}\ }\textbf
  {\bibinfo {volume} {106}},\ \bibinfo {pages} {250601} (\bibinfo {year}
  {2011})}\BibitemShut {NoStop}%
\bibitem [{\citenamefont {Sivak}\ and\ \citenamefont
  {Crooks}(2012)}]{Crooks2012}%
  \BibitemOpen
  \bibfield  {author} {\bibinfo {author} {\bibfnamefont {D.~A.}\ \bibnamefont
  {Sivak}}\ and\ \bibinfo {author} {\bibfnamefont {G.~E.}\ \bibnamefont
  {Crooks}},\ }\bibfield  {title} {\bibinfo {title} {Thermodynamic metrics and
  optimal paths},\ }\href {https://doi.org/10.1103/PhysRevLett.108.190602}
  {\bibfield  {journal} {\bibinfo  {journal} {Phys. Rev. Lett.}\ }\textbf
  {\bibinfo {volume} {108}},\ \bibinfo {pages} {190602} (\bibinfo {year}
  {2012})}\BibitemShut {NoStop}%
\bibitem [{\citenamefont {Terr\'en~Alonso}\ \emph {et~al.}(2022)\citenamefont
  {Terr\'en~Alonso}, \citenamefont {Abiuso}, \citenamefont {Perarnau-Llobet},\
  and\ \citenamefont {Arrachea}}]{Geometric_Optimization}%
  \BibitemOpen
  \bibfield  {author} {\bibinfo {author} {\bibfnamefont {P.}~\bibnamefont
  {Terr\'en~Alonso}}, \bibinfo {author} {\bibfnamefont {P.}~\bibnamefont
  {Abiuso}}, \bibinfo {author} {\bibfnamefont {M.}~\bibnamefont
  {Perarnau-Llobet}},\ and\ \bibinfo {author} {\bibfnamefont {L.}~\bibnamefont
  {Arrachea}},\ }\bibfield  {title} {\bibinfo {title} {Geometric optimization
  of nonequilibrium adiabatic thermal machines and implementation in a qubit
  system},\ }\href {https://doi.org/10.1103/PRXQuantum.3.010326} {\bibfield
  {journal} {\bibinfo  {journal} {PRX Quantum}\ }\textbf {\bibinfo {volume}
  {3}},\ \bibinfo {pages} {010326} (\bibinfo {year} {2022})}\BibitemShut
  {NoStop}%
\bibitem [{\citenamefont {Gu\'ery-Odelin}\ \emph {et~al.}(2019)\citenamefont
  {Gu\'ery-Odelin}, \citenamefont {Ruschhaupt}, \citenamefont {Kiely},
  \citenamefont {Torrontegui}, \citenamefont {Mart\'{\i}nez-Garaot},\ and\
  \citenamefont {Muga}}]{Shortcuts_to_adiabaticity}%
  \BibitemOpen
  \bibfield  {author} {\bibinfo {author} {\bibfnamefont {D.}~\bibnamefont
  {Gu\'ery-Odelin}}, \bibinfo {author} {\bibfnamefont {A.}~\bibnamefont
  {Ruschhaupt}}, \bibinfo {author} {\bibfnamefont {A.}~\bibnamefont {Kiely}},
  \bibinfo {author} {\bibfnamefont {E.}~\bibnamefont {Torrontegui}}, \bibinfo
  {author} {\bibfnamefont {S.}~\bibnamefont {Mart\'{\i}nez-Garaot}},\ and\
  \bibinfo {author} {\bibfnamefont {J.~G.}\ \bibnamefont {Muga}},\ }\bibfield
  {title} {\bibinfo {title} {Shortcuts to adiabaticity: Concepts, methods, and
  applications},\ }\href {https://doi.org/10.1103/RevModPhys.91.045001}
  {\bibfield  {journal} {\bibinfo  {journal} {Rev. Mod. Phys.}\ }\textbf
  {\bibinfo {volume} {91}},\ \bibinfo {pages} {045001} (\bibinfo {year}
  {2019})}\BibitemShut {NoStop}%
\bibitem [{\citenamefont {Casert}\ and\ \citenamefont
  {Whitelam}(2024)}]{control_active}%
  \BibitemOpen
  \bibfield  {author} {\bibinfo {author} {\bibfnamefont {C.}~\bibnamefont
  {Casert}}\ and\ \bibinfo {author} {\bibfnamefont {S.}~\bibnamefont
  {Whitelam}},\ }\bibfield  {title} {\bibinfo {title} {Learning protocols for
  the fast and efficient control of active matter},\ }\href
  {https://doi.org/10.1038/s41467-024-52878-2} {\bibfield  {journal} {\bibinfo
  {journal} {Nature Communications}\ }\textbf {\bibinfo {volume} {15}},\
  \bibinfo {pages} {9128} (\bibinfo {year} {2024})}\BibitemShut {NoStop}%
\bibitem [{\citenamefont {Ye}\ \emph {et~al.}(2022)\citenamefont {Ye},
  \citenamefont {Cerisola}, \citenamefont {Abiuso}, \citenamefont {Anders},
  \citenamefont {Perarnau-Llobet},\ and\ \citenamefont {Holubec}}]{Viktor2022}%
  \BibitemOpen
  \bibfield  {author} {\bibinfo {author} {\bibfnamefont {Z.}~\bibnamefont
  {Ye}}, \bibinfo {author} {\bibfnamefont {F.}~\bibnamefont {Cerisola}},
  \bibinfo {author} {\bibfnamefont {P.}~\bibnamefont {Abiuso}}, \bibinfo
  {author} {\bibfnamefont {J.}~\bibnamefont {Anders}}, \bibinfo {author}
  {\bibfnamefont {M.}~\bibnamefont {Perarnau-Llobet}},\ and\ \bibinfo {author}
  {\bibfnamefont {V.}~\bibnamefont {Holubec}},\ }\bibfield  {title} {\bibinfo
  {title} {Optimal finite-time heat engines under constrained control},\ }\href
  {https://doi.org/10.1103/PhysRevResearch.4.043130} {\bibfield  {journal}
  {\bibinfo  {journal} {Phys. Rev. Res.}\ }\textbf {\bibinfo {volume} {4}},\
  \bibinfo {pages} {043130} (\bibinfo {year} {2022})}\BibitemShut {NoStop}%
\bibitem [{\citenamefont {Holubec}\ and\ \citenamefont
  {Ryabov}(2018)}]{HolubecMaxPower2018}%
  \BibitemOpen
  \bibfield  {author} {\bibinfo {author} {\bibfnamefont {V.}~\bibnamefont
  {Holubec}}\ and\ \bibinfo {author} {\bibfnamefont {A.}~\bibnamefont
  {Ryabov}},\ }\bibfield  {title} {\bibinfo {title} {Cycling tames power
  fluctuations near optimum efficiency},\ }\href
  {https://doi.org/10.1103/PhysRevLett.121.120601} {\bibfield  {journal}
  {\bibinfo  {journal} {Phys. Rev. Lett.}\ }\textbf {\bibinfo {volume} {121}},\
  \bibinfo {pages} {120601} (\bibinfo {year} {2018})}\BibitemShut {NoStop}%
\bibitem [{\citenamefont {Polettini}\ and\ \citenamefont
  {Esposito}(2017)}]{Polettini_2017}%
  \BibitemOpen
  \bibfield  {author} {\bibinfo {author} {\bibfnamefont {M.}~\bibnamefont
  {Polettini}}\ and\ \bibinfo {author} {\bibfnamefont {M.}~\bibnamefont
  {Esposito}},\ }\bibfield  {title} {\bibinfo {title} {Carnot efficiency at
  divergent power output},\ }\href
  {https://doi.org/10.1209/0295-5075/118/40003} {\bibfield  {journal} {\bibinfo
   {journal} {Europhysics Letters}\ }\textbf {\bibinfo {volume} {118}},\
  \bibinfo {pages} {40003} (\bibinfo {year} {2017})}\BibitemShut {NoStop}%
\bibitem [{\citenamefont {Schmiedl}\ and\ \citenamefont
  {Seifert}(2007)}]{Schmiedl-Seifert2007}%
  \BibitemOpen
  \bibfield  {author} {\bibinfo {author} {\bibfnamefont {T.}~\bibnamefont
  {Schmiedl}}\ and\ \bibinfo {author} {\bibfnamefont {U.}~\bibnamefont
  {Seifert}},\ }\bibfield  {title} {\bibinfo {title} {Efficiency at maximum
  power: An analytically solvable model for stochastic heat engines},\ }\href
  {https://doi.org/10.1209/0295-5075/81/20003} {\bibfield  {journal} {\bibinfo
  {journal} {Europhysics Letters}\ }\textbf {\bibinfo {volume} {81}},\ \bibinfo
  {pages} {20003} (\bibinfo {year} {2007})}\BibitemShut {NoStop}%
\bibitem [{\citenamefont {Loos}\ \emph {et~al.}(2024)\citenamefont {Loos},
  \citenamefont {Monter}, \citenamefont {Ginot},\ and\ \citenamefont
  {Bechinger}}]{viscoelastic_transport_2024}%
  \BibitemOpen
  \bibfield  {author} {\bibinfo {author} {\bibfnamefont {S.~A.~M.}\
  \bibnamefont {Loos}}, \bibinfo {author} {\bibfnamefont {S.}~\bibnamefont
  {Monter}}, \bibinfo {author} {\bibfnamefont {F.}~\bibnamefont {Ginot}},\ and\
  \bibinfo {author} {\bibfnamefont {C.}~\bibnamefont {Bechinger}},\ }\bibfield
  {title} {\bibinfo {title} {Universal symmetry of optimal control at the
  microscale},\ }\href {https://doi.org/10.1103/PhysRevX.14.021032} {\bibfield
  {journal} {\bibinfo  {journal} {Phys. Rev. X}\ }\textbf {\bibinfo {volume}
  {14}},\ \bibinfo {pages} {021032} (\bibinfo {year} {2024})}\BibitemShut
  {NoStop}%
\bibitem [{\citenamefont {Blaber}\ and\ \citenamefont
  {Sivak}(2023)}]{Blaber_2023}%
  \BibitemOpen
  \bibfield  {author} {\bibinfo {author} {\bibfnamefont {S.}~\bibnamefont
  {Blaber}}\ and\ \bibinfo {author} {\bibfnamefont {D.~A.}\ \bibnamefont
  {Sivak}},\ }\bibfield  {title} {\bibinfo {title} {Optimal control in
  stochastic thermodynamics},\ }\href
  {https://doi.org/10.1088/2399-6528/acbf04} {\bibfield  {journal} {\bibinfo
  {journal} {Journal of Physics Communications}\ }\textbf {\bibinfo {volume}
  {7}},\ \bibinfo {pages} {033001} (\bibinfo {year} {2023})}\BibitemShut
  {NoStop}%
\bibitem [{\citenamefont {Then}\ and\ \citenamefont {Engel}(2008)}]{Then2008}%
  \BibitemOpen
  \bibfield  {author} {\bibinfo {author} {\bibfnamefont {H.}~\bibnamefont
  {Then}}\ and\ \bibinfo {author} {\bibfnamefont {A.}~\bibnamefont {Engel}},\
  }\bibfield  {title} {\bibinfo {title} {Computing the optimal protocol for
  finite-time processes in stochastic thermodynamics},\ }\href
  {https://doi.org/10.1103/PhysRevE.77.041105} {\bibfield  {journal} {\bibinfo
  {journal} {Phys. Rev. E}\ }\textbf {\bibinfo {volume} {77}},\ \bibinfo
  {pages} {041105} (\bibinfo {year} {2008})}\BibitemShut {NoStop}%
\bibitem [{\citenamefont {Plata}\ \emph {et~al.}(2019)\citenamefont {Plata},
  \citenamefont {Gu\'ery-Odelin}, \citenamefont {Trizac},\ and\ \citenamefont
  {Prados}}]{Plata2019}%
  \BibitemOpen
  \bibfield  {author} {\bibinfo {author} {\bibfnamefont {C.~A.}\ \bibnamefont
  {Plata}}, \bibinfo {author} {\bibfnamefont {D.}~\bibnamefont
  {Gu\'ery-Odelin}}, \bibinfo {author} {\bibfnamefont {E.}~\bibnamefont
  {Trizac}},\ and\ \bibinfo {author} {\bibfnamefont {A.}~\bibnamefont
  {Prados}},\ }\bibfield  {title} {\bibinfo {title} {Optimal work in a harmonic
  trap with bounded stiffness},\ }\href
  {https://doi.org/10.1103/PhysRevE.99.012140} {\bibfield  {journal} {\bibinfo
  {journal} {Phys. Rev. E}\ }\textbf {\bibinfo {volume} {99}},\ \bibinfo
  {pages} {012140} (\bibinfo {year} {2019})}\BibitemShut {NoStop}%
\bibitem [{\citenamefont {Majumdar}\ \emph {et~al.}(2025)\citenamefont
  {Majumdar}, \citenamefont {Chatterjee},\ and\ \citenamefont
  {Marathe}}]{MAJUMDAR2025130278}%
  \BibitemOpen
  \bibfield  {author} {\bibinfo {author} {\bibfnamefont {R.}~\bibnamefont
  {Majumdar}}, \bibinfo {author} {\bibfnamefont {M.}~\bibnamefont
  {Chatterjee}},\ and\ \bibinfo {author} {\bibfnamefont {R.}~\bibnamefont
  {Marathe}},\ }\bibfield  {title} {\bibinfo {title} {Optimizing power and
  efficiency of a single spin heat engine},\ }\href
  {https://doi.org/https://doi.org/10.1016/j.physa.2024.130278} {\bibfield
  {journal} {\bibinfo  {journal} {Physica A: Statistical Mechanics and its
  Applications}\ }\textbf {\bibinfo {volume} {658}},\ \bibinfo {pages} {130278}
  (\bibinfo {year} {2025})}\BibitemShut {NoStop}%
\bibitem [{\citenamefont {Dechant}\ \emph {et~al.}(2017)\citenamefont
  {Dechant}, \citenamefont {Kiesel},\ and\ \citenamefont
  {Lutz}}]{Dechant_2017}%
  \BibitemOpen
  \bibfield  {author} {\bibinfo {author} {\bibfnamefont {A.}~\bibnamefont
  {Dechant}}, \bibinfo {author} {\bibfnamefont {N.}~\bibnamefont {Kiesel}},\
  and\ \bibinfo {author} {\bibfnamefont {E.}~\bibnamefont {Lutz}},\ }\bibfield
  {title} {\bibinfo {title} {Underdamped stochastic heat engine at maximum
  efficiency},\ }\href {https://doi.org/10.1209/0295-5075/119/50003} {\bibfield
   {journal} {\bibinfo  {journal} {Europhysics Letters}\ }\textbf {\bibinfo
  {volume} {119}},\ \bibinfo {pages} {50003} (\bibinfo {year}
  {2017})}\BibitemShut {NoStop}%
\bibitem [{\citenamefont {Frim}\ and\ \citenamefont
  {DeWeese}(2022)}]{optimal_geodesic_brownian}%
  \BibitemOpen
  \bibfield  {author} {\bibinfo {author} {\bibfnamefont {A.~G.}\ \bibnamefont
  {Frim}}\ and\ \bibinfo {author} {\bibfnamefont {M.~R.}\ \bibnamefont
  {DeWeese}},\ }\bibfield  {title} {\bibinfo {title} {Optimal finite-time
  brownian carnot engine},\ }\href
  {https://doi.org/10.1103/PhysRevE.105.L052103} {\bibfield  {journal}
  {\bibinfo  {journal} {Phys. Rev. E}\ }\textbf {\bibinfo {volume} {105}},\
  \bibinfo {pages} {L052103} (\bibinfo {year} {2022})}\BibitemShut {NoStop}%
\bibitem [{\citenamefont {Brandner}\ and\ \citenamefont
  {Saito}(2020)}]{Brandner2020}%
  \BibitemOpen
  \bibfield  {author} {\bibinfo {author} {\bibfnamefont {K.}~\bibnamefont
  {Brandner}}\ and\ \bibinfo {author} {\bibfnamefont {K.}~\bibnamefont
  {Saito}},\ }\bibfield  {title} {\bibinfo {title} {Thermodynamic geometry of
  microscopic heat engines},\ }\href
  {https://doi.org/10.1103/PhysRevLett.124.040602} {\bibfield  {journal}
  {\bibinfo  {journal} {Phys. Rev. Lett.}\ }\textbf {\bibinfo {volume} {124}},\
  \bibinfo {pages} {040602} (\bibinfo {year} {2020})}\BibitemShut {NoStop}%
\bibitem [{\citenamefont {Craun}\ and\ \citenamefont
  {Bamieh}(2015)}]{Craun2015}%
  \BibitemOpen
  \bibfield  {author} {\bibinfo {author} {\bibfnamefont {M.}~\bibnamefont
  {Craun}}\ and\ \bibinfo {author} {\bibfnamefont {B.}~\bibnamefont {Bamieh}},\
  }\bibfield  {title} {\bibinfo {title} {{Optimal Periodic Control of an Ideal
  Stirling Engine Model}},\ }\href {https://doi.org/10.1115/1.4029682}
  {\bibfield  {journal} {\bibinfo  {journal} {Journal of Dynamic Systems,
  Measurement, and Control}\ }\textbf {\bibinfo {volume} {137}},\ \bibinfo
  {pages} {071002} (\bibinfo {year} {2015})}\BibitemShut {NoStop}%
\bibitem [{\citenamefont {Paul}\ and\ \citenamefont
  {Hoffmann}(2021)}]{symmetry2021}%
  \BibitemOpen
  \bibfield  {author} {\bibinfo {author} {\bibfnamefont {R.}~\bibnamefont
  {Paul}}\ and\ \bibinfo {author} {\bibfnamefont {K.~H.}\ \bibnamefont
  {Hoffmann}},\ }\href {https://www.mdpi.com/2073-8994/13/5/873} {\bibinfo
  {title} {Cyclic control optimization algorithm for stirling engines}},\
  \bibinfo {howpublished} {\emph{Symmetry},~\textbf{13}(5), 873} (\bibinfo
  {year} {2021})\BibitemShut {NoStop}%
\bibitem [{\citenamefont {For\~ao}\ \emph {et~al.}(2024)\citenamefont
  {For\~ao}, \citenamefont {Filho}, \citenamefont {Akasaki},\ and\
  \citenamefont {Fiore}}]{collisional_engines}%
  \BibitemOpen
  \bibfield  {author} {\bibinfo {author} {\bibfnamefont {G.~A.~L.}\
  \bibnamefont {For\~ao}}, \bibinfo {author} {\bibfnamefont {F.~S.}\
  \bibnamefont {Filho}}, \bibinfo {author} {\bibfnamefont {B.~A.~N.}\
  \bibnamefont {Akasaki}},\ and\ \bibinfo {author} {\bibfnamefont {C.~E.}\
  \bibnamefont {Fiore}},\ }\bibfield  {title} {\bibinfo {title} {Thermodynamics
  of underdamped brownian collisional engines: General features and resonant
  phenomena},\ }\href {https://doi.org/10.1103/PhysRevE.110.054125} {\bibfield
  {journal} {\bibinfo  {journal} {Phys. Rev. E}\ }\textbf {\bibinfo {volume}
  {110}},\ \bibinfo {pages} {054125} (\bibinfo {year} {2024})}\BibitemShut
  {NoStop}%
\bibitem [{\citenamefont {Serra-Garcia}\ \emph {et~al.}(2016)\citenamefont
  {Serra-Garcia}, \citenamefont {Foehr}, \citenamefont {Moler\'on},
  \citenamefont {Lydon}, \citenamefont {Chong},\ and\ \citenamefont
  {Daraio}}]{cantilever}%
  \BibitemOpen
  \bibfield  {author} {\bibinfo {author} {\bibfnamefont {M.}~\bibnamefont
  {Serra-Garcia}}, \bibinfo {author} {\bibfnamefont {A.}~\bibnamefont {Foehr}},
  \bibinfo {author} {\bibfnamefont {M.}~\bibnamefont {Moler\'on}}, \bibinfo
  {author} {\bibfnamefont {J.}~\bibnamefont {Lydon}}, \bibinfo {author}
  {\bibfnamefont {C.}~\bibnamefont {Chong}},\ and\ \bibinfo {author}
  {\bibfnamefont {C.}~\bibnamefont {Daraio}},\ }\bibfield  {title} {\bibinfo
  {title} {Mechanical autonomous stochastic heat engine},\ }\href
  {https://doi.org/10.1103/PhysRevLett.117.010602} {\bibfield  {journal}
  {\bibinfo  {journal} {Phys. Rev. Lett.}\ }\textbf {\bibinfo {volume} {117}},\
  \bibinfo {pages} {010602} (\bibinfo {year} {2016})}\BibitemShut {NoStop}%
\bibitem [{\citenamefont {Jarzynski}(1997)}]{Jarzynski1997}%
  \BibitemOpen
  \bibfield  {author} {\bibinfo {author} {\bibfnamefont {C.}~\bibnamefont
  {Jarzynski}},\ }\bibfield  {title} {\bibinfo {title} {Nonequilibrium equality
  for free energy differences},\ }\href
  {https://doi.org/10.1103/PhysRevLett.78.2690} {\bibfield  {journal} {\bibinfo
   {journal} {Phys. Rev. Lett.}\ }\textbf {\bibinfo {volume} {78}},\ \bibinfo
  {pages} {2690} (\bibinfo {year} {1997})}\BibitemShut {NoStop}%
\bibitem [{\citenamefont {Wiese}\ \emph {et~al.}(2024)\citenamefont {Wiese},
  \citenamefont {Kroy},\ and\ \citenamefont {Holubec}}]{wiese2024}%
  \BibitemOpen
  \bibfield  {author} {\bibinfo {author} {\bibfnamefont {R.}~\bibnamefont
  {Wiese}}, \bibinfo {author} {\bibfnamefont {K.}~\bibnamefont {Kroy}},\ and\
  \bibinfo {author} {\bibfnamefont {V.}~\bibnamefont {Holubec}},\ }\bibfield
  {title} {\bibinfo {title} {Modeling the efficiency and effective temperature
  of bacterial heat engines},\ }\href
  {https://doi.org/10.1103/PhysRevE.110.064608} {\bibfield  {journal} {\bibinfo
   {journal} {Phys. Rev. E}\ }\textbf {\bibinfo {volume} {110}},\ \bibinfo
  {pages} {064608} (\bibinfo {year} {2024})}\BibitemShut {NoStop}%
\bibitem [{\citenamefont {Ginot}\ and\ \citenamefont
  {Bechinger}(2024)}]{Ginot2024}%
  \BibitemOpen
  \bibfield  {author} {\bibinfo {author} {\bibfnamefont {F.}~\bibnamefont
  {Ginot}}\ and\ \bibinfo {author} {\bibfnamefont {C.}~\bibnamefont
  {Bechinger}},\ }\href@noop {} {\bibinfo {title} {Energy recuperation of
  driven colloids in non-equilibrium baths}},\ \bibinfo {howpublished}
  {\url{https://doi.org/10.21203/rs.3.rs-5261412/v1}} (\bibinfo {year}
  {2024}),\ \bibinfo {note} {pREPRINT (Version 1)}\BibitemShut {NoStop}%
\bibitem [{\citenamefont {Arold}\ \emph {et~al.}(2018)\citenamefont {Arold},
  \citenamefont {Dechant},\ and\ \citenamefont {Lutz}}]{heat_leakage}%
  \BibitemOpen
  \bibfield  {author} {\bibinfo {author} {\bibfnamefont {D.}~\bibnamefont
  {Arold}}, \bibinfo {author} {\bibfnamefont {A.}~\bibnamefont {Dechant}},\
  and\ \bibinfo {author} {\bibfnamefont {E.}~\bibnamefont {Lutz}},\ }\bibfield
  {title} {\bibinfo {title} {Heat leakage in overdamped harmonic systems},\
  }\href {https://doi.org/10.1103/PhysRevE.97.022131} {\bibfield  {journal}
  {\bibinfo  {journal} {Phys. Rev. E}\ }\textbf {\bibinfo {volume} {97}},\
  \bibinfo {pages} {022131} (\bibinfo {year} {2018})}\BibitemShut {NoStop}%
\bibitem [{\citenamefont {Brunt}(2006)}]{Brunt_COV}%
  \BibitemOpen
  \bibfield  {author} {\bibinfo {author} {\bibfnamefont {B.}~\bibnamefont
  {Brunt}},\ }\href {https://doi.org/10.1007/b97436} {\emph {\bibinfo {title}
  {The Calculus of Variations}}}\ (\bibinfo  {publisher} {Springer New York,
  NY},\ \bibinfo {year} {2006})\ pp.\ \bibinfo {pages} {73--77}\BibitemShut
  {NoStop}%
\bibitem [{\citenamefont {Horn}\ and\ \citenamefont {Lin}(1967)}]{Horn1967}%
  \BibitemOpen
  \bibfield  {author} {\bibinfo {author} {\bibfnamefont {F.~J.~M.}\
  \bibnamefont {Horn}}\ and\ \bibinfo {author} {\bibfnamefont {R.~C.}\
  \bibnamefont {Lin}},\ }\bibfield  {title} {\bibinfo {title} {Periodic
  processes: A variational approach},\ }\href
  {https://doi.org/10.1021/i260021a005} {\bibfield  {journal} {\bibinfo
  {journal} {Industrial {\&} Engineering Chemistry Process Design and
  Development}\ }\textbf {\bibinfo {volume} {6}},\ \bibinfo {pages} {21}
  (\bibinfo {year} {1967})}\BibitemShut {NoStop}%
\bibitem [{\citenamefont {Ib{\'a}{\~{n}}ez}\ \emph {et~al.}(2024)\citenamefont
  {Ib{\'a}{\~{n}}ez}, \citenamefont {Dieball}, \citenamefont {Lasanta},
  \citenamefont {Godec},\ and\ \citenamefont
  {Rica}}]{heating_cooling_asymmetry}%
  \BibitemOpen
  \bibfield  {author} {\bibinfo {author} {\bibfnamefont {M.}~\bibnamefont
  {Ib{\'a}{\~{n}}ez}}, \bibinfo {author} {\bibfnamefont {C.}~\bibnamefont
  {Dieball}}, \bibinfo {author} {\bibfnamefont {A.}~\bibnamefont {Lasanta}},
  \bibinfo {author} {\bibfnamefont {A.}~\bibnamefont {Godec}},\ and\ \bibinfo
  {author} {\bibfnamefont {R.~A.}\ \bibnamefont {Rica}},\ }\bibfield  {title}
  {\bibinfo {title} {Heating and cooling are fundamentally asymmetric and
  evolve along distinct pathways},\ }\href
  {https://doi.org/10.1038/s41567-023-02269-z} {\bibfield  {journal} {\bibinfo
  {journal} {Nature Physics}\ }\textbf {\bibinfo {volume} {20}},\ \bibinfo
  {pages} {135} (\bibinfo {year} {2024})}\BibitemShut {NoStop}%
\bibitem [{\citenamefont {Esposito}\ \emph {et~al.}(2009)\citenamefont
  {Esposito}, \citenamefont {Lindenberg},\ and\ \citenamefont {Van~den
  Broeck}}]{Seifert_CA_eff}%
  \BibitemOpen
  \bibfield  {author} {\bibinfo {author} {\bibfnamefont {M.}~\bibnamefont
  {Esposito}}, \bibinfo {author} {\bibfnamefont {K.}~\bibnamefont
  {Lindenberg}},\ and\ \bibinfo {author} {\bibfnamefont {C.}~\bibnamefont
  {Van~den Broeck}},\ }\bibfield  {title} {\bibinfo {title} {Universality of
  efficiency at maximum power},\ }\href
  {https://doi.org/10.1103/PhysRevLett.102.130602} {\bibfield  {journal}
  {\bibinfo  {journal} {Phys. Rev. Lett.}\ }\textbf {\bibinfo {volume} {102}},\
  \bibinfo {pages} {130602} (\bibinfo {year} {2009})}\BibitemShut {NoStop}%
\bibitem [{\citenamefont {Abiuso}\ and\ \citenamefont
  {Perarnau-Llobet}(2020)}]{low_dissipation_2020}%
  \BibitemOpen
  \bibfield  {author} {\bibinfo {author} {\bibfnamefont {P.}~\bibnamefont
  {Abiuso}}\ and\ \bibinfo {author} {\bibfnamefont {M.}~\bibnamefont
  {Perarnau-Llobet}},\ }\bibfield  {title} {\bibinfo {title} {Optimal cycles
  for low-dissipation heat engines},\ }\href
  {https://doi.org/10.1103/PhysRevLett.124.110606} {\bibfield  {journal}
  {\bibinfo  {journal} {Phys. Rev. Lett.}\ }\textbf {\bibinfo {volume} {124}},\
  \bibinfo {pages} {110606} (\bibinfo {year} {2020})}\BibitemShut {NoStop}%
\bibitem [{\citenamefont {Barros}\ \emph {et~al.}(2025)\citenamefont {Barros},
  \citenamefont {Whitelam}, \citenamefont {Ciliberto},\ and\ \citenamefont
  {Bellon}}]{CilibertoPRE2025}%
  \BibitemOpen
  \bibfield  {author} {\bibinfo {author} {\bibfnamefont {N.}~\bibnamefont
  {Barros}}, \bibinfo {author} {\bibfnamefont {S.}~\bibnamefont {Whitelam}},
  \bibinfo {author} {\bibfnamefont {S.}~\bibnamefont {Ciliberto}},\ and\
  \bibinfo {author} {\bibfnamefont {L.}~\bibnamefont {Bellon}},\ }\bibfield
  {title} {\bibinfo {title} {Learning efficient erasure protocols for an
  underdamped memory},\ }\href {https://doi.org/10.1103/PhysRevE.111.044114}
  {\bibfield  {journal} {\bibinfo  {journal} {Phys. Rev. E}\ }\textbf {\bibinfo
  {volume} {111}},\ \bibinfo {pages} {044114} (\bibinfo {year}
  {2025})}\BibitemShut {NoStop}%
\bibitem [{\citenamefont {Luchinsky}\ \emph {et~al.}(1999)\citenamefont
  {Luchinsky}, \citenamefont {Mannella}, \citenamefont {McClintock},\ and\
  \citenamefont {Stocks}}]{Luchinsky1999}%
  \BibitemOpen
  \bibfield  {author} {\bibinfo {author} {\bibfnamefont {D.}~\bibnamefont
  {Luchinsky}}, \bibinfo {author} {\bibfnamefont {R.}~\bibnamefont {Mannella}},
  \bibinfo {author} {\bibfnamefont {P.}~\bibnamefont {McClintock}},\ and\
  \bibinfo {author} {\bibfnamefont {N.}~\bibnamefont {Stocks}},\ }\bibfield
  {title} {\bibinfo {title} {Stochastic resonance in electrical circuits. i.
  conventional stochastic resonance},\ }\href
  {https://doi.org/10.1109/82.793710} {\bibfield  {journal} {\bibinfo
  {journal} {IEEE Transactions on Circuits and Systems II: Analog and Digital
  Signal Processing}\ }\textbf {\bibinfo {volume} {46}},\ \bibinfo {pages}
  {1205} (\bibinfo {year} {1999})}\BibitemShut {NoStop}%
\bibitem [{\citenamefont {Freitas}\ \emph {et~al.}(2020)\citenamefont
  {Freitas}, \citenamefont {Delvenne},\ and\ \citenamefont
  {Esposito}}]{CircuitsEsposito2020}%
  \BibitemOpen
  \bibfield  {author} {\bibinfo {author} {\bibfnamefont {N.}~\bibnamefont
  {Freitas}}, \bibinfo {author} {\bibfnamefont {J.-C.}\ \bibnamefont
  {Delvenne}},\ and\ \bibinfo {author} {\bibfnamefont {M.}~\bibnamefont
  {Esposito}},\ }\bibfield  {title} {\bibinfo {title} {Stochastic and quantum
  thermodynamics of driven rlc networks},\ }\href
  {https://doi.org/10.1103/PhysRevX.10.031005} {\bibfield  {journal} {\bibinfo
  {journal} {Phys. Rev. X}\ }\textbf {\bibinfo {volume} {10}},\ \bibinfo
  {pages} {031005} (\bibinfo {year} {2020})}\BibitemShut {NoStop}%
\bibitem [{\citenamefont {Solon}\ and\ \citenamefont
  {Horowitz}(2018)}]{min_fluctuations2018}%
  \BibitemOpen
  \bibfield  {author} {\bibinfo {author} {\bibfnamefont {A.~P.}\ \bibnamefont
  {Solon}}\ and\ \bibinfo {author} {\bibfnamefont {J.~M.}\ \bibnamefont
  {Horowitz}},\ }\bibfield  {title} {\bibinfo {title} {Phase transition in
  protocols minimizing work fluctuations},\ }\href
  {https://doi.org/10.1103/PhysRevLett.120.180605} {\bibfield  {journal}
  {\bibinfo  {journal} {Phys. Rev. Lett.}\ }\textbf {\bibinfo {volume} {120}},\
  \bibinfo {pages} {180605} (\bibinfo {year} {2018})}\BibitemShut {NoStop}%
\bibitem [{\citenamefont {Saha}\ and\ \citenamefont
  {Marathe}(2019)}]{Saha_2019}%
  \BibitemOpen
  \bibfield  {author} {\bibinfo {author} {\bibfnamefont {A.}~\bibnamefont
  {Saha}}\ and\ \bibinfo {author} {\bibfnamefont {R.}~\bibnamefont {Marathe}},\
  }\bibfield  {title} {\bibinfo {title} {Stochastic work extraction in a
  colloidal heat engine in the presence of colored noise},\ }\href
  {https://doi.org/10.1088/1742-5468/ab39d4} {\bibfield  {journal} {\bibinfo
  {journal} {Journal of Statistical Mechanics: Theory and Experiment}\ }\textbf
  {\bibinfo {volume} {2019}},\ \bibinfo {pages} {094012} (\bibinfo {year}
  {2019})}\BibitemShut {NoStop}%
\bibitem [{\citenamefont {Holubec}\ \emph {et~al.}(2020)\citenamefont
  {Holubec}, \citenamefont {Steffenoni}, \citenamefont {Falasco},\ and\
  \citenamefont {Kroy}}]{ViktorPRR2020}%
  \BibitemOpen
  \bibfield  {author} {\bibinfo {author} {\bibfnamefont {V.}~\bibnamefont
  {Holubec}}, \bibinfo {author} {\bibfnamefont {S.}~\bibnamefont {Steffenoni}},
  \bibinfo {author} {\bibfnamefont {G.}~\bibnamefont {Falasco}},\ and\ \bibinfo
  {author} {\bibfnamefont {K.}~\bibnamefont {Kroy}},\ }\bibfield  {title}
  {\bibinfo {title} {Active brownian heat engines},\ }\href
  {https://doi.org/10.1103/PhysRevResearch.2.043262} {\bibfield  {journal}
  {\bibinfo  {journal} {Phys. Rev. Res.}\ }\textbf {\bibinfo {volume} {2}},\
  \bibinfo {pages} {043262} (\bibinfo {year} {2020})}\BibitemShut {NoStop}%
\bibitem [{\citenamefont {Holubec}\ and\ \citenamefont
  {Marathe}(2020)}]{Rahul2020}%
  \BibitemOpen
  \bibfield  {author} {\bibinfo {author} {\bibfnamefont {V.}~\bibnamefont
  {Holubec}}\ and\ \bibinfo {author} {\bibfnamefont {R.}~\bibnamefont
  {Marathe}},\ }\bibfield  {title} {\bibinfo {title} {Underdamped active
  brownian heat engine},\ }\href {https://doi.org/10.1103/PhysRevE.102.060101}
  {\bibfield  {journal} {\bibinfo  {journal} {Phys. Rev. E}\ }\textbf {\bibinfo
  {volume} {102}},\ \bibinfo {pages} {060101} (\bibinfo {year}
  {2020})}\BibitemShut {NoStop}%
\bibitem [{\citenamefont {Di~Bello}\ \emph {et~al.}(2024)\citenamefont
  {Di~Bello}, \citenamefont {Majumdar}, \citenamefont {Marathe}, \citenamefont
  {Metzler},\ and\ \citenamefont {Rold\'an}}]{Rahul2024}%
  \BibitemOpen
  \bibfield  {author} {\bibinfo {author} {\bibfnamefont {C.}~\bibnamefont
  {Di~Bello}}, \bibinfo {author} {\bibfnamefont {R.}~\bibnamefont {Majumdar}},
  \bibinfo {author} {\bibfnamefont {R.}~\bibnamefont {Marathe}}, \bibinfo
  {author} {\bibfnamefont {R.}~\bibnamefont {Metzler}},\ and\ \bibinfo {author}
  {\bibfnamefont {E.}~\bibnamefont {Rold\'an}},\ }\bibfield  {title} {\bibinfo
  {title} {Brownian particle in a poisson-shot-noise active bath: Exact
  statistics, effective temperature, and inference},\ }\href
  {https://doi.org/https://doi.org/10.1002/andp.202300427} {\bibfield
  {journal} {\bibinfo  {journal} {Annalen der Physik}\ }\textbf {\bibinfo
  {volume} {536}},\ \bibinfo {pages} {2300427} (\bibinfo {year}
  {2024})}\BibitemShut {NoStop}%
\bibitem [{\citenamefont {Lee}\ \emph {et~al.}(2022)\citenamefont {Lee},
  \citenamefont {Lee}, \citenamefont {Kwon},\ and\ \citenamefont
  {Park}}]{HYParkPRL22}%
  \BibitemOpen
  \bibfield  {author} {\bibinfo {author} {\bibfnamefont {J.~S.}\ \bibnamefont
  {Lee}}, \bibinfo {author} {\bibfnamefont {S.}~\bibnamefont {Lee}}, \bibinfo
  {author} {\bibfnamefont {H.}~\bibnamefont {Kwon}},\ and\ \bibinfo {author}
  {\bibfnamefont {H.}~\bibnamefont {Park}},\ }\bibfield  {title} {\bibinfo
  {title} {Speed limit for a highly irreversible process and tight finite-time
  landauer's bound},\ }\href {https://doi.org/10.1103/PhysRevLett.129.120603}
  {\bibfield  {journal} {\bibinfo  {journal} {Phys. Rev. Lett.}\ }\textbf
  {\bibinfo {volume} {129}},\ \bibinfo {pages} {120603} (\bibinfo {year}
  {2022})}\BibitemShut {NoStop}%
\end{thebibliography}%

\end{document}